\documentclass[
    aip,
    jcp,
    amsmath,
    amssymb,
    reprint,
    floatfix,showkeys
]{revtex4-2}
\usepackage[utf8]{inputenc}
\usepackage{graphicx}
\usepackage{mathtools}
\usepackage{booktabs}
\usepackage{hyperref}
\usepackage{subcaption}
\usepackage{braket}
\hypersetup{hidelinks}
\graphicspath{{./figures/}}
\usepackage{placeins}
\usepackage{xcolor}

\AtBeginDocument{\setcitestyle{numbers,square}}

\makeatletter
\renewcommand{\p@subsection}{}
\renewcommand{\p@subsubsection}{}
\makeatother

\makeatletter

\renewcommand{\figurename}{Fig.}

\renewcommand{\fnum@figure}{\textbf{\figurename~\thefigure}}

\renewcommand{\@caption@fignum@sep}{\space}

\makeatother
\begin{document}

\title{Bogoliubov coupled-cluster theory for $\mathfrak{su}(2)$ Hamiltonians}

\author{Swarnamoy Ghosh}
\email[Author to whom correspondence should be addressed: ]{swarnamoy.ghosh@rice.edu}
\affiliation{Department of Chemistry, Rice University, Houston, Texas 77005, USA}

\author{Thomas M. Henderson}
\affiliation{Department of Chemistry, Rice University, Houston, Texas 77005, USA}
\affiliation{Department of Physics and Astronomy, Rice University, Houston, Texas 77005, USA}

\author{Gustavo E. Scuseria}
\email[Author to whom correspondence should be addressed: ]{guscus@rice.edu}
\affiliation{Department of Chemistry, Rice University, Houston, Texas 77005, USA}
\affiliation{Department of Physics and Astronomy, Rice University, Houston, Texas 77005, USA}

\begin{abstract}
Coupled-cluster theory is the method of choice for weakly correlated systems, but in strongly correlated systems where the mean-field reference is qualitatively poor, it can break down badly.  These failures can be ameliorated by using symmetry-broken mean-field references. For systems that spontaneously break spin symmetry, we might favor an unrestricted reference. In systems that instead break number symmetry, we turn to Bogoliubov coupled cluster theory. In this work, we apply Bogoliubov coupled-cluster to Hamiltonians with an underlying $\mathfrak{su}(2)$ algebra to study the pairing Hamiltonian and the spin XXZ and $J_1$--$J_2$ models by exploiting the correspondence between fermionic-pair operators and spin-$1/2$ operators.  For the pairing problem, the reference is the usual Bardeen--Cooper--Schrieffer (BCS) quasiparticle vacuum, while for spin systems we use the analogous spin-BCS reference, which breaks $S_z$ symmetry and provides a flexible starting point for strongly correlated regimes.  Correlation is incorporated through a hierarchy of coupled-cluster approximations. Beyond energies, we compute the response density matrices for the evaluation of spin--spin correlation functions for the XXZ and $J_1$--$J_2$ Heisenberg models, and the pairing parameter and particle-number variance for the pairing Hamiltonian.  We also derive the corresponding relaxed density matrices, including the effects of orbital relaxation. Benchmark calculations for the half-filled pairing Hamiltonian and spin models demonstrate that Bogoliubov
coupled cluster provides a systematically improvable wave-function-based approach for $\mathfrak{su}(2)$ Hamiltonians, including geometrically
frustrated spin systems requiring complex and noncollinear BCS references.
\end{abstract}


\maketitle

\section{Introduction}
The coupled cluster\cite{Coester1958,Cizek1966,Cizek1969,PurvisBartlett1982CCSD,Paldus1999,ShavittBartlett2009} (CC) method provides one of the most accurate and systematically improvable approaches for weakly correlated many-body systems, with a computational cost that scales polynomially with system size. Unfortunately, the coupled-cluster method fails for strongly correlated systems, where the reference wave function is not ideal.

Strongly correlated systems, such as spin-model Hamiltonians, have direct relevance to chemistry and condensed matter physics. Magnetic materials, transition-metal compounds, and molecular spin systems, where the essential physics is governed by competing spin interactions~\cite{Sharma2014FeSClusters,VanOosten1995La2CuO4}, can be modelled as spin systems. Despite their apparent simplicity, spin Hamiltonians can display strong quantum fluctuations, magnetic frustration, and competing ordered or
disordered phases. 
The study of quantum spin Hamiltonians has therefore motivated a broad range of complementary numerical and variational approaches. 

Exact diagonalization (ED) provides important benchmark data but is limited to small lattices due to the exponential growth of the Hilbert space.  The density-matrix renormalization group (DMRG) and related tensor-network methods have been remarkably successful in one-dimensional and quasi-one-dimensional systems, and have also been applied to frustrated two-dimensional models using cylinder geometries \cite{Schollwock2005,Schollwock2011,Gong2014J1J2DMRG,Sandvik2010,Jiang2012J1J2DMRG}.  Quantum Monte Carlo methods are powerful for many unfrustrated spin systems; variational quantum Monte Carlo (VQMC) and projected parton wave functions have therefore been widely used to study frustrated magnets and spin-liquid candidates \cite{Sandvik2010,Hu2013J1J2VMC,Morita2015J1J2VMC}.  However, frustrated Hamiltonians often suffer from the sign problem~\cite{Sandvik2010, TroyerWiese2005SignProblem}. Other approaches, including Schwinger-boson mean-field theory, spin-wave theory, and parton mean-field theories, provide useful physical pictures of magnetically ordered and spin-liquid phases \cite{Auerbach1994,Bauer2017,Sachdev1992}. 

The coupled-cluster method has also been developed extensively for quantum magnetism by Bishop, Farnell, and coworkers, and has been applied to a variety of Heisenberg and frustrated $J_1$--$J_2$ models \cite{Bishop1998J1J2,FarnellBishop2004,Richter2015J1J2CCM,Darradi2008J1J2CCM,Richter2015J1J2CCM,Farnell2001TriangleKagomeCCM,Bishop2009SquareTriangleCCM,Bishop2010KagomeSquareCCM,Farnell2011KagomeArbitrarySpin}. Cluster mean-field, hierarchical mean-field, and block-correlated approaches provide another route for treating strongly correlated lattice Hamiltonians \cite{Isaev2009HMFJ1J2,Ren2014ClusterMFJ1J2,JimenezHoyos2015CMF}. 

These methods have greatly advanced our understanding of spin Hamiltonians, but they also illustrate the difficulty of constructing a compact, systematically improvable wave-function ansatz that can describe both energies and correlation functions across weakly and strongly correlated regimes.

Previous work has shown that Bogoliubov coupled-cluster theory\cite{Henderson2014PRC,Signoracci2015BCCOpenShell} and symmetry-projected Bogoliubov coupled-cluster theory \cite{Qiu2019PBCC} can extend the coupled-cluster framework toward strongly correlated pairing Hamiltonians while retaining a systematically improvable exponential ansatz\cite{Henderson2014PRC,Qiu2019PBCC}. The Bogoliubov CC theory has also been developed for applications in nuclear physics, where a particle-number breaking Bogoliubov reference is used to describe open-shell nuclei. In particular, Signoracci, Duguet, Hagen, and Jansen formulated and applied ab initio Bogoliubov coupled-cluster (BCC) theory to open-shell nuclei, and more recent work by Tichai, Demol, and Duguet extended BCC calculations to open-shell Ca, Ni, and Sn isotopic chains \cite{Signoracci2015BCCOpenShell,Tichai2024HeavyMassBCC}. 

In this paper, we develop and benchmark BCC theory for $\mathfrak{su}(2)$ Hamiltonians. 
For the pairing Hamiltonian, we use the number-symmetry-broken Bardeen--Cooper--Schrieffer~\cite{Bardeen1957BCS} (BCS) wave
function as a reference and work at half filling. For the XXZ\cite{Yang1966XXZ,YangYang1966XXZChainI,YangYang1966XXZChainII,
Cuccoli1995XXZSquare,Bishop2017XXZSquareRevisited} and $J_1$--$J_2$ Heisenberg models, we use the analogous spin-BCS (sBCS)~\cite{Liu2023SpinAGP} wave function that breaks $S_z$ symmetry.  There is no fundamental distinction between these two references at the algebraic level: the spin-$1/2$ algebra is isomorphic to the seniority-zero pairing algebra, and the sBCS state is the spin representation of the same Bogoliubov/Nambu quasiparticle construction used in ordinary BCS~\cite{Nambu1960GaugeInvariance}. We consider CCD, CCSD, CCSDT, and CCSDTQ approximations to assess the systematic convergence of the Bogoliubov CC hierarchy.

In addition to ground-state energies, we compute response properties. For the spin Hamiltonians, we evaluate two-body correlation functions $\braket{\mathbf{S}_p \cdot \mathbf{S}_q}$, using coupled-cluster response equations. For the pairing Hamiltonian, we compute the pairing parameter and particle-number variance.  These observables provide more insights into the CC method, since accurate energies do not necessarily imply accurate reduced density matrices (RDMs) or property expectation values.  We also examine orbital relaxation effects for the pairing Hamiltonian.

The remainder of this paper is organized as follows.  In Sec.~2, we introduce the $\mathfrak{su}(2)$ spin--pairing algebra correspondence, define the BCS references, and formulate the Bogoliubov coupled-cluster equations.  We also describe the linear-response formalism and separately include the orbital-relaxation contribution to compute spin-correlation and pairing properties. In Sec.~3, we present numerical results for the half-filled reduced BCS Hamiltonian and for XXZ and $J_1$--$J_2$ spin models on square and triangular lattices. 
We compare the performance of different Bogoliubov CC truncation levels against mean-field, projected mean-field, and exact diagonalization benchmarks, where available.  Finally, Sec.~4 summarizes our conclusions.

\section{Theory} 
\subsection{$S_z$-Conserving Spin Hamiltonians}
A generic global $S_z$-conserving two-body spin Hamiltonian is given by
\begin{equation}
\label{Canonical_Ham}
\begin{split}
H_{S} = h_0 + \sum_p h^{010}_p S^z_p &+ \sum_{pq} h^{020}_{pq} S^z_pS^z_q
\\
&+  \sum_{pq} h_{pq}^{101} S^{+}_pS^{-}_q.
\end{split}
\end{equation}
By appropriately choosing the coefficients $h_0,h^{010}_p,h^{101}_{pq}$ and $h^{020}_{pq}$, a variety of Hamiltonians can be written within this formalism. For example, in the nearest-neighbor Heisenberg model, we would set $h_0 = h^{010}_p =0$ and $h^{020}_{pq} = h^{101}_{pq} = J/2$ for nearest neighbors $p$ and $q$ where the sums in Eq.~\eqref{Canonical_Ham} run over ordered pairs. Although Eq.~\eqref{Canonical_Ham} does not represent the most general $\mathfrak{su}(2)$ Hamiltonian, many models of physical interest can be written in this form.

The operators $S^{+}, S^{-}$ and $S^z$ obey the  $\mathfrak{su}(2)$ commutation algebra
\begin{subequations}
\begin{align}
\big[S^{+}_p,S^{-}_q\big] &= 2 \, \delta_{pq} \, S^z_p,
\\
\big[S^{z}_p,S^{\pm}_q] &= \pm\delta_{pq} \, S^{\pm}_p.
\end{align}
\end{subequations}
The  spin-raising and -lowering operators $S^{+}_p$ and $S^{-}_p$ act on local spin-1/2 basis as
\begin{subequations}
\begin{align}
S^{+}_p\ket{\downarrow}_p &= \ket{\uparrow}_p, \qquad S^{+}_p\ket{\uparrow}_p = 0,
\\
S^{-}_p\ket{\uparrow}_p &= \ket{\downarrow}_p, \qquad S^{-}_p\ket{\downarrow}_p = 0
\end{align}
\end{subequations}

For implementational simplicity, we work in the $\mathfrak{su}(2)$ pairing algebra representation instead. This representation is obtained by a Nambu particle--hole transformation\cite{Nambu1960GaugeInvariance}, under which the spin operators are mapped to pair operators as
\begin{subequations}
\begin{align}
S^{+}_p &\mapsto P_p^{\dagger},
\\
S^{-}_p &\mapsto P_p,
\\
S^{z}_p &\mapsto \frac{N_p-1}{2},
\end{align}
\end{subequations}
where we define pair-creation and annihilation operators $P_p^\dagger$ and $P_p$, while the number operator $N_p$ measures (twice) the number of pairs in the level.  These pair operators also satisfy an $\mathfrak{su}(2)$ algebra, with
\begin{subequations}
\begin{align}
\big[P^{\dagger}_p,P_q\big] &= \delta_{pq} \big(N_p - 1\big),
\\
\big[N_p,P_q] &= -2 \, \delta_{pq}\, P_q,
\\
\big[N_p,P^{\dagger}_q\big] &= 2 \, \delta_{pq} \, P^{\dagger}_p.
\end{align}
\end{subequations}

Using the above mapping, any expression in the spin representation can be written in the pairing representation, and vice versa.  Therefore, we can write a generic two-body $S_z$ symmetry conserving spin-Hamiltonian (Eq.\ref{Canonical_Ham}), in terms of a generic two-body number-conserving pairing Hamiltonian as
\begin{equation}
\begin{split}
\label{Pairing_Ham}
H = H_0 + \sum_p \tilde{h}_{p}^{010} \, N_{p} &+\sum_{pq} W_{pq} N_p \, N_q 
\\
&+ \sum_{pq} V_{pq} \, P^{\dagger}_p \, P_q.
\end{split}
\end{equation}
Note that the coefficients in Eq. \eqref{Pairing_Ham} and Eq. \eqref{Canonical_Ham} differ.

In this work, we consider short-range Heisenberg spin systems like the XXZ and $J_1-J_2$ models. The formalism, however, applies to any $\mathfrak{su}(2)$ Hamiltonian of the form Eq.\eqref{Canonical_Ham} (or Eq.\ref{Pairing_Ham}).

\subsection{Bardeen-Cooper-Schrieffer Ansatz}
The Bardeen-Cooper-Schrieffer (BCS) wave function can be written as
\begin{equation}
\ket{\mathrm{BCS}} = \prod_p\big(u_p + v_p^{*} \, P^{\dagger}_p\big) \ket{-}
\end{equation}
where $u_p$ and $v_p$ are the variational parameters with the normalization condition
\begin{equation}
|u_p|^2 + |v_p|^2 = 1
\end{equation}
and where we observe that
\begin{equation}
P_p|-\rangle =0.
\end{equation}
The BCS wave function, originally introduced to describe superconductivity in solids, has also been used to explain the large pairing gaps observed in even–even nuclei ~\cite{BohrMottelsonPines1958}. The BCS state breaks particle-number symmetry $N$. While this has little impact in macroscopic systems such as solids, it can strongly affect the correlation energy in finite systems such as nuclei. As a result, particle-number projection and related symmetry-restoration techniques within mean-field theories have become increasingly important in nuclear and quantum chemistry~\cite{Sheikh2021SymmetryRestoration,Scuseria2011ProjectedQP}. In this work, we study the $\mathfrak{su}(2)$ pairing Hamiltonian with BCS as a reference.

After the Nambu transformation to the spin representation, the BCS wave function becomes
\begin{equation}
 \ket{\mathrm{BCS}} = \prod_p \big(u_p + v^{*}_p \, S^{+}_p\big) \ket{\Downarrow}
\end{equation}
where $\ket{\Downarrow}$ is the spin ``vacuum'' state
\begin{equation}
\ket{\Downarrow} = \otimes_p \ket{\downarrow}_p
\end{equation}
and is annihilated by $S_p^-$. Each local spinor
$u_p\ket{\downarrow}_p+v_p^*\ket{\uparrow}_p$ can equivalently be
viewed as a local $\mathfrak{su}(2)$ rotation of $\ket{\downarrow}_p$. In other words, the local spins in BCS are oriented in arbitrary directions. This is
closely related to the model-state construction used in conventional
coupled-cluster treatments of quantum spin systems, where the local
spin axes are rotated so that all spins in the reference state point
along a common local $-z$ direction
\cite{FarnellBishop2004,Farnell2019NoncoplanarCCM}. Complex local rotations of this type have also been considered in
spin-lattice coupled-cluster calculations
\cite{Farnell2019NoncoplanarCCM}.

In the spin representation, the symmetry-broken BCS state is a superposition of configurations with different $S_z$ quantum numbers. Since the spin Hamiltonians considered here conserve $S_z$, the exact calculations are performed in the $S_z=0$ sector, while the BCS reference is constrained to have zero average magnetization, $\langle S_z \rangle=0$. In the pairing representation, the corresponding calculations are performed at fixed particle number $N=N_{0}$, with $N_{0}$ chosen at half filling, while the BCS reference is constrained to satisfy \(\langle N\rangle=N_0\).

\subsection{Bogoliubov Coupled Cluster Theory}
The BCS wave function is the vacuum state with respect to operators in a quasiparticle basis, which is obtained by the basis transformation\cite{Bogoliubov1958Superconductivity,Valatin1958Superconductivity,Anderson1958Pseudospin,Dukelsky2004RichardsonGaudin} 
\begin{subequations}
\label{BCS_transformation}
\begin{align}
P^{\dagger}_p &=  u_p \, v_p \, \left(1 - \mathcal{N}_p\right) + u_p^{2} \, \mathcal{P}_p^{\dagger} - v^{2}_p \, \mathcal{P}_p,
\\
N_p &= 2 \, v_p^{*} \, v_p + \big(u_p^{*} \, u_p - v_p^{*} \, v_p\big) \, \mathcal{N}_p
\\
&+ 2 \, u_p \, v_p^{*} \, \mathcal{P}^{\dagger}_p + 2 \, u_p^{*} \, v_p \, \mathcal{P}_p,
\nonumber
\end{align}
\end{subequations}
where the quasiparticle operators $\mathcal{P}_p^\dagger$, $\mathcal{P}_p$, and $\mathcal{N}_p$ respectively create a pair, destroy a pair, or measure the number of particles in the transformed level $p$.  We have
\begin{equation}
\mathcal{P}_p \ket{\mathrm{BCS}} = \mathcal{N}_p \ket{\mathrm{BCS}}  = 0.
\end{equation}

The Hamiltonian in Eq.\eqref{Pairing_Ham} can be expressed in this quasiparticle basis as 
\begin{equation}
\label{quasiparticle_Hamiltonian}
\begin{split}
     H = E_0 + \sum_{p} H^{010}_p{\mathcal{N}}_p + \sum_{p}\big( H^{100}_{p} {\mathcal{P}}^{\dagger}_p + H^{001}_p {\mathcal{P}}_p\big)   \\
+ \sum_{pq}H^{020}_{pq} {\mathcal{N}}_p{\mathcal{N}}_q + \sum_{pq} {H}^{101}_{pq} {\mathcal{P}}^{\dagger}_p{\mathcal{P}}_q \\+
\sum_{pq} \big(H^{200}_{pq} {\mathcal{P}}^{\dagger}_p{\mathcal{P}}^{\dagger}_q + H^{002}_{pq} {\mathcal{P}}_p{\mathcal{P}}_q\big)
\\+\sum_{pq} \big(H^{110}_{pq}{\mathcal{P}}^{\dagger}_{p} {\mathcal{N}}_{q} + H^{011}_{pq} {\mathcal{N}}_{p}{\mathcal{P}}_q \big).
\end{split}
\end{equation}
Our notation is not standard, but $H^{pqr}$ is meant to denote $p$ powers of $\mathcal{P}^\dagger$, $q$ powers of $\mathcal{N}$, and $r$ powers of $\mathcal{P}$, in the normal order $\mathcal{P}^\dagger \to \mathcal{N} \to \mathcal{P}$.

In this quasiparticle basis, the CC equations are easier to evaluate, and since this basis transformation is unitary in nature, the Hamiltonian in the quasiparticle basis still has the same original spectrum. In Eq.\eqref{quasiparticle_Hamiltonian}, most of the matrix elements are symmetric, but the order of the indices needs to be viewed with care in the non-symmetric matrices $H^{110}$ and $H^{011}$. The diagonal entries of $H^{200}$ and $H^{002}$ are undefined, since $\mathcal{P}^{\dagger}_p\mathcal{P}^{\dagger}_p = \mathcal{P}_p\mathcal{P}_p=0$, so we take them to be zero.  

Dynamic correlations beyond the mean-field $\ket{\mathrm{BCS}}$ state are incorporated via the Bogoliubov coupled-cluster (BCC) formalism. In BCC, as in general coupled cluster wave functions, the correlated wave function is written as
\begin{equation}
\ket{\Psi} = \mathrm{e}^T \ket{0},
\end{equation}
where $\ket{0}$ is the mean-field reference (in this case, $\ket{\mathrm{BCS}}$) and $T$ is the cluster operator, written generically as
\begin{equation}
T = \sum_{Q} T_{Q} = \sum_Q \sum_{q \in Q}  t_{q} \, \mathcal{A}_{q}^{\dagger},
\end{equation}
where $\mathcal{A}^{\dagger}_q$ creates one or more excitations out of the mean-field state, and $t_q$ is the amplitude of the excitation; $Q$ is the excitation rank. In our case, we have
\begin{subequations}
\begin{align}
T_1 &= \sum_{p}t_p \, \mathcal{P}^{\dagger}_p,
\\
T_2 &= \frac{1}{2!} \, \sum_{pq} t_{pq} \, \mathcal{P}^{\dagger}_p \, \mathcal{P}^{\dagger}_q,
\\
T_3 &= \frac{1}{3!} \, \sum_{pqr} t_{pqr} \, \mathcal{P}^{\dagger}_p \, \mathcal{P}^{\dagger}_q \, \mathcal{P}_r^{\dagger},
\\
T_4 &= \frac{1}{4!} \, \sum_{pqrs} t_{pqrs} \, \mathcal{P}^{\dagger}_p \, \mathcal{P}^{\dagger}_q \, \mathcal{P}_r^{\dagger} \, \mathcal{P}^{\dagger}_s,
\end{align}
\end{subequations}
where the amplitude coefficients are fully symmetric under interchange of indices (explaining the $1/k!$ in $T_k$) and where we choose amplitudes with repeated indices to be zero. In the spin representation, $\mathcal{P}_p^\dagger$ creates a spin flip
relative to the local reference spinor. Thus, $T_n$ contains
$n$-spin-flip clusters, and truncating the Bogoliubov cluster operator
at excitation rank $n$ is analogous to a SUB$n$-type truncation in
conventional spin coupled-cluster theory
\cite{FarnellBishop2004,Farnell2019NoncoplanarCCM}. Our truncation does not restrict the spatial extent of the retained
clusters and therefore differs from the commonly used LSUB$n$ hierarchy,
in which clusters are restricted to contiguous lattice sites. For
spin-$1/2$ systems, LSUB$n$ corresponds to the localized
SUB$n$--$n$ truncation
\cite{FarnellBishop2004,Farnell2019NoncoplanarCCM}.

Inserting the CC wave function into the Schr\"odinger equation, 
\begin{equation}
H \, \mathrm{e}^{T} \ket{0} = E_{\mathrm{CC}} \, \mathrm{e}^{T} \ket{0}
\end{equation}
and applying a similarity transformation to the Hamiltonian, we projectively obtain the CC energy $E_{\mathrm{CC}}$ and the amplitudes $t_{q}$ from
\begin{subequations}
\label{CC_Hbar}
\begin{align}
E_{\mathrm{CC}} &= \bra{0} \mathrm{e}^{-T} \, H \, \mathrm{e}^T \ket{0},
\\
0 &=R_{q} = \bra{0} \, \mathcal{A}_q \, \mathrm{e}^{-T} \, H \, \mathrm{e}^T \ket{0}.
\end{align}
\end{subequations}

With the cluster operator truncated at singles and doubles (CCSD), $T=T_1+T_2$, the energy reads
\begin{equation}
\label{cc_energy}
E_{\mathrm{CCSD}} =  E_{0} + \sum_{p} t_p H^{001}_p + \sum_{pq} \left(t_{pq} + t_p \, t_q\right) \, H^{002}_{pq}.
\end{equation}
Since $H$ is two-body in nature, the CC energy expression in Eq. \ref{cc_energy} is the same for any higher-order CC.

\subsubsection{Response properties and orbital relaxation}
\label{relax_response}

In this work, we calculate several observables in addition to energy using CC theory. Here, we explain the procedure to compute the observables using CC. 

Typically, properties in quantum mechanics are obtained from energy derivatives.  The dipole moment, for example, is the derivative of the energy with respect to the strength of an applied dipole field.  For variational methods, the Hellmann-Feynman theorem permits us to calculate such energy derivatives as expectation values.  Since the CC energy is not variational with respect to cluster amplitudes and $\bar{H} = e^{-T}He^{T}$, the Hellmann-Feynman theorem does not apply; instead, we must generalize it using the standard CC response formalism\cite{ShavittBartlett2009,Monkhorst1977CCProperties,KochJorgensen1990CCResponse}.

To understand this formalism, we define a Hamiltonian perturbed by an operator $O$ whose expectation value is desired,
\begin{equation}
H(x) = H_0 + x \, O ,
\end{equation}
where $H_0$ is the unperturbed piece of $H(x)$ and $x$ is the perturbation strength, and define a CC functional
\begin{align}
\mathcal{E}(x,C(x),T(x),Z(x))
   &= \bra{0(\mathbf{C})} \mathrm{e}^{-T(x)} \, H(x) \, \mathrm{e}^{T(x)} \ket{0(\mathbf{C})}
\nonumber
\\
   &+ \sum_\mu z_\mu(x) R_\mu(x).
\end{align}
Here we assume the reference state $|0\rangle$ depends explicitly on the orbital parameters $\mathbf{C}$ which in turn depends on $x$.  Equivalently, the functional can be written as
\begin{equation}
\mathcal{E} = \bra{0(\mathbf{C})} (1+Z(x)) \, \mathrm{e}^{-T(x)} \, H(x) \, \mathrm{e}^{T(x)} \ket{0(\mathbf{C})},
\end{equation}
where
\begin{equation}
Z(x) = \sum_\mu z_\mu(x) \, \mathcal{A}_\mu(x)
\end{equation}
is a de-excitation operator acting to the right, and thus an excitation operator acting on the left. 

Stationarity of $\mathcal{E}$ with respect to the $Z$ amplitudes gives the ordinary CC equations,
\begin{equation}
\left.\frac{\partial \mathcal{E}}{\partial z_\mu}\right|_{x=0} = R_\mu = 0 .
\end{equation}
The $Z$ amplitudes are determined by requiring stationarity with respect to the cluster amplitudes,
\begin{equation}
\left.\frac{\partial \mathcal{E}}{\partial t_\mu}\right|_{x=0} = \bra{0} \left(1+Z\right) \, [\bar{H},A_\mu^\dagger] \ket{0} = 0,
\label{eq:z_equations}
\end{equation}
where $\mathcal{A}_{\mu}^{\dagger}$ is the excitation operator associated with the amplitude $t_\mu$.  Equation \eqref{eq:z_equations} defines the $Z$ equations.

The first-order derivative of $\mathcal{E}$ with respect to $x$ can be expressed using the chain-rule
\begin{equation}
\label{dL_dx}
\frac{\mathrm{d}\mathcal{E}}{\mathrm{d}x}
    =   \frac{\partial\mathcal{E}}{\partial x}
    +   \sum_\mu \frac{\partial \mathcal{E}}{\partial t_\mu} \, \frac{\mathrm{d}t_\mu}{\mathrm{d}x}
    +   \sum_\mu \frac{\partial \mathcal{E}}{\partial z_\mu} \, \frac{\mathrm{d}z_\mu}{\mathrm{d}x}
    +   \sum_a \frac{\partial \mathcal{E}}{\partial C_a} \, \frac{\mathrm{d} C_a}{\mathrm{d} x}.
\end{equation}
Satisfaction of the $T$ and $Z$ equations means stationarity of $\mathcal{E}$ with respect to $z$ and $t$ so Eq.\eqref{dL_dx} reduces to
\begin{equation}
\label{response_equation}
\frac{\mathrm{d}\mathcal{E}}{\mathrm{d}x} = \frac{\partial\mathcal{E}}{\partial x} +  \sum_a \frac{\partial \mathcal{E}}{\partial C_a} \, \frac{\mathrm{d} C_a}{\mathrm{d} x},
\end{equation}
where the last term accounts for the orbital dependence of the coupled cluster energy.  Thus, this $Z$-vector formulation removes the need to compute $\mathrm{d}t/\mathrm{d}x$ or $\mathrm{d}z/\mathrm{d}x$.  Equivalently, one does not need to solve the CC equations separately for each perturbed Hamiltonian $H(x)$ in order to obtain the response property.

Note that $\mathcal{E} = E_{\mathrm{CC}}$ after the cluster equations are solved. If we freeze the reference $|0\rangle$ so that the orbital relaxation term vanishes, we get the CC response expectation value of the operator $O$ at $x=0$:
\begin{equation}
\label{analytic_response}
    \left.\frac{\mathrm{d}\mathcal{E}}{\mathrm{d}x}\right|_{x=0} = \bra{0} \left(1+Z\right) \, \mathrm{e}^{-T} \, O \, \mathrm{e}^T \ket{0}.
\end{equation}

This expression is used to construct the response reduced density matrices (RDM). 
In practice, these RDMs are obtained by inserting the elementary operators appearing in the Hamiltonian in Eq. \eqref{Pairing_Ham} into Eq. \eqref{analytic_response}. 
For example, the one-body response RDMs are defined as
\begin{subequations}
\begin{align}
\label{Response_RDMS}
(\gamma^{010}_{p})_{res} &= \bra{0} \left(1+Z\right) \, \mathrm{e}^{-T} \, N_p \, \mathrm{e}^T \ket{0},
\\
(\gamma^{100}_{p})_{res} &= \bra{0} \left(1+Z\right) \, \mathrm{e}^{-T} \, P_p^\dagger \, \mathrm{e}^T \ket{0},
\\
(\gamma^{001}_{p})_{res} &= \bra{0} \left(1+Z\right) \, \mathrm{e}^{-T} \, P_p \, \mathrm{e}^T \ket{0}.
\end{align}
\end{subequations}
Note that because the BCC wave function breaks number symmetry, we have non-zero symmetry-breaking density matrices such as $\gamma^{001}_p$.

In this work, we use this linear response procedure to evaluate spin--spin correlation functions for the XXZ and $J_1-J_2$ Hamiltonians, and the pairing parameter $\Delta$ and particle-number variance for the pairing Hamiltonian.

The response RDMs defined above are unrelaxed with respect to the reference.  That is, they include the response of the cluster amplitudes through the $Z$ vector, but they do not include the response of the reference itself.   If the reference orbitals or quasiparticle parameters are allowed to change with the perturbation, the last term in Eq.~\eqref{response_equation} gives an additional orbital-relaxation contribution,
\begin{equation}
\left.\frac{\mathrm{d}\mathcal{E}}{\mathrm{d}x}\right|
= \left. \frac{\partial\mathcal{E}}{\partial x} \right|_{T,Z,C} 
+ \sum_a \frac{\partial \mathcal{E}}{\partial C_a} \, \frac{\mathrm{d} C_a}{\mathrm{d} x}.
\label{eq:relaxed_response}
\end{equation}
The first term is the unrelaxed response contribution of Eq.~\eqref{analytic_response} at fixed $T,Z$ and $\mathbf{C}$, while the second term accounts for the first-order change of the reference induced by the perturbation. The orbital response $\mathrm{d} \mathbf{C}/\mathrm{d} x$ is obtained from the coupled-perturbed BCS equations, obtained by differentiating the mean-field stationarity conditions with respect to the perturbation strength.  The explicit derivation of these equations is given in Appendix~\ref{relaxed_RDMs}.

\section{Results}
In this section, we present the CC energies and response properties for the reduced BCS Hamiltonian, XXZ, and $J_1-J_2$ Heisenberg spin models. These Hamiltonians can be used to model certain physical systems \cite{Sharma2014FeSClusters,VanOosten1995La2CuO4} and also exhibit spin frustration and strong pairing physics. 

In this work, we
constrain the BCS to have zero average magnetization,
$\langle S_z\rangle=0$, corresponding to half filling in the pairing
representation, while allowing the wave function to break $S_z$
symmetry. The AGP results are obtained by variation after
projection \cite{ JimenezHoyos2012PHF, Scuseria2011ProjectedQP}(VAP), in which the projected energy is minimized directly. We solve both the CC amplitude equations and the left-state ($Z$-vector) equations using a modified Broyden method,\cite{Johnson1988ModifiedBroyden,Baran2008BroydenNuclear,Qiu2018ProjectedCCOptimization} with convergence defined by a maximum residual smaller than $10^{-10}$. The amplitude equations were obtained by using the symbolic algebra package \texttt{drudge} and its accompanying
automatic code generation facility \texttt{gristmill}\cite{ZhaoScuseriaDrudgeGristmill}.  

We also report the orbital-relaxed RDMs for the pairing model. For the spin models, the orbital relaxation contribution to the spin--spin correlation functions vanishes to numerical precision by the symmetry of the BCS references considered here, with the relaxed--response differences of order $10^{-11}$. We therefore report unrelaxed response correlations for spin models.

\subsection{The Pairing Hamiltonian}
The reduced BCS Hamiltonian is a Hamiltonian consisting of electron pair creation, pair annihilation, and number operators, and can be written as 
\begin{equation}
\label{eq:pairing_ham}
 H = \sum_p \epsilon_p \, N_p - G \, \sum_{pq} P^{\dagger}_p \, P_q
\end{equation}
where $\epsilon_p$ are the single-particle energy levels (we take $\epsilon_p = p$), and $G$ is the interaction strength. Although the Hamiltonian has number symmetry, for $G$ greater than a critical value $G_c$, number symmetry spontaneously breaks at the mean-field level. We thus add a chemical potential $\lambda$ so that we work with the grand potential $H - \lambda \, \left(N - N_0\right)$ where $N_0$ is the target number. 

In our case, we work at half filling which simplifies the calculation significantly. Generally speaking, the chemical potential $\lambda$ that gives the right average particle number at the mean-field level does not do so at the coupled cluster level. At half filling, the Hamiltonian has a particle-hole symmetry \cite{Hirsch2002AnnPhys} which fixes the choice of chemical potential around the center of the spectrum,
\begin{equation}
\lambda=\frac{\epsilon_{\min}+\epsilon_{\max}-G}{2}.
\end{equation}
Since this value is symmetry-determined, both the BCS reference and CC wave function built upon it use the same chemical potential. 

The reduced BCS Hamiltonian, originally formulated as a phenomenological model for superconductivity in solids \cite{Bardeen1957BCS}, can be solved exactly with a form of a Bethe ansatz \cite{Richardson1963, Richardson1966}, for any choice of $\epsilon_p$ and $G$. This allows us to generate exact results for systems far larger than are accessible via exact diagonalization. 

\begin{figure}[!htbp]
\includegraphics[width=\columnwidth]{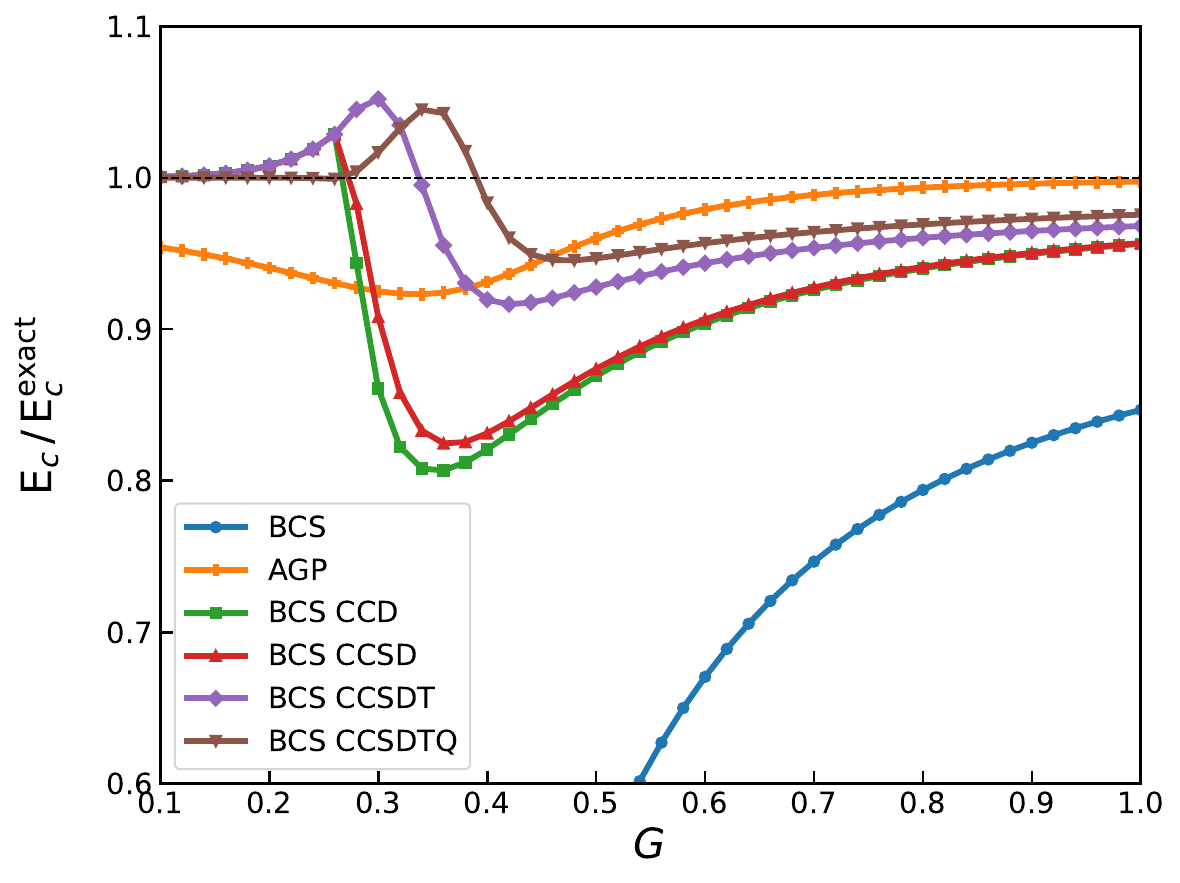}
\caption{Fraction of correlation energy recovered in the half-filled pairing Hamiltonian with 20 levels. The CC methods are based on the BCS reference. In this case, $G_c \approx 0.268$.}
\label{fig:pairing_20_Ec}
\end{figure}

Figure \ref{fig:pairing_20_Ec} shows the fraction of correlation energy recovered for the half-filled pairing Hamiltonian with 20 levels. For $G<G_c$, the optimized BCS preserves the number symmetry, or in other words, it reduces to a Hartree-Fock (HF) state. Hence, it does not recover any correlation energy below $G_c$. The BCS reference breaks number symmetry spontaneously at the critical $G_c\approx 0.268$. Above $G_c$, the BCS correlation energy increases smoothly with $G$.

One interesting point to note is that when the optimized BCS reduces to a HF solution, odd quasiparticle excitations would change the particle number and thus do not contribute: their coefficients are zero by symmetry. Hence, CCD, CCSD, and CCSDT yield identical energies. Above $G_c$, odd excitations are allowed, and the CCD, CCSD, and CCSDT curves split.

Note that CCD and CCSD display a pronounced dip in the recovered correlation energy, reaching a minimum near $G\approx 0.35$.  This indicates that the low-order excitations are insufficient to describe the rapid change in correlation structure following the onset of number-symmetry breaking. The inclusion of higher-rank excitations substantially improves the description. Although both CCSDT and CCSDTQ slightly overcorrelate in the transition region, the maximum overcorrelation occurs at different interaction strengths for CCSDT and CCSDTQ, suggesting that triples and quadruples redistribute the correlation energy differently across the symmetry-breaking crossover. 

The derivative discontinuity in the curve in Figure \ref{fig:pairing_20_Ec} is exaggerated by the way in which the data is plotted. The transition from HF to BCS is second-order in nature, while the transition in the correlated result is first-order. The exact energy is of course smooth, and as the truncated CC becomes more complete, the derivative discontinuity in the total energy decreases. The plots showing these discontinuities are provided in the supplementary material.

\begin{figure}[!htbp]
\includegraphics[width=\columnwidth]{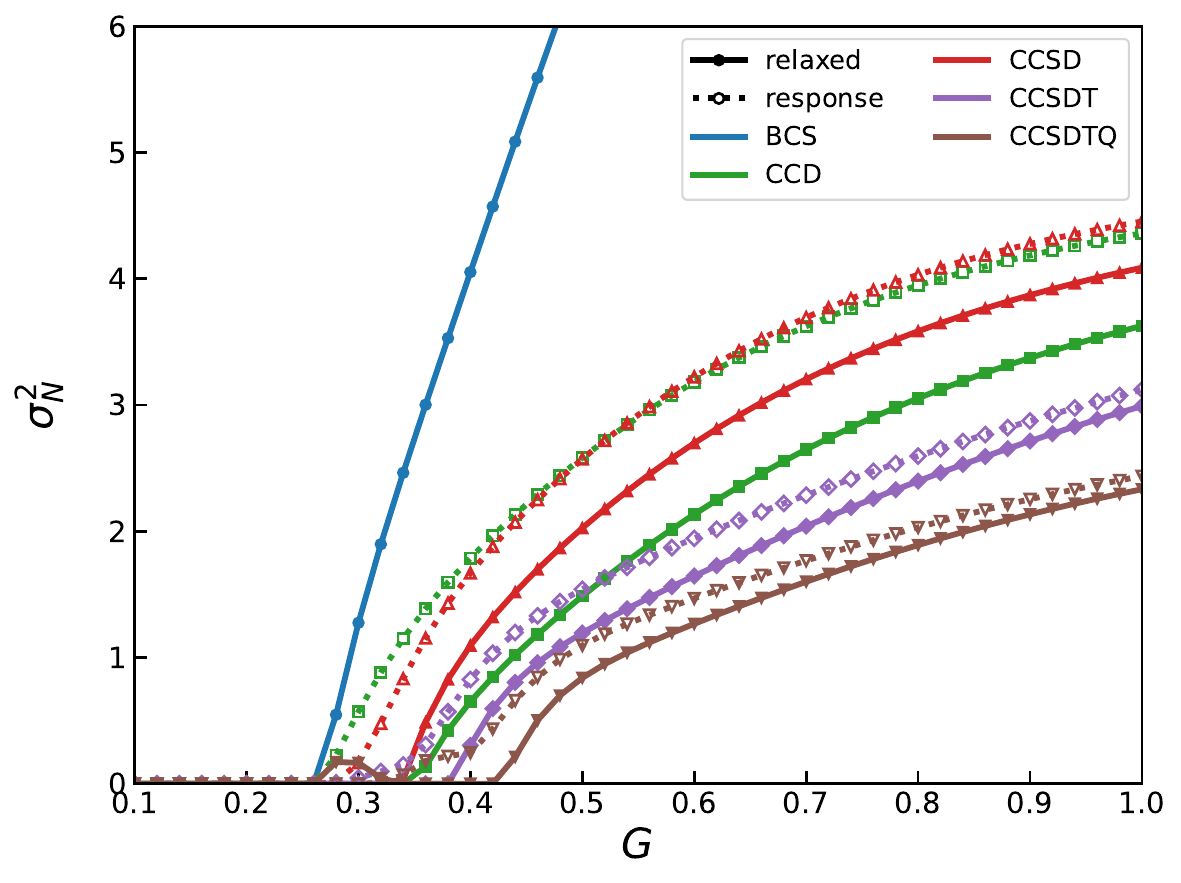}
\caption{Fluctuations in particle number in the BCS and BCS-based CC wave functions for the half-filled pairing Hamiltonian
with 20 levels. The response and relaxed results are shown in dashed and solid lines, respectively.}
\label{fig:pairing_20_Variance}
\end{figure} 

We also evaluate properties using the procedure outlined in Sec.~\ref{relax_response}. We report results obtained using both linear-response and orbital-relaxed RDMs. The relaxed and the response numbers for each CC method are denoted with solid  and dashed lines, respectively.
Figure \ref{fig:pairing_20_Variance} shows the particle number fluctuation $\sigma_N^2 = \braket{N^2} - \braket{N}^2$. Below $G_c$, the BCS reference preserves the number symmetry, and the particle-number fluctuation is zero.  For $G\ge G_c$, the BCS reference breaks particle-number symmetry, leading to a rapid increase in $\sigma^2_N$. Correlating BCS with CC substantially reduces this number fluctuation, indicating that the correlated wave function tends to restore the fixed particle number of the exact state.  The reduction becomes more pronounced as higher excitation ranks are included, with CCSDTQ giving the smallest particle-number fluctuation among the methods considered. Including orbital relaxation further lowers $\sigma_N^2$ relative to the corresponding unrelaxed response values. It is more pronounced for CCSDT and CCSDTQ, where the fluctuation remains nearly zero over a finite range of $G>G_c$, despite BCS breaking the number symmetry. As $G$ increases further, the variance grows smoothly for all methods.   The BCS reference has by far the largest number fluctuation, while CC suppresses the fluctuation as higher excitation ranks are included. 

However, the relaxed and response curves for CCD and CCSD exhibit a peculiar trend. The CCD and CCSD responses are very similar, except near $G_c$. However, the difference between the CCSD response and CCSD relaxed is small compared to CCD, and the CCSD relaxed is worse than the CCD relaxed. We attribute this to the Thouless theorem,\cite{Thouless1960} which shows that single excitations amount to a change of mean-field reference.  This means that once we include single excitations, the results in CC theory are more weakly dependent on the reference wave function.  The upshot is that the CCSD relaxed properties should be close to the CCSD response properties, while this is not necessarily true of CCD.

We must mention a slight difficulty for $G \gtrsim G_c$: the non-Hermitian CC form of the expectation value means that expectation values of positive-definite operators are not guaranteed to be positive, so we may have $\braket{N^2} - \braket{N}^2 < 0$ in the immediate vicinity of the point of symmetry breaking. 


\begin{figure}[!htbp]
    \centering

    \begin{subfigure}{\columnwidth}
        \centering
        \includegraphics[width=\linewidth]{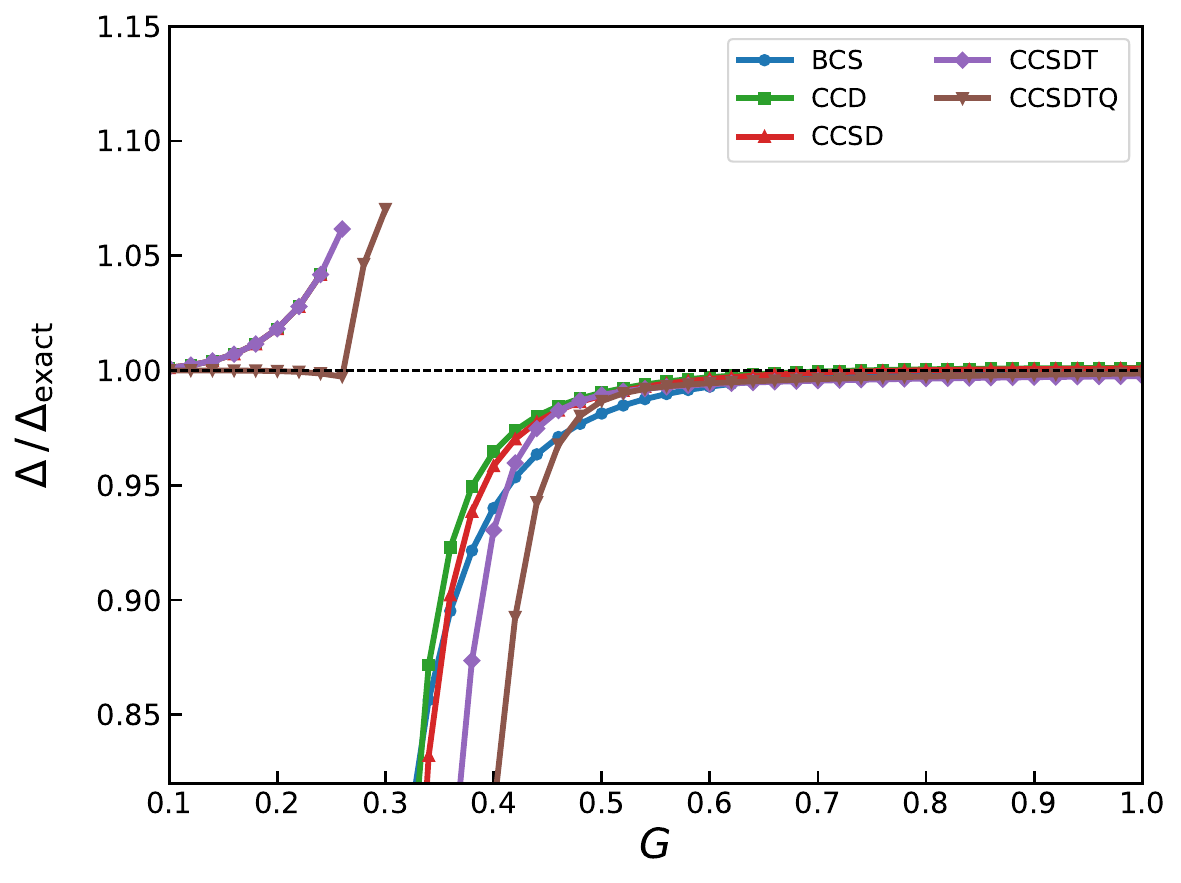}
        \caption{}
        \label{fig:pairing_20_delta_ratio_relaxed}
    \end{subfigure}

    \vspace{0.5em}

    \begin{subfigure}{\columnwidth}
        \centering
        \includegraphics[width=\linewidth]{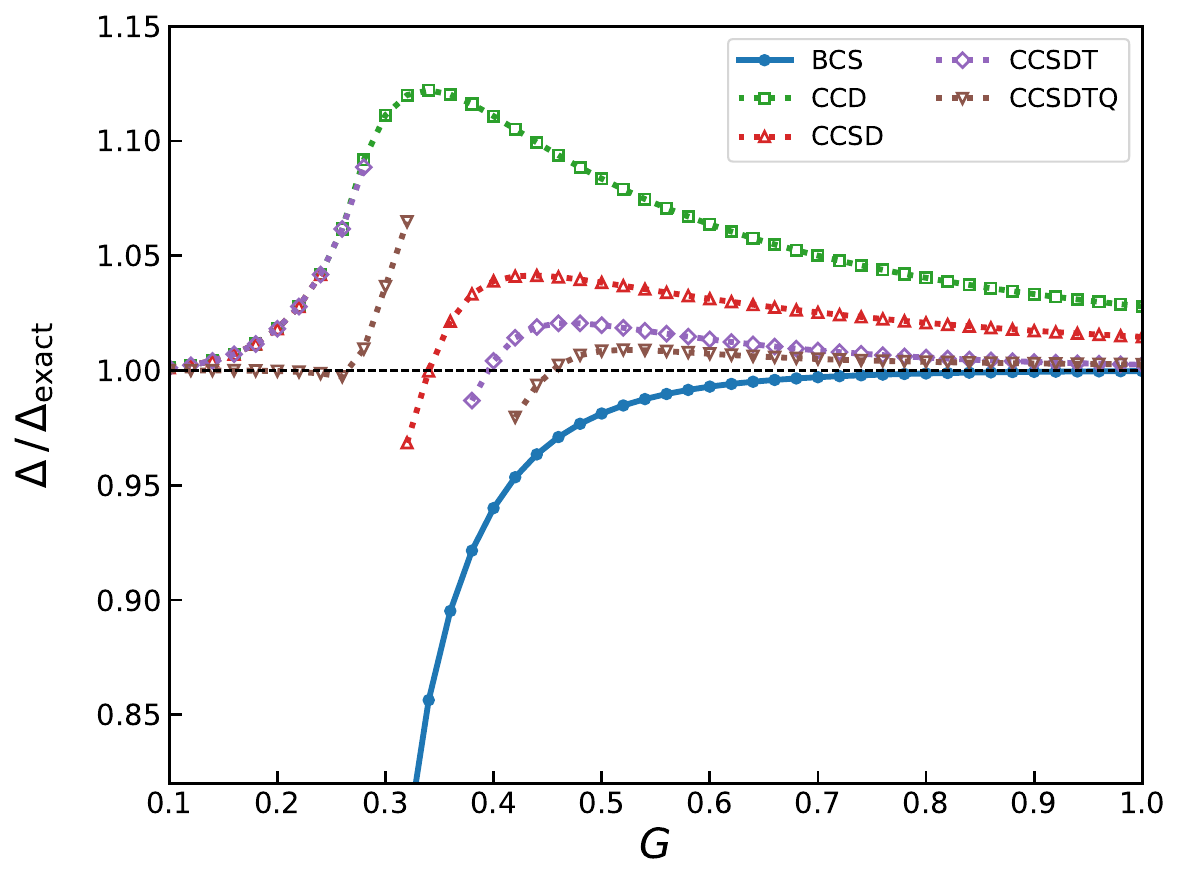}
        \caption{}
        \label{fig:pairing_20_delta_ratio_response}
    \end{subfigure}

    \caption{The pairing parameter $\Delta$ for the half-filled pairing Hamiltonian with 20 levels evaluated using (a) relaxed and (b) response density matrices.}
    \label{fig:pairing_20_delta_ratio}
\end{figure}
 
We also evaluate the pairing parameter $\Delta$ for the CC wave functions. We use the pairing parameter \cite{Braun1998FixedN,VonDelft1996ParityAffected}
\begin{equation}
\label{Delta_gap}
\Delta = G\sum_p C_p
\end{equation}
where 
\begin{equation}
C_p^2 = \braket{P^{\dagger}_p \, P_p} - \frac{1}{4} \, \braket{N_p}^2 = \frac{1}{2} \, \braket{N_p} - \frac{1}{4} \, \braket{N_p}^2.
\end{equation}
For the BCS case, pairing parameter $\Delta$ reduces to the usual BCS superconducting gap  $\Delta_{\text{BCS}} = G \, \sum u_p \, v_p$, and this gap vanishes in the number-symmetry-preserving sector. However, for correlated wave functions, this gap is usually non-zero.

Figure~\ref{fig:pairing_20_delta_ratio} shows the pairing parameter ratio  $\Delta/\Delta_{\mathrm{exact}}$, for the half-filled pairing Hamiltonian.  The BCS reference remains in the number-symmetry-preserving phase below $G_c$, and the BCS gap is therefore zero in this region. Above $G_c$, BCS breaks particle-number symmetry, and the gap becomes finite. For clarity, points affected by reference-transition artifacts near the BCS symmetry-breaking point have been omitted from the plot. 

The response values generally overestimate the $\Delta_{\text{exact}}$, whereas the orbital-relaxed values generally underestimate it. Thus, orbital relaxation partially counteracts the overestimation of the response, but can slightly overcorrect $\Delta$ in the intermediate-coupling regime.  At larger $G$, both response and relaxed results become smoother, with the CCSDT and CCSDTQ curves remaining closest to unity.

\subsection{The XXZ Hamiltonian}
The XXZ Heisenberg Hamiltonian is given by
\begin{equation}
H_\mathrm{XXZ} =\sum_{\langle pq\rangle} \Big(S^x_{p} \, S^{x}_{q} + S^y_p \, S^y_q + \Delta \, S^{z}_{p} \, S^{z}_{q} \Big )
\end{equation}
where the notation $\langle pq\rangle$ denotes nearest-neighbor pairs. The physics of the XXZ model depends on the parameter $\Delta$, and we can study different phases of the XXZ model by tuning the value of $\Delta$.

This Hamiltonian has global $S_z$ symmetry, and we choose to perform the exact calculations in the $S_z=0$ sector. For the symmetry-broken BCS reference, we impose $\langle S_z \rangle=0$, corresponding to zero average magnetization.

\subsubsection{Square lattice}
In the thermodynamic limit, the square XXZ model exhibits three distinct phases:\cite{Yang1966XXZ,YangYang1966XXZChainI,YangYang1966XXZChainII,YangYang1966XXZChainIII,Liu2023SpinAGP,MikeskaKolezhuk2004OneDimensional} a ferromagnetic phase for $\Delta \le -1$, a gapless XY phase for $-1\le \Delta \le 1$, and an antiferromagnetic (AFM) phase for $\Delta > 1$.

In the square lattice, the $S_z = 0$ sector is the ground state for $\Delta > -1$. At $\Delta = -1$, the ground states of different $S_z$ sectors are degenerate, while for $\Delta < -1$ the ground state has maximal $S_z$ instead\cite{Liu2023SpinAGP,Bishop2017XXZSquareRevisited}.

\begin{figure*}[t]
\begin{subfigure}{0.48\textwidth}
    \includegraphics[width=\linewidth]{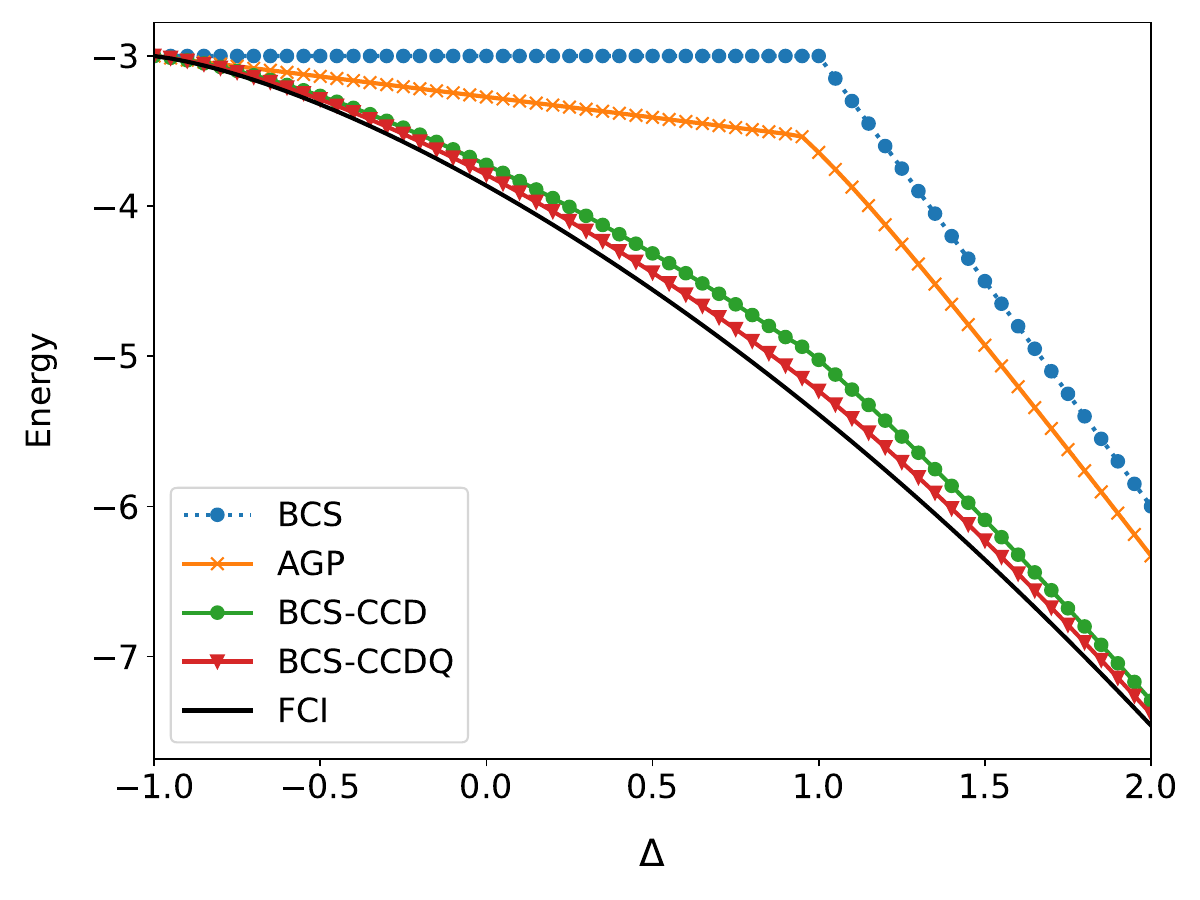}
    \caption{}
    \label{fig:myplot1a}
\end{subfigure}
\hfill
\begin{subfigure}{0.48\textwidth}
    \includegraphics[width=\linewidth]{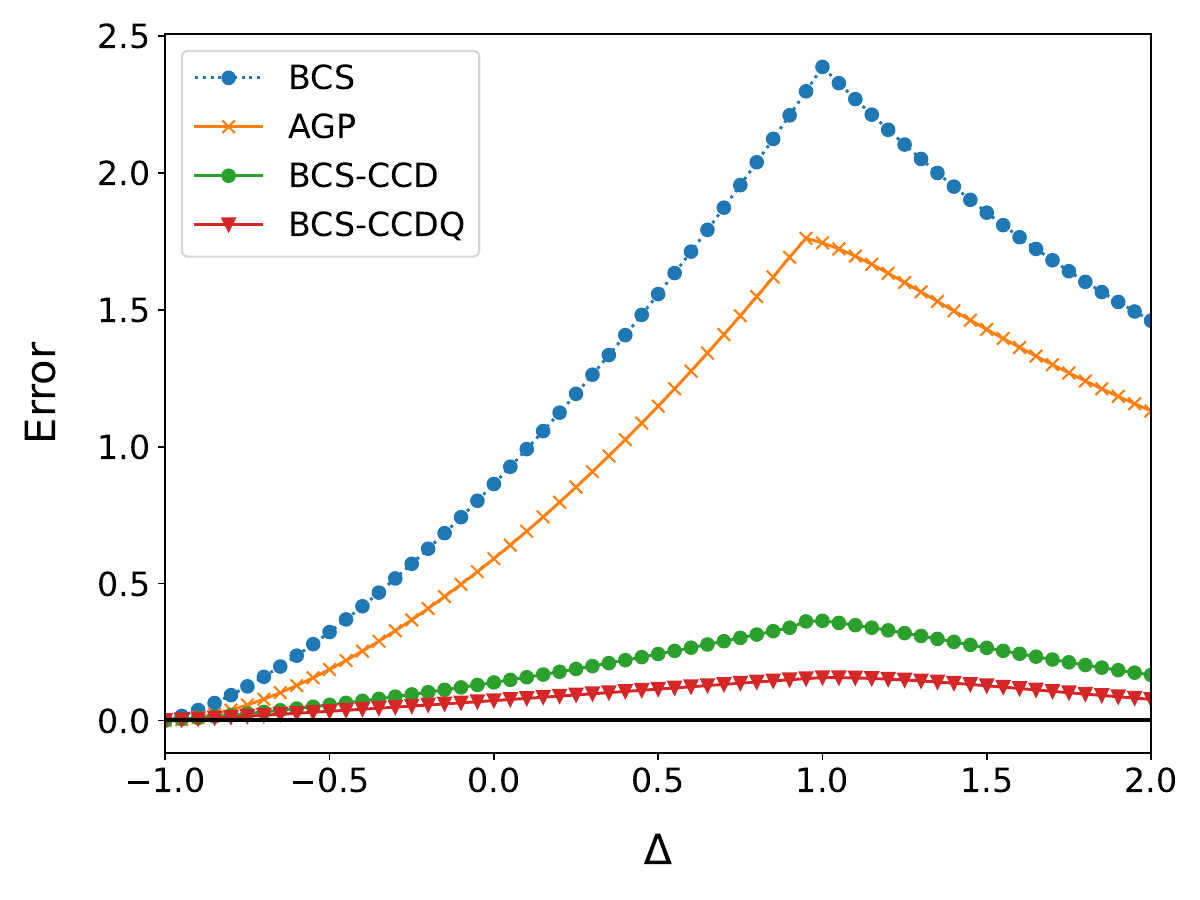}
    \caption{}
    \label{fig:myplot1b}
\end{subfigure}
\caption{Total energy and energy error for the 12-site one-dimensional XXZ model with periodic boundary conditions. (a) Total energy. (b) Deviation of the total energy from the exact energy.}
\label{fig:myplot1}
\end{figure*}

\begin{figure*}[t]
\begin{subfigure}{0.48\textwidth}
    \includegraphics[width=\linewidth]{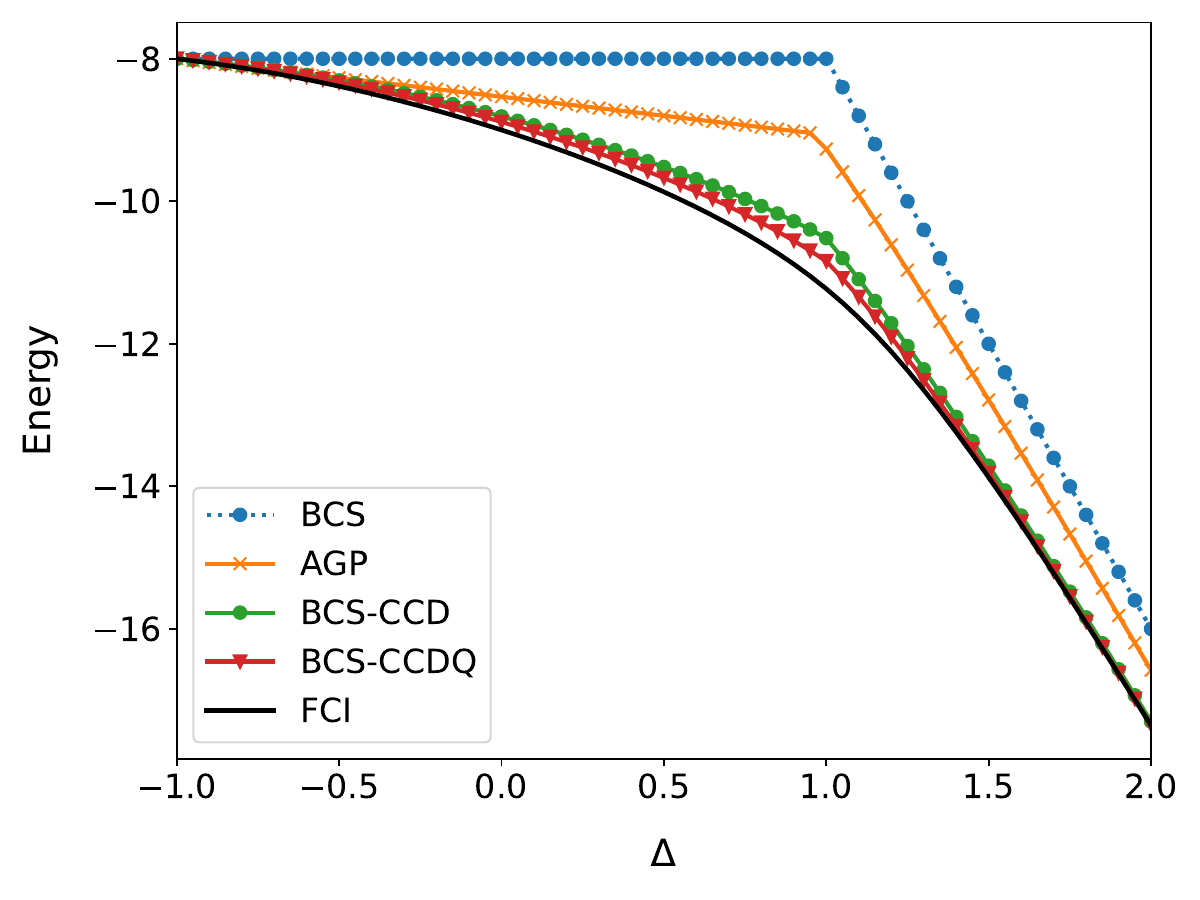}
    \caption{}
    \label{fig:xxz_4x4_pbc_energy}
\end{subfigure}
\hfill
\begin{subfigure}{0.48\textwidth}
    \includegraphics[width=\linewidth]{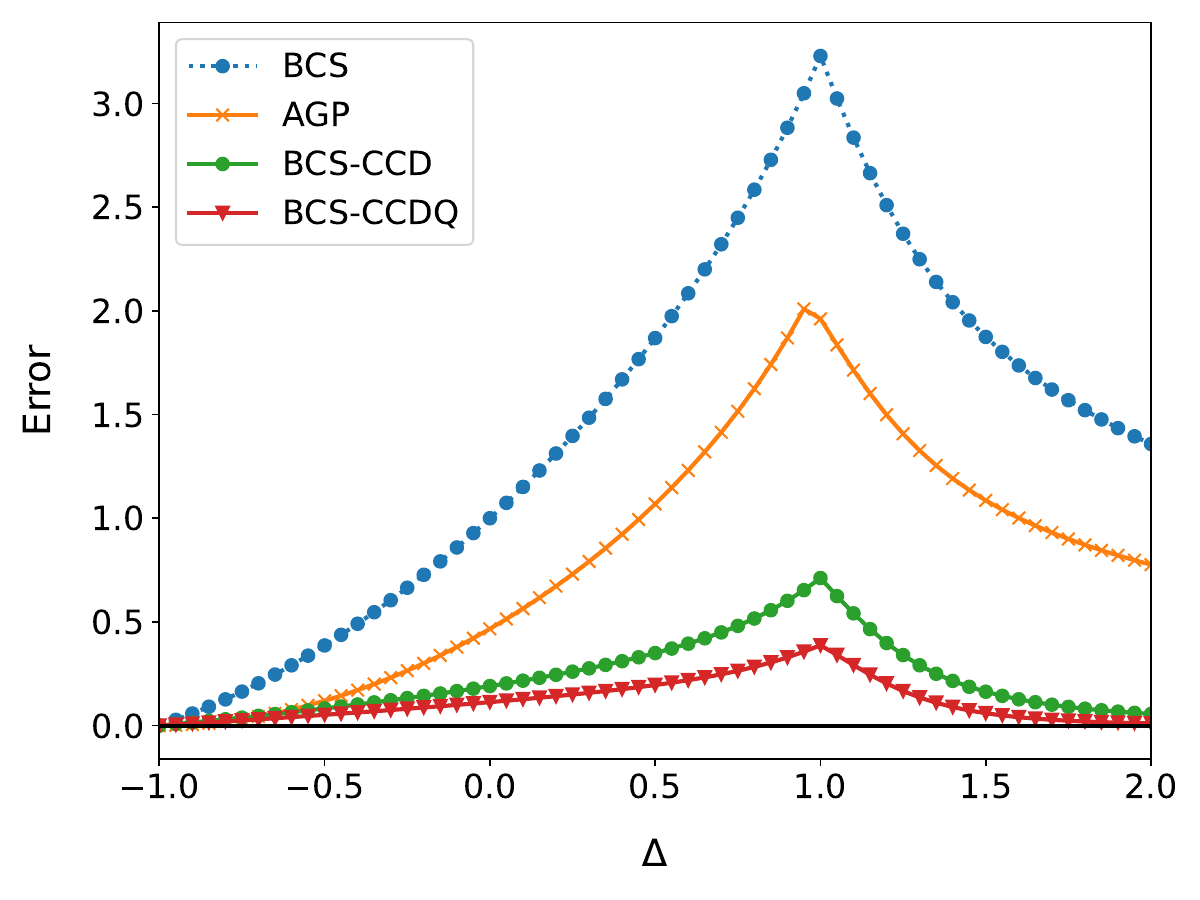}
    \caption{}
    \label{fig:xxz_4x4_pbc_error}
\end{subfigure}
\caption{Two-dimensional $4\times4$ XXZ model with periodic boundary conditions. (a) Total energy. (b) Deviation of the total energy from the exact energy.}
\label{fig:xxz_4x4_pbc}
\end{figure*}

In addition to the symmetry-broken BCS reference, we performed mean-field calculations using the $S_z$-projected BCS wave function, referred to here as the antisymmetrized geminal power (AGP) ansatz \cite{Liu2023SpinAGP}. The calculations we performed here are restricted to finite systems, but while true phase transitions are absent, the existence of the phase transitions in the thermodynamic limit is reflected by the existence of multiple mean-field solutions.  This results in curves that are not entirely smooth due to crossings between distinct solutions. 

For the XXZ Hamiltonian, the orbital-relaxation contribution to the properties of interest turns out to be negligible. We therefore report spin--spin correlation functions obtained from the unrelaxed response RDMs. 

Figure \ref{fig:myplot1} shows the result for the 12-site one-dimensional model with periodic boundary conditions. At $\Delta = -1$, the BCS reference is exact, and therefore both AGP and the coupled-cluster methods built upon it are also exact. As $\Delta$ increases, the deviation of the total energy from the exact result increases slightly. The BCS reference exhibits the largest error, and the $S_z$ projection improves the description but remains insufficient for quantitative agreement. By contrast, adding coupled-cluster correlations on top of BCS yields a systematic improvement, with the error decreasing as the excitation level increases. Among the methods considered, BCS-CCDQ provides the closest agreement with the exact result across the full parameter range. However, in the large $\Delta$ limit, the energy again approaches the exact limit, since the XXZ Hamiltonian reduces to an Ising-like model in this regime, for which the BCS reference becomes exact.

\begin{figure*}[htbp]
    \centering

    \begin{subfigure}{0.48\textwidth}
        \centering
        \includegraphics[width=\linewidth]{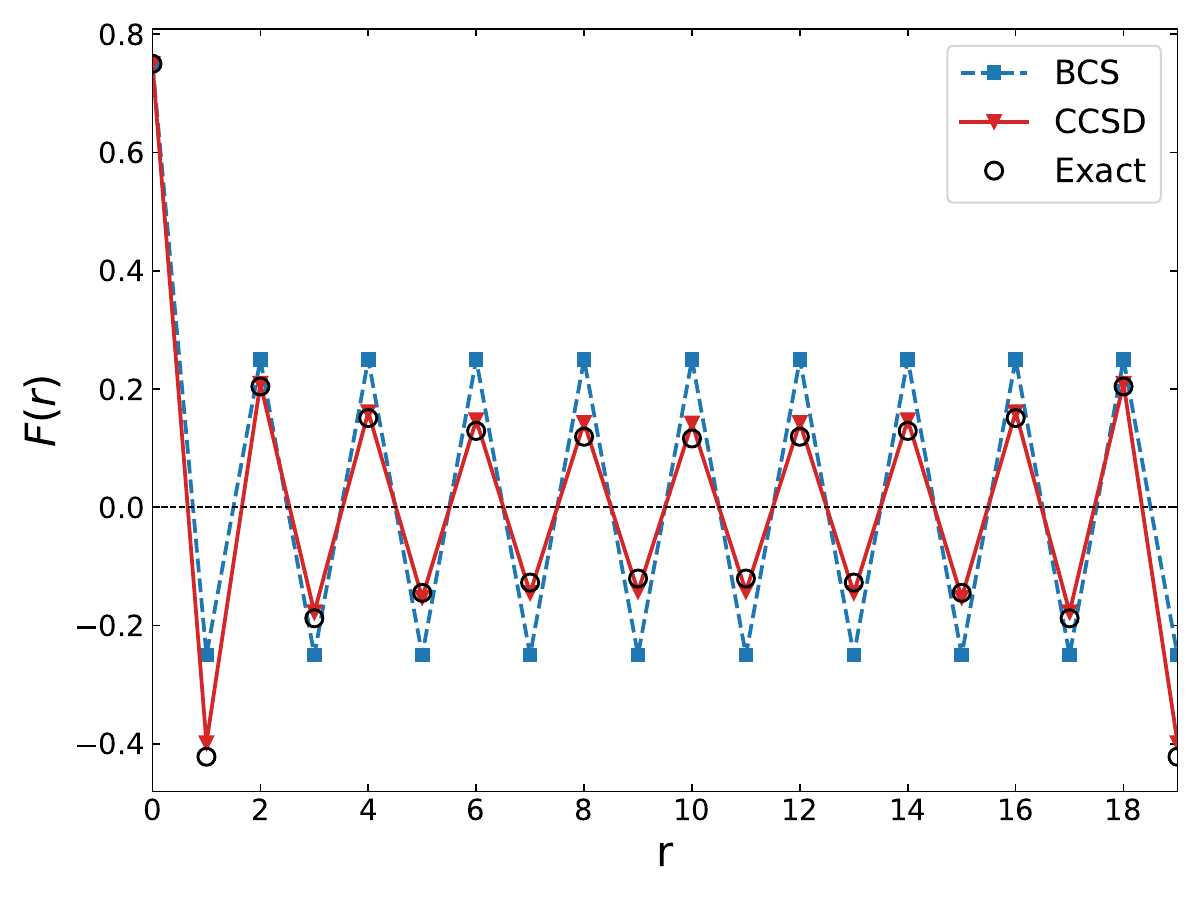}
        \caption{}
        \label{fig:1dxxz_20_spsq_0}
    \end{subfigure}
    \hfill
    \begin{subfigure}{0.48\textwidth}
        \centering
        \includegraphics[width=\linewidth]{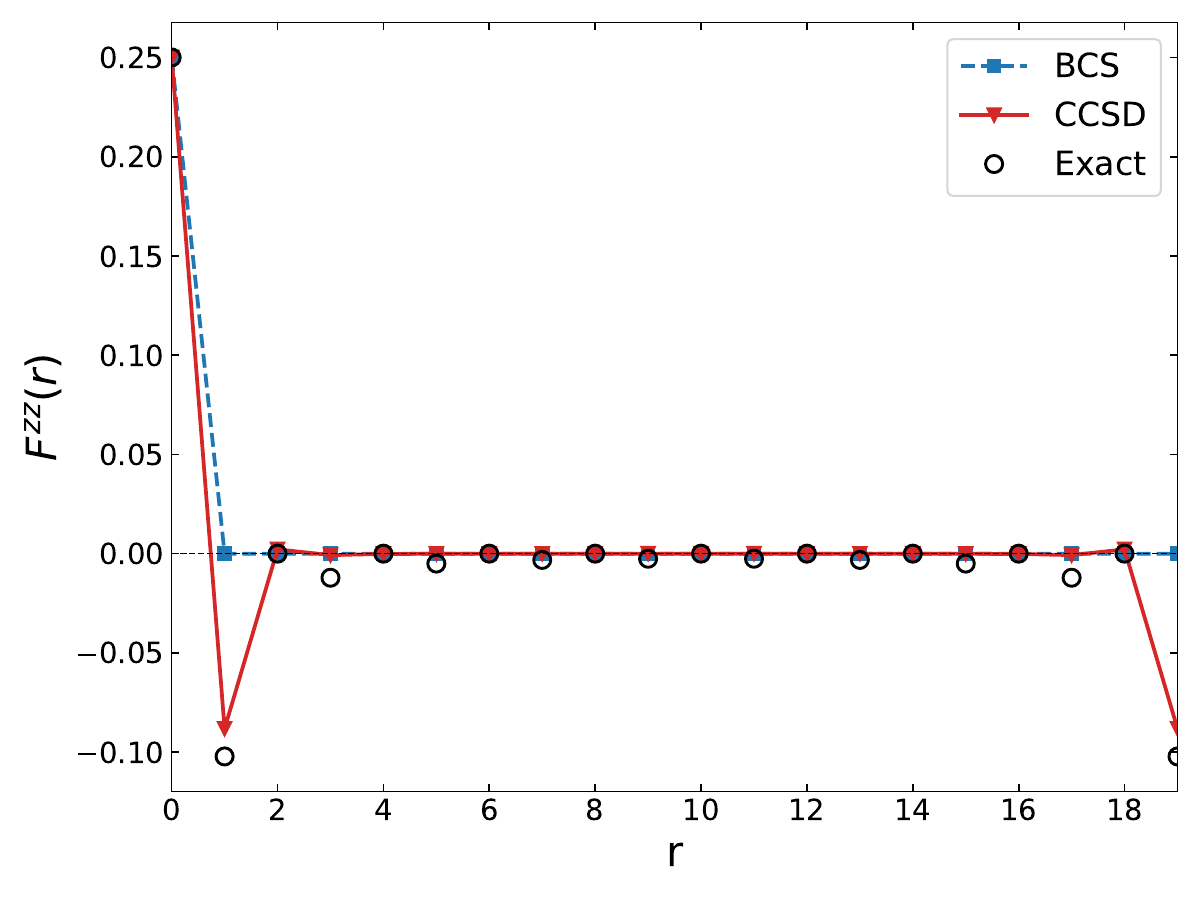}
        \caption{}
        \label{fig:1dxxz_20_szsz_0}
    \end{subfigure}

    \vspace{0.5cm}

    \begin{subfigure}{0.48\textwidth}
        \centering
        \includegraphics[width=\linewidth]{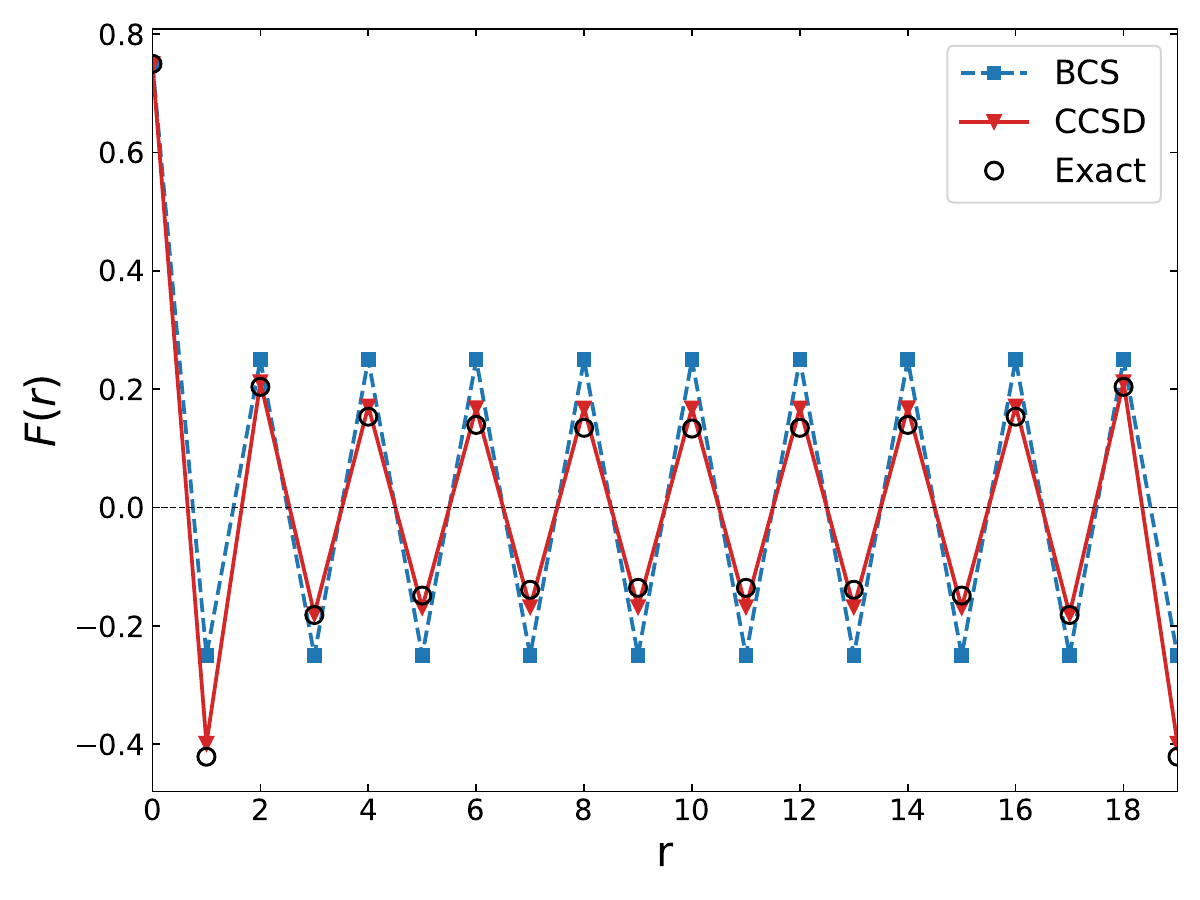}
        \caption{}
        \label{fig:1dxxz_20_spsq_2}
    \end{subfigure}
    \hfill
    \begin{subfigure}{0.48\textwidth}
        \centering
        \includegraphics[width=\linewidth]{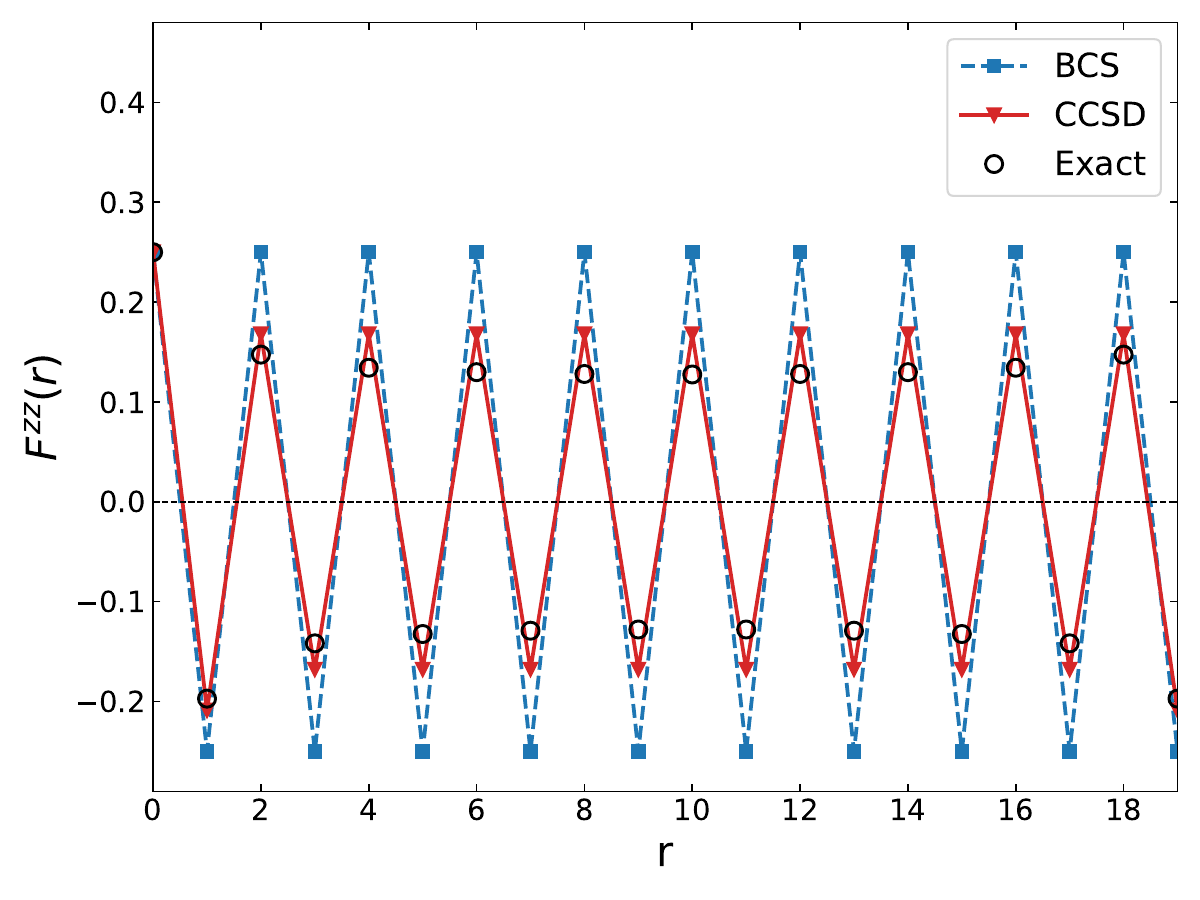}
        \caption{}
        \label{fig:1dxxz_20_szsz_2}
    \end{subfigure}

    \caption{
    Spin-spin correlation functions for the 20-site XXZ model with periodic boundary conditions. The left column shows $F(r)$ and the right column shows $F^{zz}(r)$.
    The top row corresponds to $\Delta=0$, while the bottom row corresponds to $\Delta=2$.
    }
    \label{fig:1dxxz_20_correlations}
\end{figure*}

Near $\Delta \approx 1$, there is a crossing of two BCS solutions, and correspondingly, between two AGP solutions. For the range $-1 \le \Delta \lesssim 1 $, which is the XY phase, the ground-state BCS wave function is what we call an extreme bimodal state, where all $u$ and $v$ have the same magnitude: $|u_p| = |v_p| = 1/\sqrt{2}$ for all $p$, and we may take $v_p$ positive while $u_p$ alternates signs from site to site.  For an extreme-bimodal reference, the odd excitations do not contribute to the CC because the Hamiltonian has a global parity symmetry, and any odd excitation changes the parity of the wave function. Thus, the odd excitations are symmetry-forbidden for this reference. The proof is given in appendix~\ref{parity_symmetry}. For $\Delta \gtrsim 1$, the exact ground state is more AFM-like, and the BCS reference reduces to a single spin configuration in which each site has a definite $S_z$; odd excitations thus change the global $S_z$ and are symmetry forbidden in this regime. Thus, CCD, CCSD, and CCSDT yield the same energy here as well.  In other words, due to the character of the mean-field reference, only even excitations contribute. We should emphasize that if we use a different reference, odd excitations do contribute to the CC method. 

For the 2D XXZ model~\cite{OkabeKikuchi1988XXZQMC,Bishop1996XXZPlanar,Bishop2000XXZHighOrder,Bishop2017XXZSquareRevisited} (Fig. \ref{fig:xxz_4x4_pbc}), the energetics are similar to those for the 1D case. As with the 1D case, the extreme-bimodal BCS is the ground state for the square lattice in the XY phase and odd excitations do not contribute in the CC method. The crossing of the CC solutions in the 2D case is more prominent than in the 1D case. The errors across all methods are largest around $\Delta=1$. Of course, all of the methods approach the exact limit at large $\Delta$; however, from both Fig. \ref{fig:myplot1} and Fig. \ref{fig:xxz_4x4_pbc}, we see that the CC methods approach the exact limit faster than BCS and AGP do. We should also note that in the thermodynamic limit, results from BCS and AGP become indistinguishable per particle~\cite{Liu2023SpinAGP}.

Figure~\ref{fig:1dxxz_20_correlations} shows the spin--spin correlation function for the 20-site XXZ chain with periodic boundary conditions. The correlation functions are calculated as 
\begin{subequations}
\begin{align}
F(r) &=\frac{1}{N} \, \sum_{p}\braket{\mathbf{S}_{p}\cdot \mathbf{S}_{p+r}},
\\
F^{zz}(r) &=\frac{1}{N} \, \sum_{p}\braket{S^{z}_{p} S^{z}_{p+r}},
\end{align}
\end{subequations}
where $N$ is the number of sites. Exact results are shown as open circles, while CCSD results are shown as a solid line and BCS in a dotted line. Owing to translational invariance, the correlation functions depend only on the separation $r$, and the plotted results exhibit the expected reflection symmetry, i.e. $F(r)=F(N-r)$. For $\Delta=0.0$ and $\Delta=2.0$, the exact correlation functions display the characteristic staggered structure of the non-ferromagnetic regime, with an approximately period-two alternation in sign as a function of distance. In both cases, CCSD reproduces the exact results very well over the full range of separations, indicating that the dominant short and intermediate range correlations are well captured by the BCS-based coupled-cluster treatment.

For the $S_z-S_z$ correlation function $F^{zz}(r)$, it vanishes for even $r$ at $\Delta=0$. At the BCS level, the correlation function is zero for all $r\ge 1$, since the spins are in the $xy$ plane. CCSD recovers the nearest-neighbor correlations well and remains near zero for separations $r>1$.   At $\Delta=2.0$ the exact $F^{zz}(r)$ displays the staggered structure of the antiferromagnetic (AFM) regime, with CCSD reproducing the N\'eel AFM structure.

\begin{figure*}[htbp]
    \centering

    \begin{subfigure}{0.48\textwidth}
        \centering
        \includegraphics[width=\linewidth]{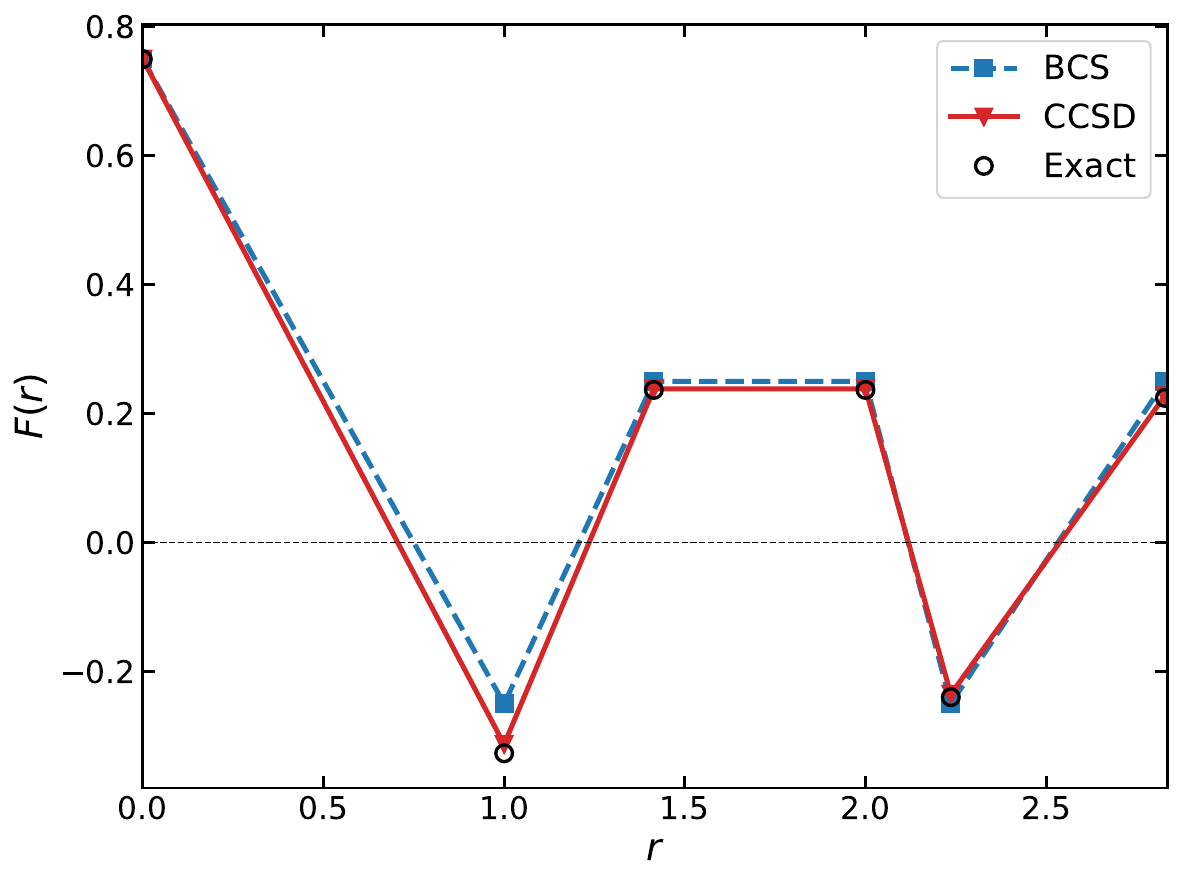}
        \caption{}
        \label{fig:xxz_2d_spsq_0}
    \end{subfigure}
    \hfill
    \begin{subfigure}{0.48\textwidth}
        \centering
        \includegraphics[width=\linewidth]{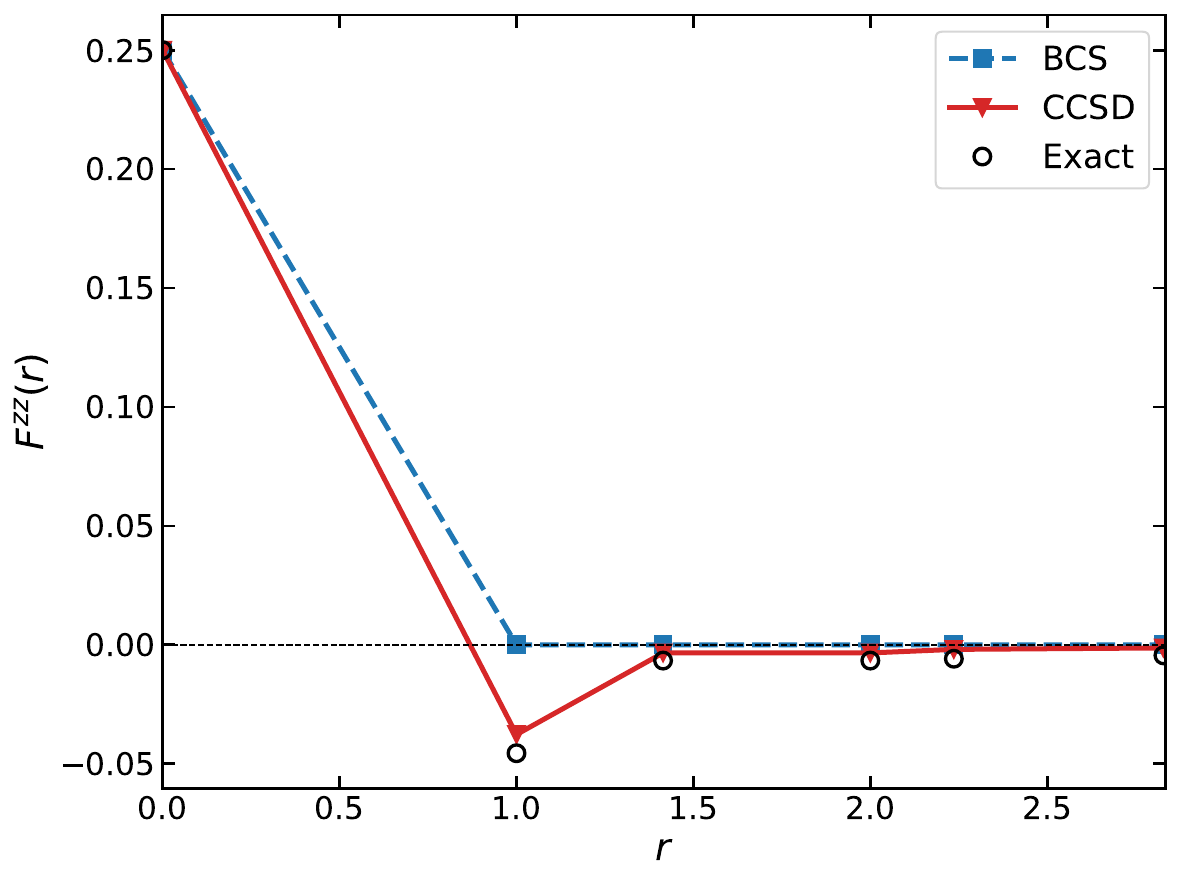}
        \caption{}
        \label{fig:xxz_2d_szsz_0}
    \end{subfigure}

    \vspace{0.5cm}

    \begin{subfigure}{0.48\textwidth}
        \centering
        \includegraphics[width=\linewidth]{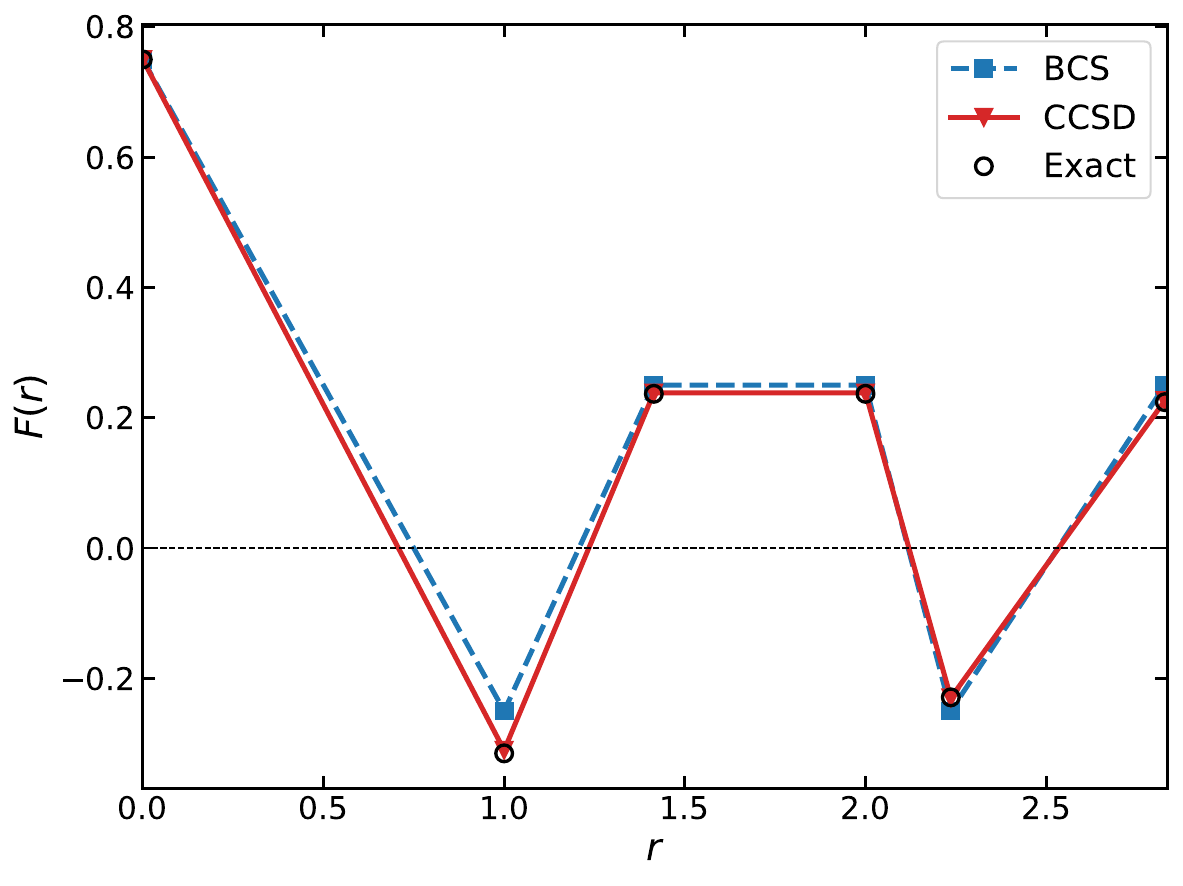}
        \caption{}
        \label{fig:xxz_2d_spsq_2}
    \end{subfigure}
    \hfill
    \begin{subfigure}{0.48\textwidth}
        \centering
        \includegraphics[width=\linewidth]{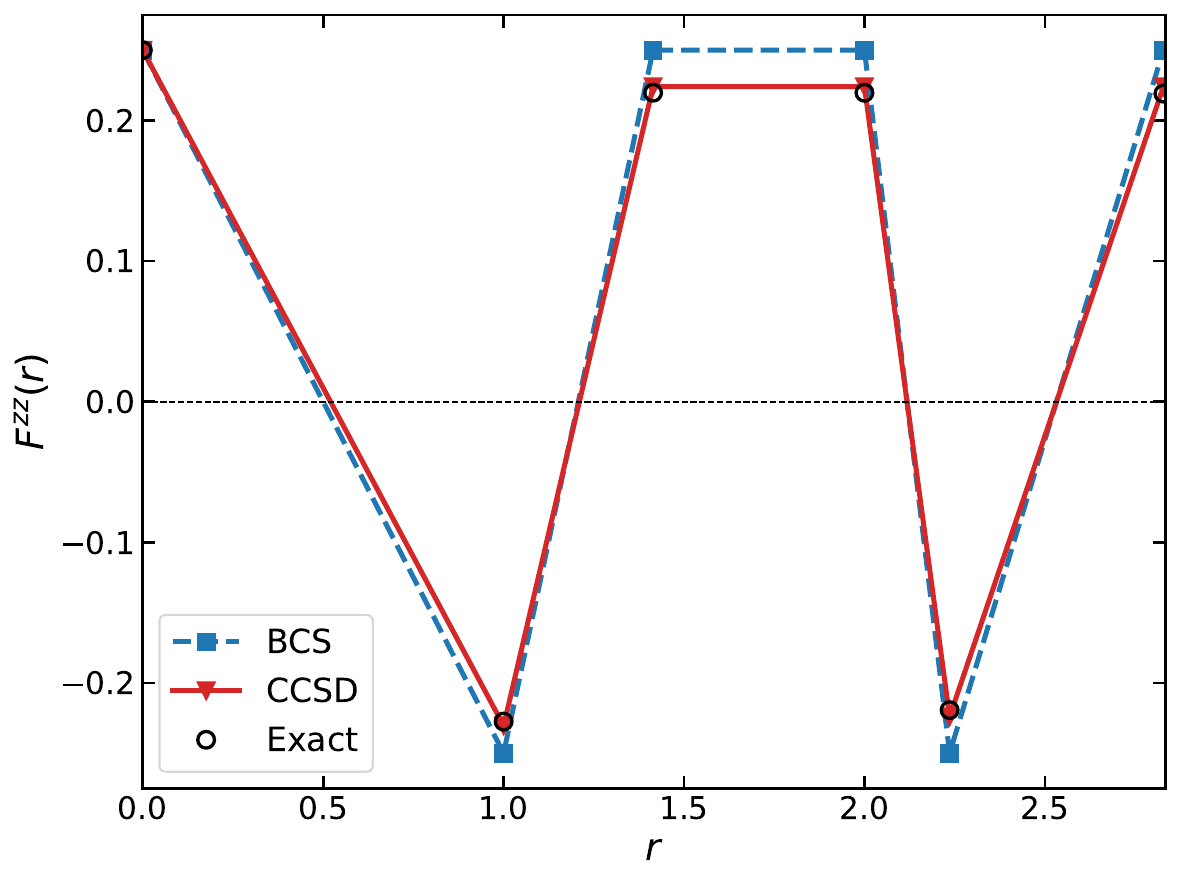}
        \caption{}
        \label{fig:xxz_2d_szsz_2}
    \end{subfigure}

    \caption{
    Spin correlation functions for the $4\times4$ XXZ model with periodic boundary conditions.
    The left column shows $F(r)$ and the right column shows $F^{zz}(r)$.
    The top row corresponds to $\Delta=0$, while the bottom row corresponds to $\Delta=2$.
    }
    \label{fig:xxz_2d_correlations}
\end{figure*}

Figure~\ref{fig:xxz_2d_correlations} shows the spin--spin correlation function $F(r)=\frac{1}{N_r}\sum_{(dx,dy)\in \mathcal{S}_r}
\braket{\mathbf{S}_{0,0}\cdot \mathbf{S}_{dx,dy}}$ for the $4\times 4$ XXZ model with periodic boundary conditions (PBC). Here, $\mathcal{S}_r$ denotes the set of lattice displacements $(dx,dy)$ whose minimum-image distance from site $(0,0)$ is $r$, and $N_r=|\mathcal{S}_r|$ is the number of sites in that distance shell. For $\Delta=0$ and $\Delta=2$, CCSD captures the characteristic antiferromagnetic sign structure, which is in very good agreement with the exact results.  

In the same spirit, Fig.~\ref{fig:xxz_2d_correlations} also shows the spin-spin correlation $F^{zz}$, where CCSD captures most of the correlation and is close to exact. $F^{zz}$ correlations at $\Delta=0$ indicate the spins being on the $xy$ plane and a N\'eel structure at $\Delta=2$ where the nearest-neighbor correlations are negative and second-nearest-neighbor correlations are positive.

\subsubsection{Triangular lattice}
\label{sec:Triangular_lattice}
We next consider the XXZ Hamiltonian on the triangular lattice. 
The triangular lattice is a canonical example of geometric frustration: for antiferromagnetic interactions, the three bonds of an elementary triangle cannot all be simultaneously satisfied by a collinear spin arrangement. 
As a result, triangular-lattice spin models naturally favor noncollinear or multisublattice magnetic structures. 
The XXZ anisotropy provides an additional tuning parameter, allowing one to interpolate between easy-plane and easy-axis regimes and making the triangular XXZ model a useful benchmark for strongly correlated spin methods \cite{White2007TriangularHeisenberg,Sellmann2015TriangularXXZ}.

This model is particularly relevant for the present BCS-based formulation because the natural mean-field reference is no longer restricted to a real bimodal pattern, as the spins are preferred to be noncollinear. 
Previous works showed that triangular and kagome XXZ lattices support an extreme-trimodal reference as the ground state in the region $- 0.5 \le \Delta \le 1 $\cite{Liu2023SpinAGP,Pal2021XXZColorful,Chertkov2021MotifScars}. For $\Delta>1$, the system remains geometrically
frustrated and has been discussed in terms of spin-supersolid behavior,
with coexisting longitudinal three-sublattice order and transverse
correlations
\cite{White2007TriangularHeisenberg,Sellmann2015TriangularXXZ,
Wessel2005SupersolidTriangular,Heidarian2005PersistentSupersolid}. 
Thus, the triangular XXZ model provides a first test of the present Bogoliubov coupled-cluster framework in a setting where complex BCS parameters are required to represent the underlying frustrated spin structure.
\begin{figure}[htbp]
    \centering
    \includegraphics[width=\columnwidth]{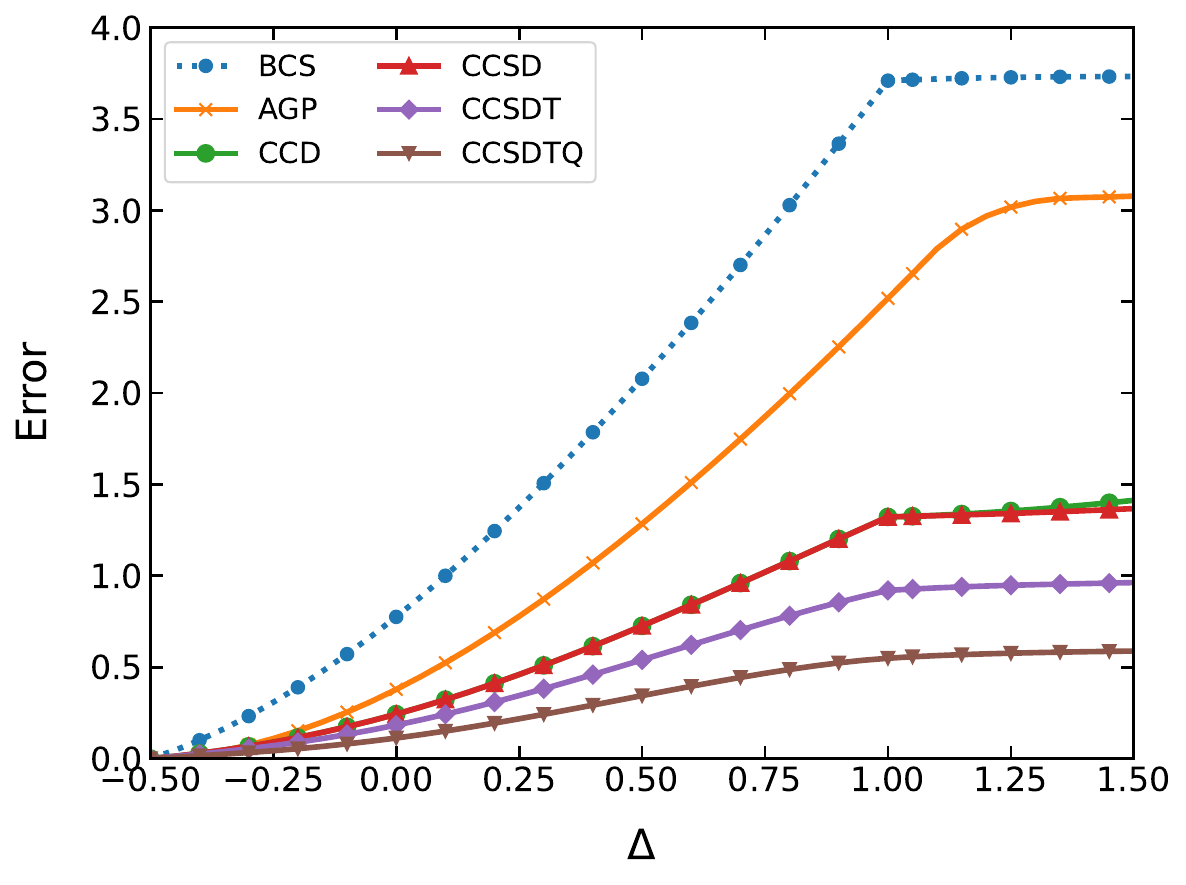}
    \caption{Energy errors for the XXZ Hamiltonian for 18-site triangular lattice with periodic boundary conditions. The BCS is constrained to \(\langle S_z\rangle=0\), while the AGP and the exact are evaluated in the $S_z=0$ sector.} 
    \label{fig:18_tri_XXZ}
\end{figure}

Figure~\ref{fig:18_tri_XXZ} shows the energy errors for the 18-site triangular lattice with PBC. For extreme-trimodal BCS, the $T_1$ contribution vanishes by symmetry. The extreme-trimodal BCS has amplitudes $|u_p| = |v_p| = 1/\sqrt{2}$ and three sublattice phases $1, \omega,\omega^2$ where $\omega = e^{2\pi i/3}$, and is invariant up to a global phase under combined lattice translations and global $S_z$ rotation. The converged higher-rank cluster amplitudes preserve the same symmetry, so that the similarity-transformed Hamiltonian also respects this symmetry. Consequently, single excitations transforming in nontrivial symmetry sectors cannot couple to the reference and have identically vanishing singles residuals. The remaining symmetry-preserving singles direction corresponds to a uniform variation of the BCS mixing angle,
\begin{equation}
u_p=\cos\theta,
\qquad
v_p=\eta_p\sin\theta,
\end{equation}
where \(\eta_p\in\{1,\omega,\omega^2\}\). The extreme-trimodal solution occurs at \(\theta=\pi/4\), for which
\begin{equation}
u_p=\frac{1}{\sqrt{2}},
\qquad
v_p=\frac{\eta_p}{\sqrt{2}}.
\end{equation}
Stationarity of the BCS energy at this point implies that the corresponding bare singles matrix element vanishes. For extreme-trimodal references, we verified that $T_1=0$ satisfies the full CC equations through CCSDTQ. The extreme-trimodal BCS reference therefore behaves as a generalized Brueckner reference for these truncated Bogoliubov CC calculations, although this vanishing of $T_1$ is dictated by symmetry.
For $\Delta>1$, the variationally optimal reference becomes non-extreme, and the single excitations become symmetry allowed.

\subsection{The $\mathrm{J_1-J_2}$ Hamiltonian}
The $\mathrm{J_1-J_2}$ Hamiltonian is written as
\begin{equation}
H_{\mathrm{J_1-J_2}} = J_1 \sum_{\langle p q\rangle} \vec{S}_p \cdot\vec{S}_{q} + J_2 \sum_{\langle \langle pq\rangle\rangle} \vec{S}_p \cdot\vec{S}_{q}
\end{equation}
where $\langle \langle pq\rangle\rangle$ denotes $p$ and $q$ being second nearest neighbors. The Hamiltonian preserves $S^2$ symmetry, so it is invariant under any global $\mathfrak{su}(2)$ rotation. The physics depends on the ratio $J_2/J_1$, where we take both $ J_2$ and $ J_1$ to be positive.

\subsubsection{Square lattice}

For a 2D square lattice in the thermodynamic limit, the ground state for $J_2/J_1 \lesssim 0.45$ is a N\'eel antiferromagnet, and for $J_2/J_1 > 0.6 $ is a striped antiferromagnetic phase \cite{Liu2023SpinAGP,Dagotto1989J1J2,Schulz1992J1J2,Capriotti2000J1J2Plaquette,Gong2014J1J2DMRG,Hu2013J1J2VMC,Morita2015J1J2VMC,Wang2013J1J2GaplessSL,Nomura2021J1J2Neural,SachdevBhatt1990BondOperator,Liu2022J1J2PEPS,Zhitomirsky1996J1J2VBC}. In between, the magnetic structure is in a highly spin-frustrated phase where the ground state is under debate, and possible candidates include the plaquette valence-bond state~\cite{Mambrini2006J1J2Plaquette, Zhitomirsky1996J1J2VBC}, the stripe valence-bond state~\cite{SachdevBhatt1990BondOperator}, or the gapless spin liquid state~\cite{Capriotti2001RVB}.  Again, for the symmetry-broken BCS reference, we impose $\braket{S_z}=0$.
\begin{figure}[htbp]
    \centering

    \begin{subfigure}{0.50\linewidth}
        \centering
        \includegraphics[width=0.90\linewidth]{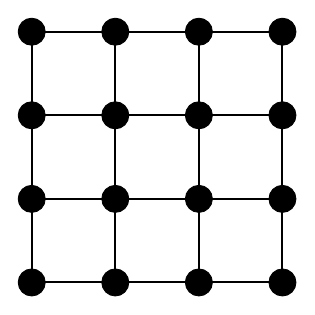}
        \caption{}
        \label{fig:Square_4x4}
    \end{subfigure}

    \par\vspace{0.1cm}

    \begin{subfigure}{0.47\linewidth}
        \centering
        \includegraphics[width=0.95\linewidth]{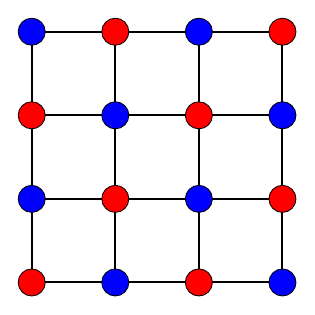}
        \caption{}
        \label{fig:Neel2D}
    \end{subfigure}
    \hfill
    \begin{subfigure}{0.47\linewidth}
        \centering
        \includegraphics[width=0.95\linewidth]{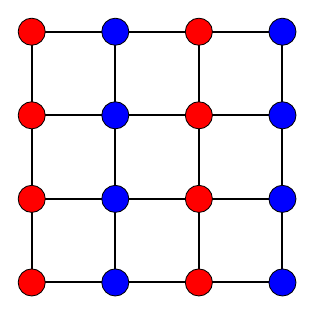}
        \caption{}
        \label{fig:Striped2D}
    \end{subfigure}

    \caption{(a) Two-dimensional $4\times4$ square lattice with periodic boundary conditions. 
Representative spin configurations for (b) N\'eel antiferromagnetic order and 
(c) stripe antiferromagnetic order. 
The blue and red sites denote opposite spin orientations.}
    \label{fig:2D_square}
\end{figure}

\begin{figure}[htbp]
    \centering

    \begin{subfigure}[t]{0.49\linewidth}
        \centering
        \includegraphics[width=\linewidth]{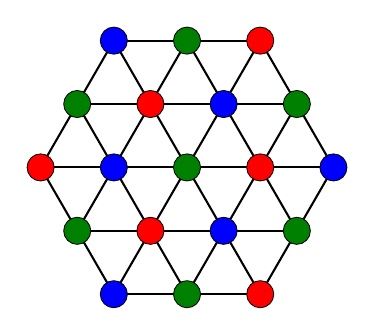}
        \caption{}
        \label{fig:tri_trimodal}
    \end{subfigure}
    \hfill
    \begin{subfigure}[t]{0.49\linewidth}
        \centering
        \includegraphics[width=\linewidth]{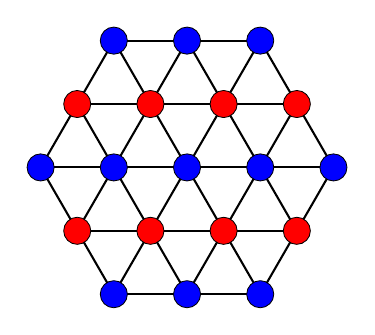}
        \caption{}
        \label{fig:tri_stripe}
    \end{subfigure}

    \caption{Triangular lattice with a hexagonal cell and periodic boundary
    conditions. (a) The color coding denotes the three sublattices associated
    with the $120^\circ$ ordered state. (b) The two-sublattice stripe pattern has spins parallel along one nearest-neighbor direction and antiparallel along
    the other two.}
    \label{fig:triangular_lattice_states}
\end{figure}
The optimized BCS references we obtain are bimodal ($\eta_p = v_p/u_p$ takes only two distinct values) or extreme bimodal ($\eta_p = \pm 1$) across all interaction ranges, with collinear antiferromagnetic spin order with spin axis not restricted to the $z$ direction. The optimized BCS bimodal reference obtained here is related to the extreme bimodal reference by a global $\mathfrak{su}(2)$ rotation. Since the Hamiltonian has $S^2$ symmetry, BCS states differing by a global $\mathfrak{su}(2)$ rotation have the same BCS mean-field and BCS-CC energies.  However, we should note that this is a special bimodal state, and a bimodal state in general is not necessarily symmetry equivalent to an extreme bimodal state, as in this case.  On the other hand, since the ground state can be chosen to be extreme bimodal, odd excitations at the BCS-CC level once again vanish by symmetry.

\begin{figure}[htbp]
    \centering
    \includegraphics[width=\columnwidth]{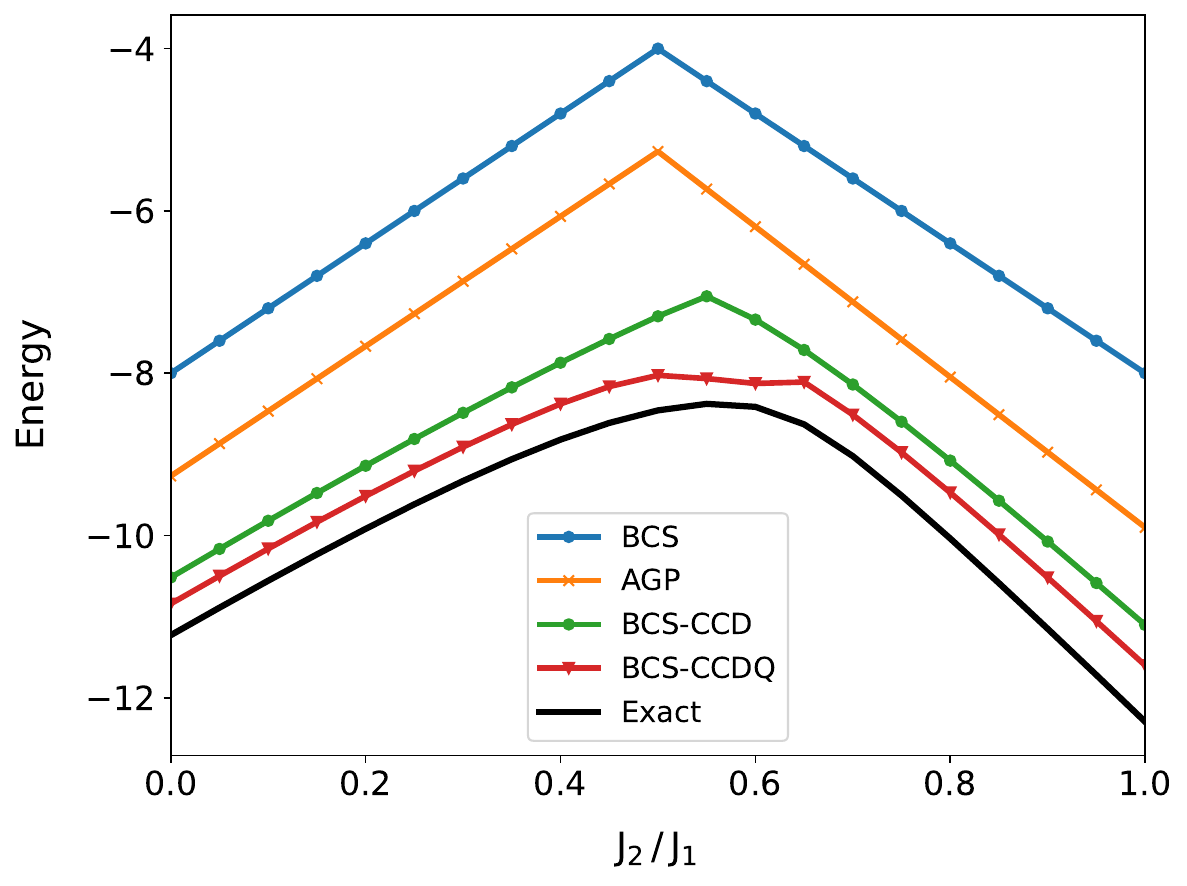}
    \caption{Total energy for the $4\times4$ square $\mathrm{J}_1-\mathrm{J}_2$ model with periodic boundary conditions. The BCS reference is constrained to $\langle S_z \rangle=0$, while AGP and exact are evaluated in the $S_z=0$ sector of the Hamiltonian. We use an extreme bimodal BCS reference and therefore exclude odd excitations in the CC treatment.}
    \label{fig:j1j2_energy}
\end{figure}

Figure~\ref{fig:j1j2_energy} shows the ground-state energies of the $4\times 4$ square $\mathrm{J_1-J_2}$ model with periodic boundary conditions. All methods display a nonmonotonic dependence on $J_2/J_1$. At the mean-field level, the symmetry projection substantially improves the energy, as seen from the improvement of AGP throughout the plot. However, the dominant recovery of correlation energy arises from the CC treatment built on the BCS reference. We have reported CCD and CCDQ results since singles and triples do not contribute. In particular, BCS-CCD lies significantly closer to FCI than either BCS or AGP, while BCS-CCDQ provides the best overall agreement with the exact results throughout the scan. These results show that the explicit inclusion of cluster correlations is essential for quantitative accuracy in the frustrated $\mathrm{J_1-J_2}$ problem.

Another notable feature is the shape of the energy profile near the frustrated region ($J_2/J_1 \approx 0.5-0.65$). The BCS and AGP curves exhibit a sharp cusp. Adding correlation through CC improves the energy, with the cusp of CCD shifting towards the maximum of the FCI curve, while CCDQ reproduces the overall shape more closely. However, the CCDQ curve develops a shallow valley in the frustrated region. We interpret this feature as a precursor or finite-size signature of the underlying phase transition, or switching between competing CC solution branches, rather than as a true phase transition, though note that while there should be three phases, we see only two at the mean-field level.

\begin{figure*}[htbp]
    \centering

    \begin{subfigure}[t]{0.48\textwidth}
        \centering
        \includegraphics[width=\linewidth]{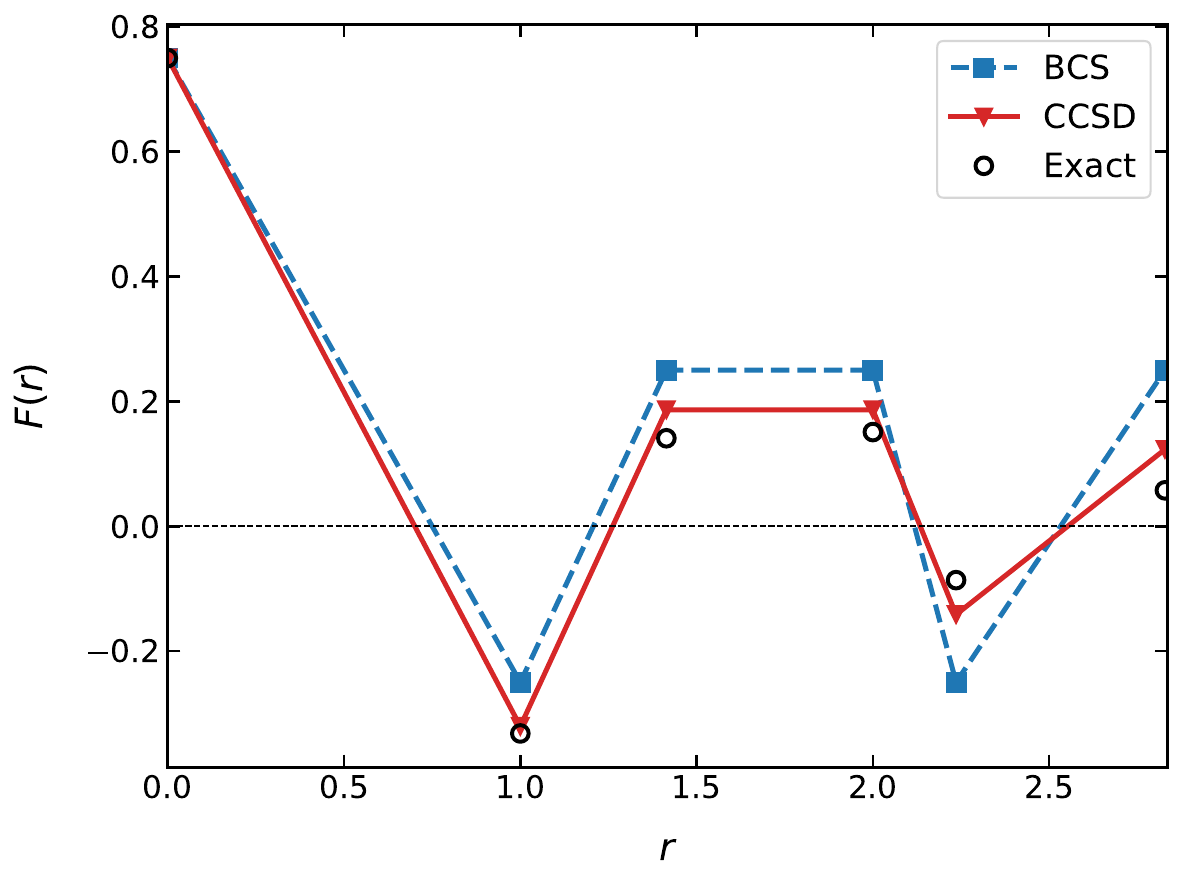}
        \caption{}
    \end{subfigure}
    \hfill
    \begin{subfigure}[t]{0.48\textwidth}
        \centering
        \includegraphics[width=\linewidth]{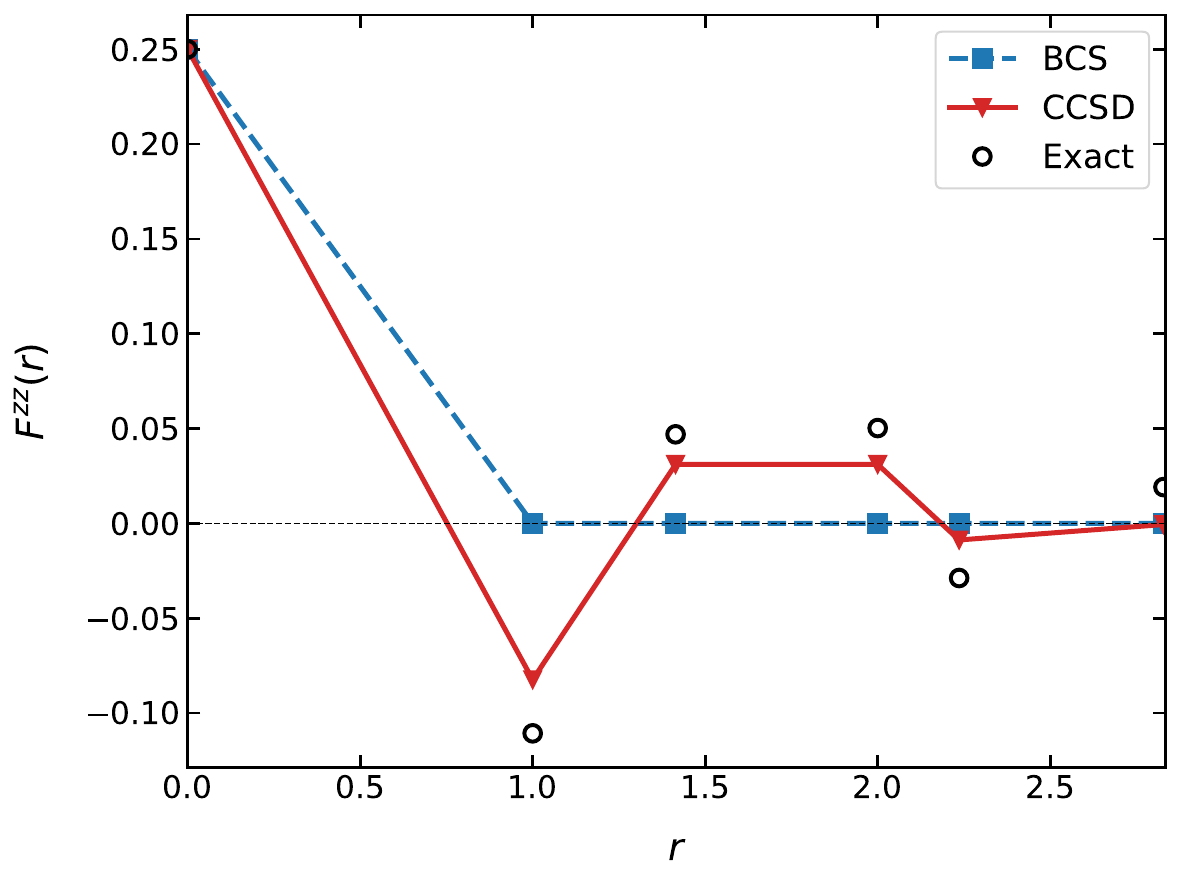}
        \caption{}
    \end{subfigure}

    \vspace{0.5em}

    \begin{subfigure}[t]{0.48\textwidth}
        \centering
        \includegraphics[width=\linewidth]{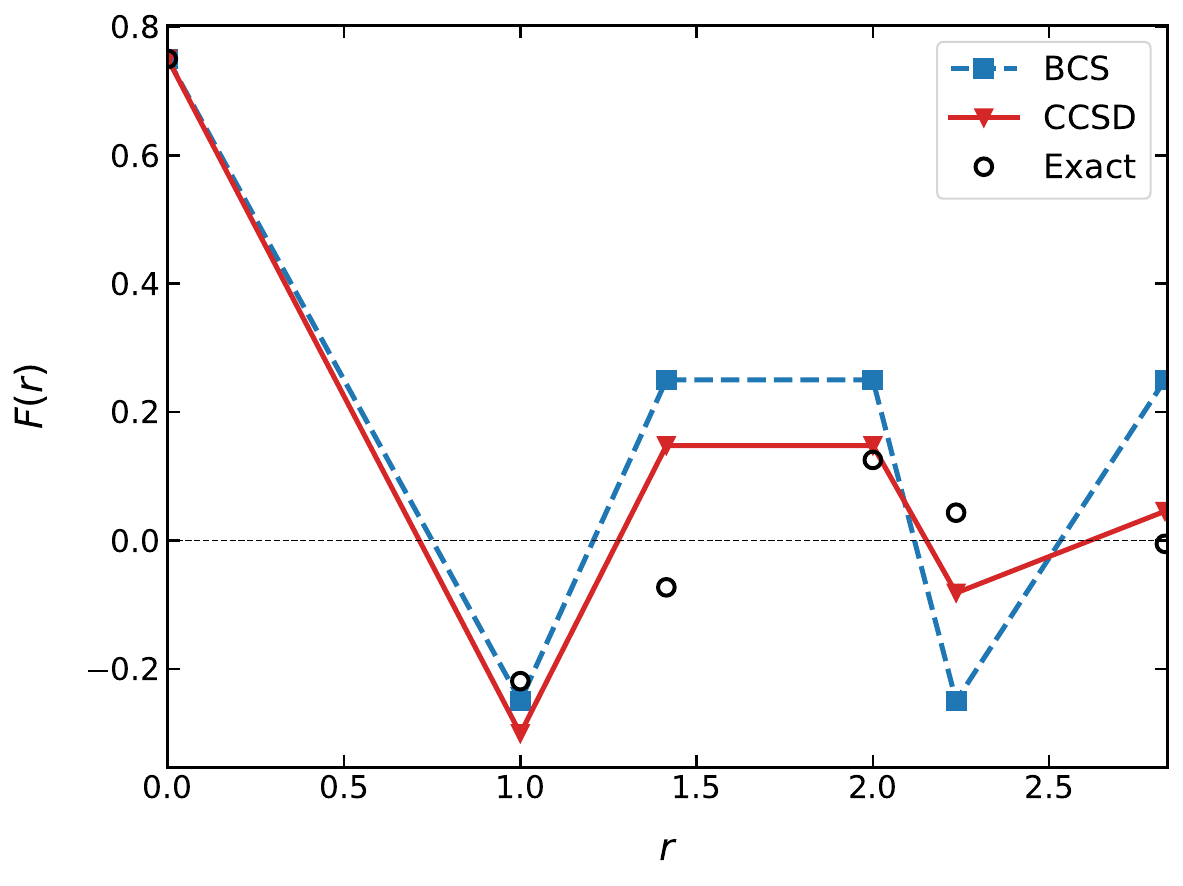}
        \caption{ }
    \end{subfigure}
    \hfill
    \begin{subfigure}[t]{0.48\textwidth}
        \centering
        \includegraphics[width=\linewidth]{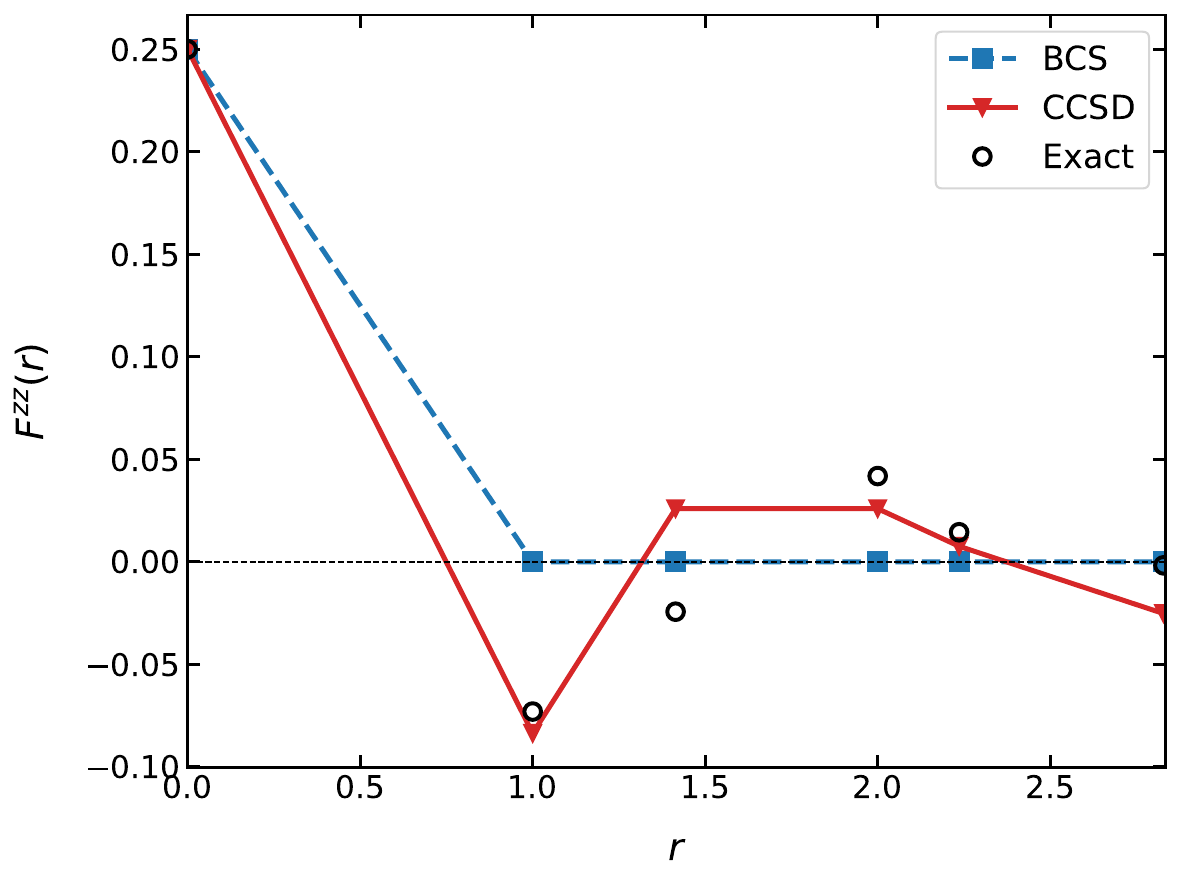}
        \caption{}
    \end{subfigure}

    \vspace{0.5em}

    \begin{subfigure}[t]{0.48\textwidth}
        \centering
        \includegraphics[width=\linewidth]{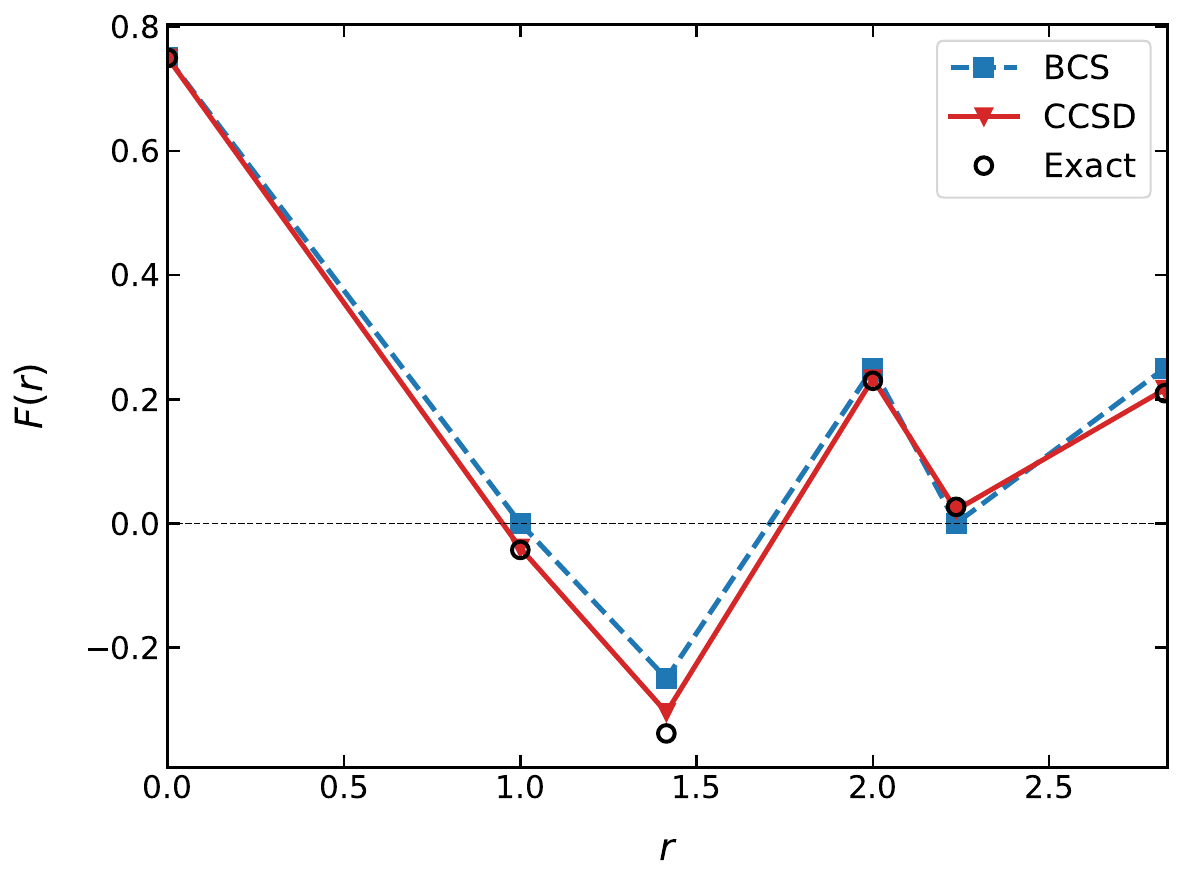}
        \caption{}
    \end{subfigure}
    \hfill
    \begin{subfigure}[t]{0.48\textwidth}
        \centering
        \includegraphics[width=\linewidth]{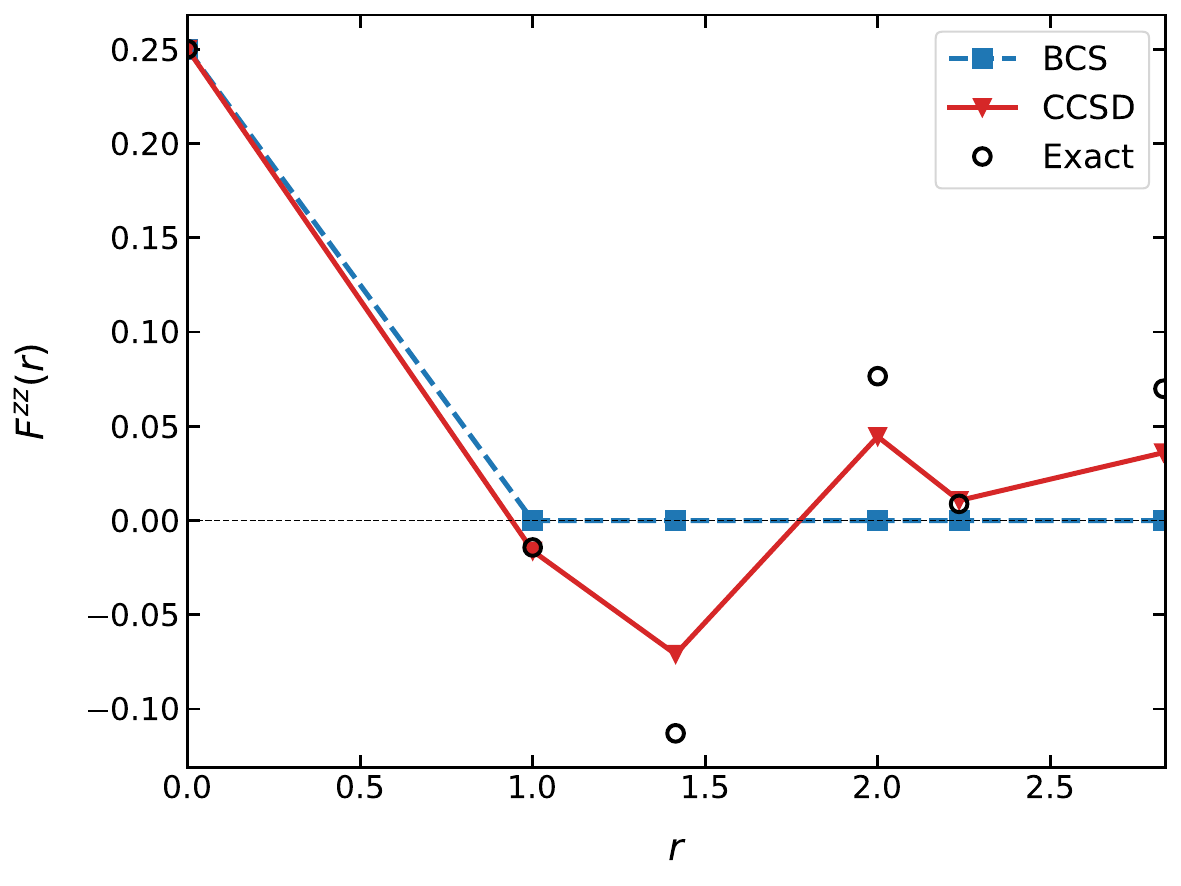}
        \caption{}
    \end{subfigure}

    \caption{
    Spin-spin correlation functions for the $4\times4$ square-lattice
    $J_1-J_2$ model with periodic boundary conditions.
    The left column shows $F(r)$ and the right column shows
    $F^{zz}(r)$ for $J_2/J_1=0.4$, $0.6$, and $0.8$ from top to bottom, respectively.
    }
    \label{fig:j1j2_2d_corr}
\end{figure*}

\begin{figure}[htbp]
    \centering
    \includegraphics[width=\columnwidth]{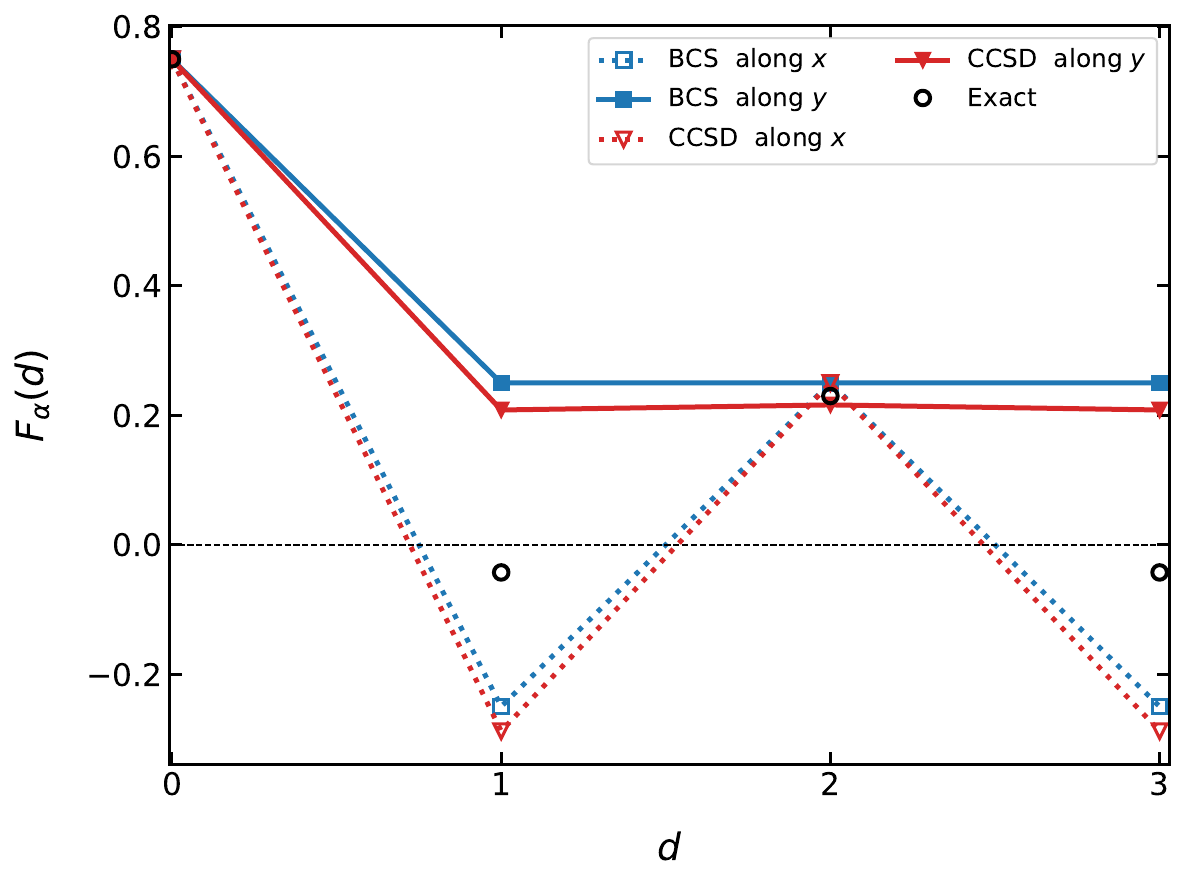}
    \caption{Direction-resolved spin correlations $F_{\alpha}(d)$ along the $x$ and $y$ lattice directions for the $4\times4$ square-lattice $J_1$--$J_2$ model with periodic boundary conditions at $J_2/J_1=0.8$.}
    \label{fig:square_j1j2_direction}
\end{figure}

Figure~\ref{fig:j1j2_2d_corr} compares the spin--spin correlation function $F(r)$ and $F^{zz}(r)$ for the $4\times4$ square $\mathrm{J_1-J_2}$ model with PBC obtained with BCS and CCSD. For $J_2=0.4$ and $J_2=0.8$, both BCS and CCSD closely resemble the exact correlations. At $J_2=0.4$, the correlations display the characteristic N\'eel-like structure.  For the stripe regime at $J_2=0.8$, radial averaging obscures the directional character of the magnetic correlations. We therefore resolve the spin correlations along the two lattice directions as
$
F_{\alpha}(d)=\frac{1}{N}\sum_i
\left\langle
\mathbf S_i\cdot\mathbf S_{i+d_{\alpha}}
\right\rangle,
$
where $\alpha=x,y$ and $d$ denotes the lattice displacement along the corresponding direction. As shown in Fig.~\ref{fig:square_j1j2_direction}, the BCS and CCSD correlations remain ferromagnetic along one direction while alternating in sign along the perpendicular direction, consistent with the stripe pattern illustrated in Fig.~\ref{fig:Striped2D}. The exact state on the finite lattice remains directionally symmetric. At $J_2=0.6$, in the strongly frustrated intermediate regime, the correlations retain a Néel-like structure, but with larger deviations from the exact correlations. 

\subsubsection{Triangular lattice} 

The triangular-lattice antiferromagnetic Heisenberg model with only nearest-neighbor interactions was originally proposed by Anderson as a
candidate realization of a resonating-valence-bond spin liquid
\cite{Anderson1973RVB}. 
Subsequent studies, however, established that the spin-$1/2$ nearest-neighbor
triangular antiferromagnet has three-sublattice $120^\circ$ magnetic order
\cite{HuseElser1988Triangular,White2007TriangularHeisenberg} as shown in Fig.~\ref{fig:triangular_lattice_states}. 
When second-neighbor antiferromagnetic interactions are included, the
triangular-lattice $J_1$--$J_2$ model becomes more strongly frustrated and
develops a spin-liquid phase around the region $J_2/J_1\approx 0.1$
\cite{Bishop2015TriangularJ1J2CCM,ZhuWhite2015TriangularJ1J2DMRG,
Hu2015TriangularJ1J2DMRG,Kaneko2014TriangularJ1J2VMC,
Iqbal2016TriangularJ1J2VMC}. 
This region lies between the $120^\circ$ ordered phase at small $J_2/J_1$
and the stripe-ordered phase at larger $J_2/J_1$.

\begin{figure}[htbp]
    \centering
    \includegraphics[width=\columnwidth]{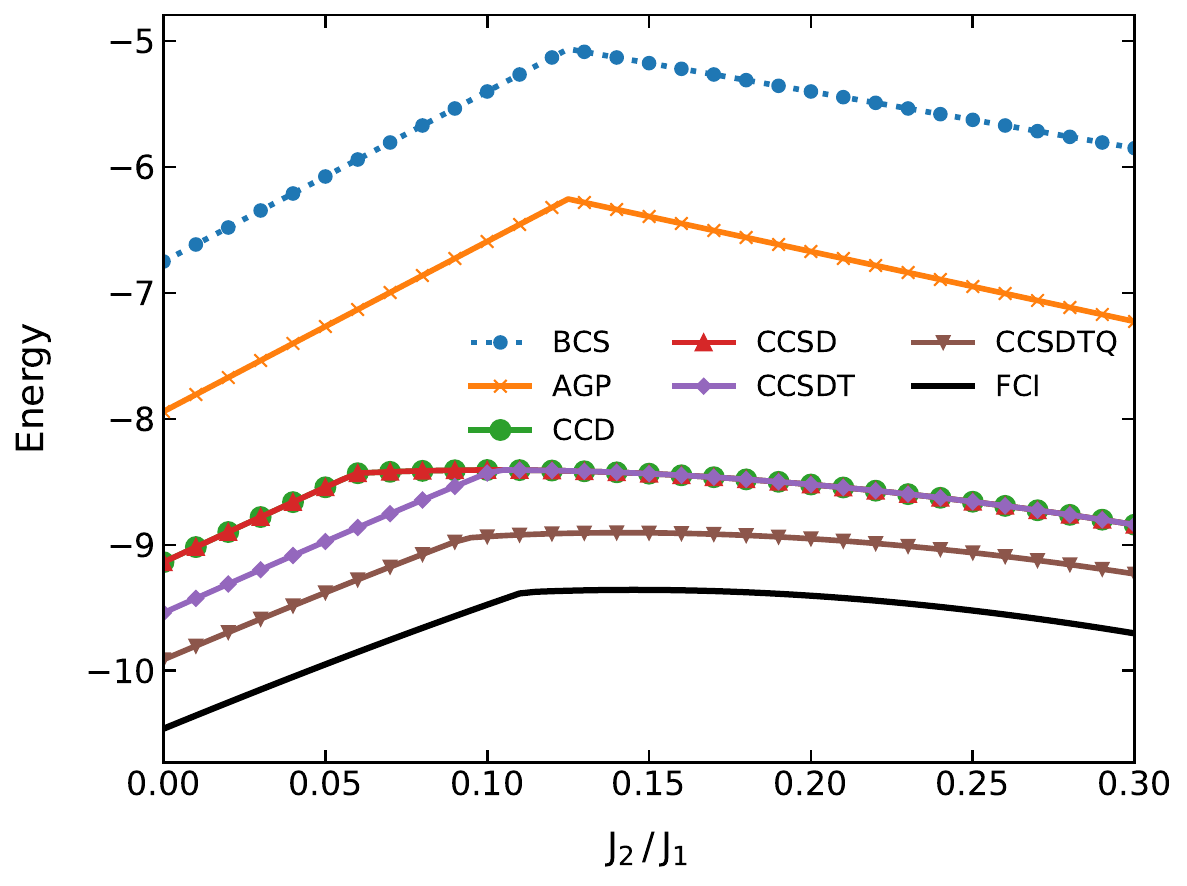}
    \caption{Total energy for the 18-site $J_1$-$J_2$ model with periodic boundary conditions. The BCS reference is constrained to $\braket{S_z}=0$, while AGP and the ED results are evaluated in the $S_z=0$ sector. We use an extreme trimodal BCS reference at small $J_2$ and an extreme-bimodal BCS reference at large $J_2$.} 
    \label{fig:18_tri_J1J2}
\end{figure}
The nature of the intermediate phase remains under debate. 
High-order coupled-cluster calculations place the loss of $120^\circ$ order
near $J_2/J_1\simeq 0.060$ and the onset of stripe order near
$J_2/J_1\simeq 0.165$
\cite{Bishop2015TriangularJ1J2CCM}. 
DMRG calculations find the spin-liquid phase to be at $0.08 \lesssim \mathrm{J}_2/\mathrm{J_1} \lesssim 0.165$, and both DMRG and Schwinger-boson mean-field theory have interpreted it
as a gapped $\mathbb{Z}_2$ spin liquid~\cite{ZhuWhite2015TriangularJ1J2DMRG,Hu2015TriangularJ1J2DMRG,Bauer2017}. 
By contrast, coupled-cluster and variational Monte Carlo (VMC) calculations favor a gapless $U(1)$
Dirac spin liquid
\cite{Kaneko2014TriangularJ1J2VMC,Iqbal2016TriangularJ1J2VMC}. 
The study of such a triangular lattice is also motivated
by triangular quantum-spin-liquid candidate materials such as
$\mathrm{YbMgGaO_4}$ and $\mathrm{Ba_3InIr_2O_9}$
\cite{Li2015YbMgGaO4,Dey2017Ba3InIr2O9}.

Figure~\ref{fig:18_tri_J1J2} shows the total energy for the $J_1$--$J_2$ model for 18-site triangular lattice with PBC. This model has two optimal BCS solutions that cross at $J_2/J_1 \approx 0.125$.
 At small $J_2/J_1$, the optimal BCS solution is an extreme-trimodal state, corresponding to the three-sublattice structure of the triangular lattice. Therefore, the singles contribution vanishes, and CCD and CCSD yield identical results. At larger $J_2/J_1$, the optimized BCS reference becomes a non-extreme bimodal state, characterized by a stripe N\'eel phase. Although the reference is non-extreme in the chosen spin quantization axis, the two local spin directions are related by a global $\mathfrak{su(2)}$ rotation to an extreme bimodal stripe reference. Because the isotropic Heisenberg Hamiltonian is invariant under global spin rotations, the extreme and non-extreme bimodal BCS descriptions are energetically equivalent. Since, at the extreme-bimodal regime, the odd excitations vanish by symmetry, the CCD, CCSD, and CCSDT yield identical energies. It is important to note that the two CC solutions are not bound to cross at the same value of $J_2/J_1$ as the two underlying BCS references.

\begin{figure*}[htbp]
    \centering

    \begin{subfigure}[t]{0.48\textwidth}
        \centering
        \includegraphics[width=\linewidth]{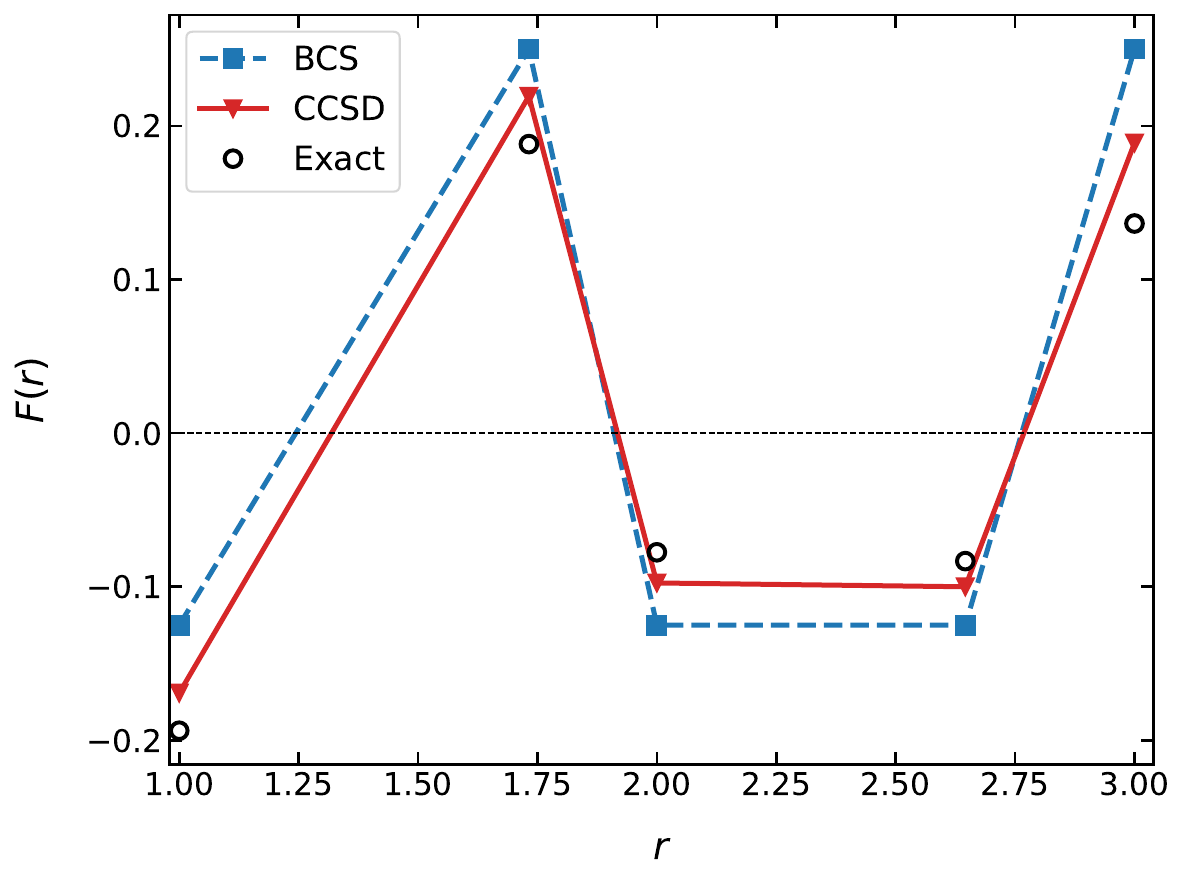}
        \caption{}
    \end{subfigure}
    \hfill
    \begin{subfigure}[t]{0.48\textwidth}
        \centering
        \includegraphics[width=\linewidth]{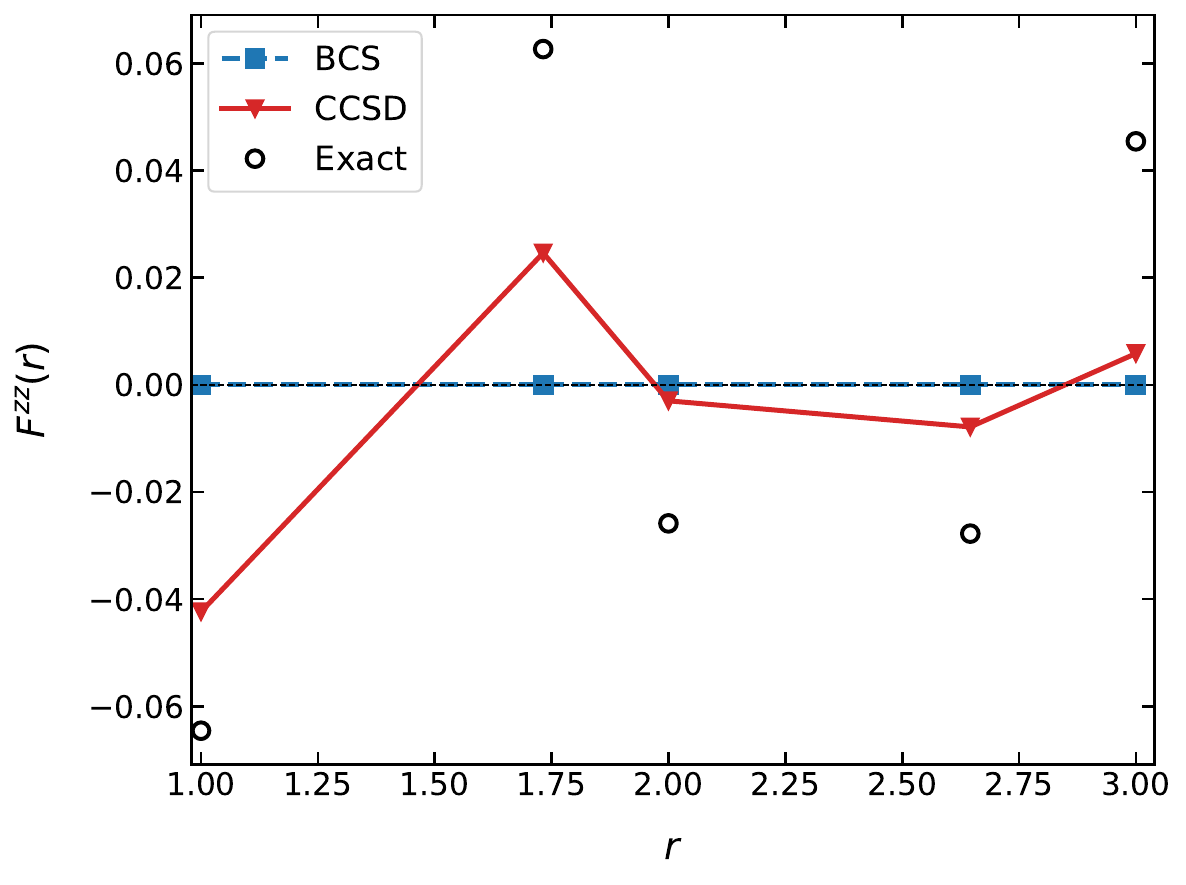}
        \caption{}
    \end{subfigure}

    \vspace{0.5em}

    \begin{subfigure}[t]{0.48\textwidth}
        \centering
        \includegraphics[width=\linewidth]{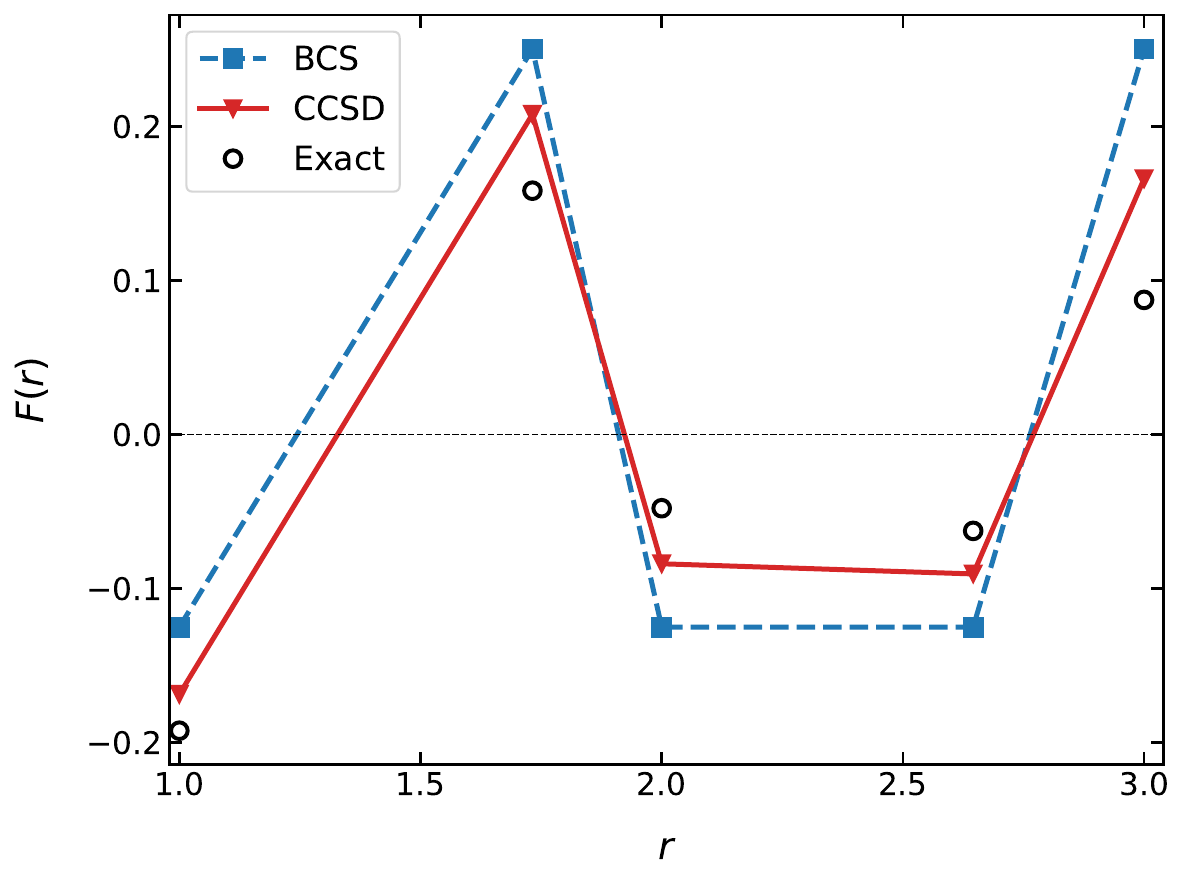}
         
        \caption{}
    \end{subfigure}
    \hfill
    \begin{subfigure}[t]{0.48\textwidth}
        \centering
        \includegraphics[width=\linewidth]{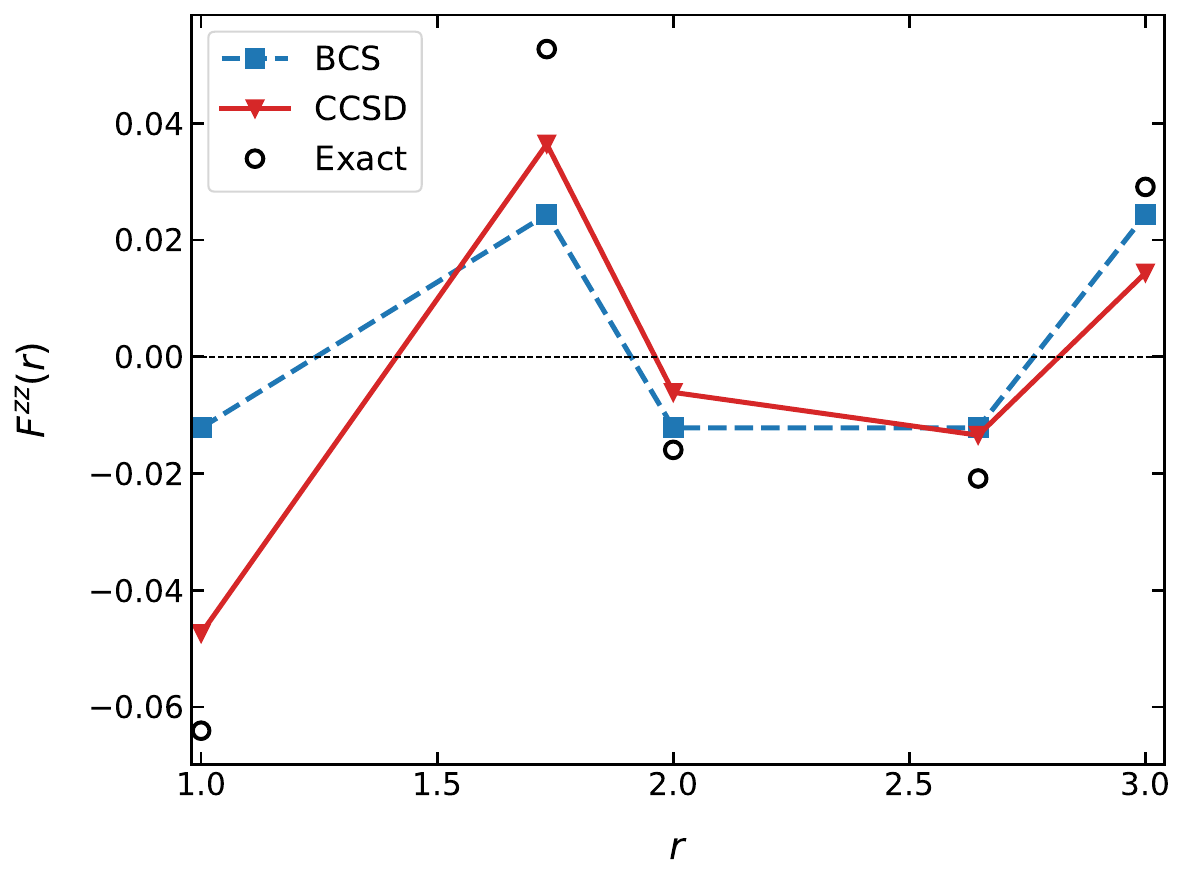}
         
        \caption{}
    \end{subfigure}

    \vspace{0.5em}

    \begin{subfigure}[t]{0.48\textwidth}
        \centering
        \includegraphics[width=\linewidth]{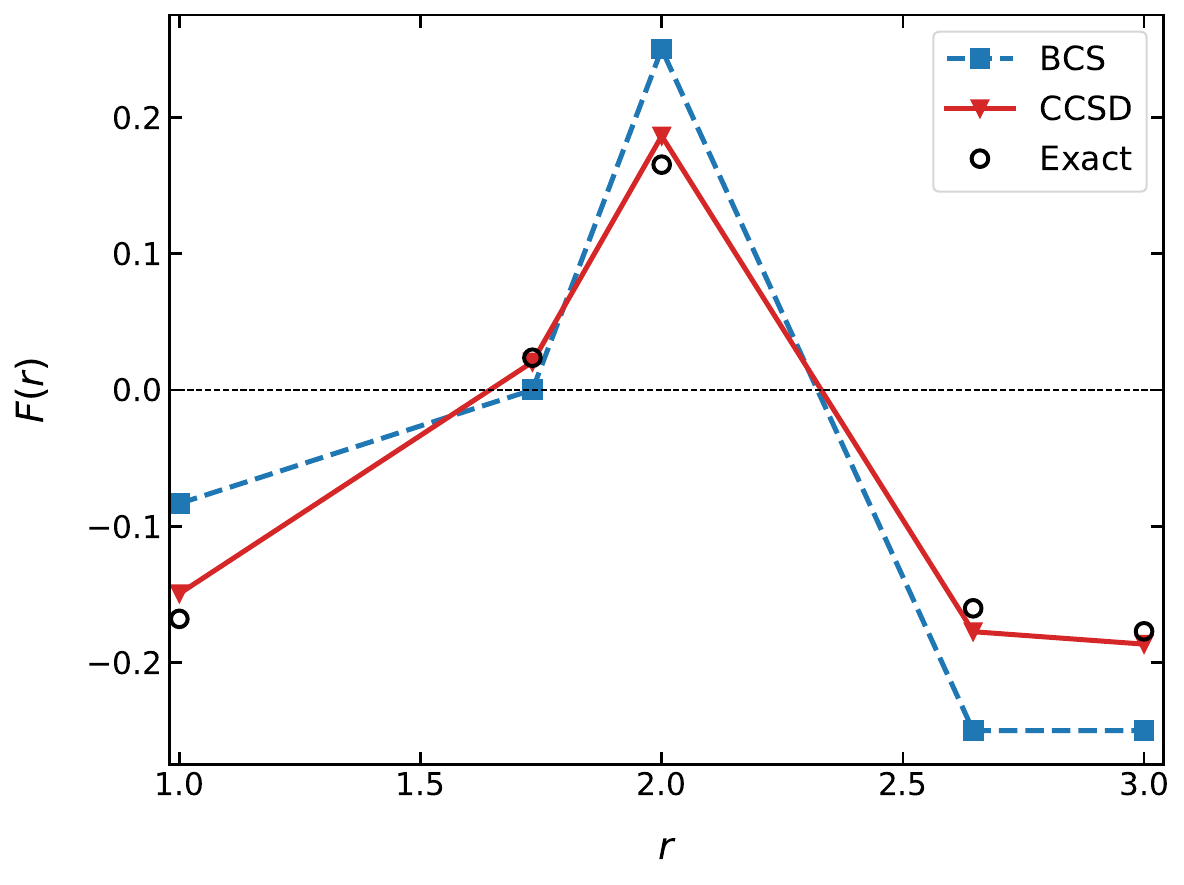}
         
        \caption{}
    \end{subfigure}
    \hfill
    \begin{subfigure}[t]{0.48\textwidth}
        \centering
        \includegraphics[width=\linewidth]{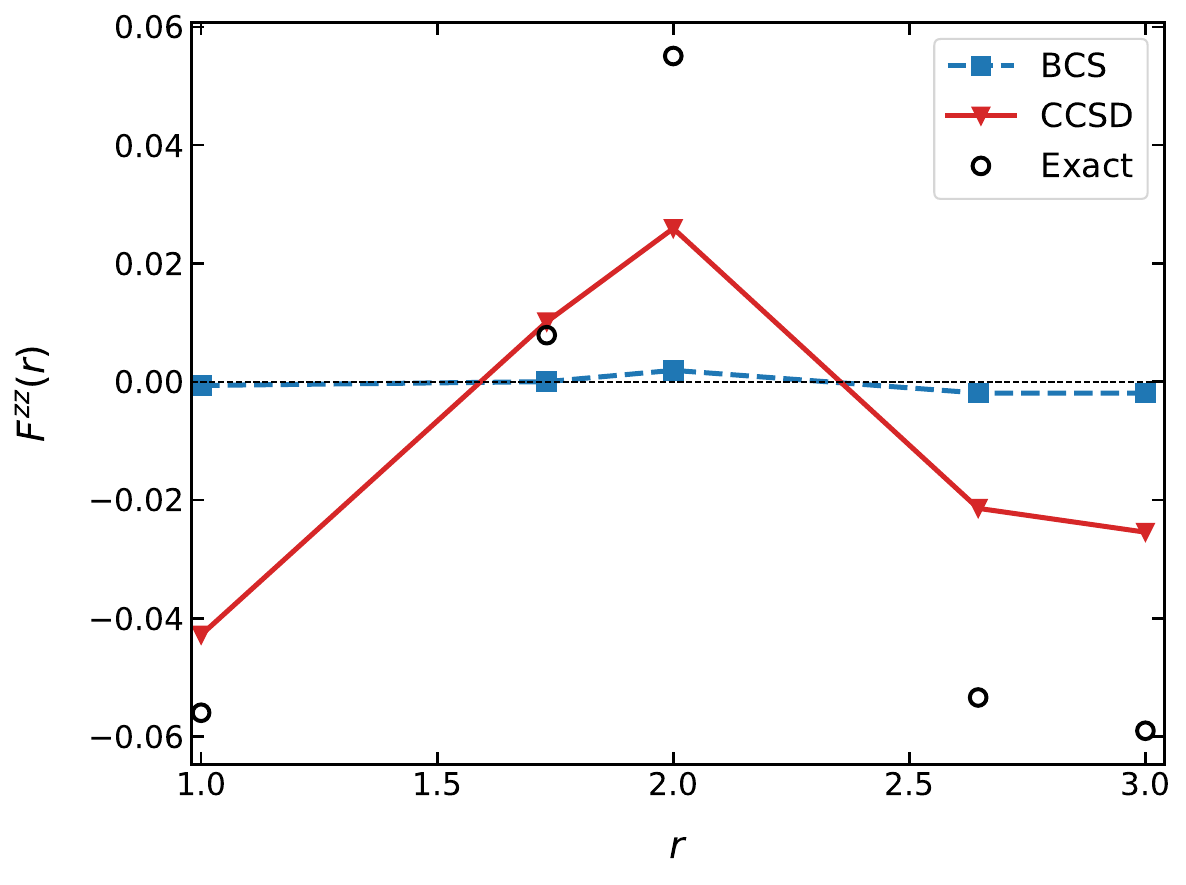}
         
        \caption{}
    \end{subfigure}

    \caption{
    Spin-spin correlation functions for the 18-site triangular lattice
    $J_1-J_2$ model with periodic boundary conditions.
    The left and right columns show $F(r)$ and $F^{zz}(r)$, respectively,
    while the first, second, and third rows correspond to
    $J_2/J_1=0.0$, $0.1$, and $0.2$, respectively.
    }
    \label{fig:triangular_j1j2_2d_corr}
\end{figure*}

\begin{figure}[htbp]
    \centering
    \includegraphics[width=\columnwidth]{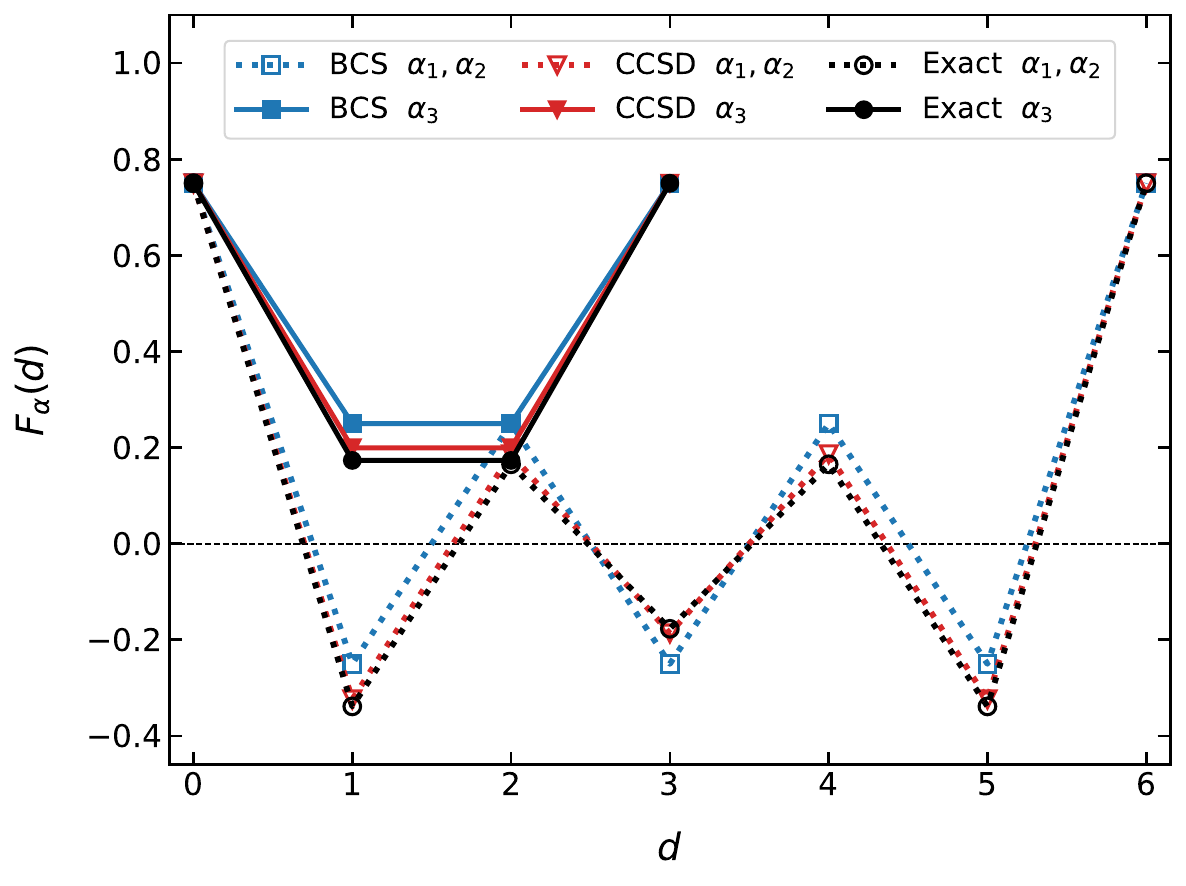}
    \caption{Direction-resolved spin-spin correlation $F_{\alpha}(d)$ along the three lattice directions for the 18-site triangular $J_1$--$J_2$ model with periodic boundary conditions at $J_2/J_1=0.2$.}
    \label{fig:tri_j1j2_direction}
\end{figure}

Figure~\ref{fig:triangular_j1j2_2d_corr} shows the spin-spin correlation functions for the 18-site triangular lattice. For $J_2/J_1 =0$ and $0.1$, the optimized BCS retains the trimodal $120^{\circ}$ structure, although at $J_2/J_1=0.1$, the CCSD $\langle S^z_iS^z_j\rangle$ acquires small imaginary components that cancel upon radial averaging. At $J_2/J_1=0.2$, the optimized BCS reference and CCSD exhibit a collinear stripe pattern. The six nearest-neighbor directions of the triangular lattice form three opposite pairs, corresponding to three bond orientations $\alpha=1,2,3$. As shown by the direction-resolved correlations $F_{\alpha}(d)$ in Fig.~\ref{fig:tri_j1j2_direction}, two directions exhibit antiferromagnetic correlations while the third is ferromagnetic, consistent with the stripe pattern illustrated in Fig.~\ref{fig:tri_stripe}. As in the square lattice, the scalar correlations $F(r)$ are generally well reproduced by CCSD, while deviations remain in $F^{zz}(r)$ at several separations.

\section{Discussion}
The Bogoliubov coupled-cluster method provides a reasonable description of the $\mathfrak{su}(2)$ Hamiltonians studied in this work. By exploiting the equivalence between the spin-$1/2$ algebra and the seniority-zero pairing algebra, the same formal structure can be used for the pairing Hamiltonian and spin models such as XXZ and $\mathrm{J_1-J_2}$. For the pairing Hamiltonian, the improvement of BCS over Hartree-Fock illustrates the usefulness of a symmetry-broken reference for describing static correlation. Our BCS-based CC method indeed yields improved energies. Above $G_c$, the BCS reference breaks particle-number symmetry, odd excitations become active, and the different CC truncations separate. The low-order CCD and CCSD methods undercorrelate in the crossover region, whereas CCSDT and CCSDTQ remove this dip but slightly overcorrelate.

The response properties reveal a more stringent test of the method. Although the particle-number fluctuation is strongly suppressed by CC, especially after orbital relaxation and at the CCSDT/CCSDTQ levels, the pairing parameter exposes limitations of the low-order relaxed density matrices. Because $\bar{H}=e^{-T}He^T$ is non-Hermitian, the CC left state defined by the $Z$ vector is not the Hermitian conjugate of the right-hand CC state. In the intermediate region near $G_c$, the CCD and CCSD left states are not accurate enough, leading to artifacts in the gap fraction. The smoother CCSDT and CCSDTQ curves indicate that higher-rank excitations improve not only the energy but also the quality of the CC bra state and response RDMs.

For the XXZ model, BCS-based CC improves upon the
broken-symmetry and projected mean-field references. On the square lattice, the method gives good agreement with the exact finite-lattice benchmarks, particularly for spin-spin correlations. On the triangular lattice,
the method can also be applied to complex, noncollinear BCS references. As discussed in Sec.~\ref{sec:Triangular_lattice}, the $T_1$ contribution vanishes in
the extreme-trimodal regime,
whereas beyond $\Delta>1$ the optimized reference becomes non-extreme and the singles become active.
For the frustrated $J_1$--$J_2$ model, CC substantially improves the energy relative to BCS and AGP, although the spin--spin correlations near the strongly frustrated region of the square lattice remain difficult. On the triangular lattice, the optimized BCS reference
changes from an extreme-trimodal three-sublattice state at small $J_2/J_1$
to a bimodal stripe state at larger $J_2/J_1$, illustrating how the symmetry
and the structure of the reference adapts to competing magnetic orders.

Overall, these results show that Bogoliubov CC is a promising wave-function-based approach for $\mathfrak{su}(2)$ Hamiltonians, including geometrically frustrated systems requiring complex and noncollinear BCS references. The results also emphasize that the quality and symmetry of the reference strongly influence the structure and performance of the CC hierarchy, and that accurate response properties require a balanced description of both the right and left CC states.  

\section*{Supplementary material}
See the supplementary material for the order of the phase transition in the BCS mean-field and the CC wave function for the pairing Hamiltonian. The supercell vectors for the triangular lattice are also defined in the supplementary material.

\section*{Data Availability}
The data that support the findings of this study are available from the corresponding author upon reasonable request.


\section*{Statements and  Declarations}
\subsection*{Conflict of interest}
The authors have no conflicts to disclose.
\subsection*{Author contribution}
\textbf{S. Ghosh}: Conceptualization (equal); Data curation (lead);
Formal analysis (lead); Investigation (lead); Methodology
(equal); Software (lead); Validation (lead); Visualization
(lead); Writing -- original draft (lead); Writing -- review
\& editing (equal).

\textbf{T. M. Henderson}: Conceptualization (equal); Formal analysis
(equal); Methodology (equal); Supervision (equal); Validation
(equal); Writing -- review \& editing (equal).

\textbf{G. E. Scuseria}: Conceptualization (equal); Funding acquisition
(lead); Methodology (equal); Project administration (lead);
Resources (lead); Supervision (lead); Writing -- review \&
editing (equal).

\begin{acknowledgments}
This work was supported by the U.S. Department of Energy, Office of Basic Energy Sciences, under Award DE-SC0001474.  G.E.S. is a Welch Foundation Chair (C-0036).
\end{acknowledgments} 
\newpage
\appendix

\section{Hamiltonian matrix elements}

Here we present the quasiparticle integrals associated with the Hamiltonian of the form

\begin{equation}
\begin{split}
     H = E_0 + \sum_{p} H^{010}_p{\mathcal{N}}_p + \sum_{p}\big( H^{100}_{p} {\mathcal{P}}^{\dagger}_p + H^{001}_p {\mathcal{P}}_p\big)   \\
+ \sum_{pq}H^{020}_{pq} {\mathcal{N}}_p{\mathcal{N}}_q + \sum_{pq} {H}^{101}_{pq} {\mathcal{P}}^{\dagger}_p{\mathcal{P}}_q \\+
\sum_{pq} \big(H^{200}_{pq} {\mathcal{P}}^{\dagger}_p{\mathcal{P}}^{\dagger}_q + H^{002}_{pq} {\mathcal{P}}_p{\mathcal{P}}_q\big)
\\+\sum_{pq} \big(H^{110}_{pq}{\mathcal{P}}^{\dagger}_{p} {\mathcal{N}}_{q} + H^{011}_{pq} {\mathcal{N}}_{p}{\mathcal{P}}_q \big),
\end{split}
\end{equation}
which is obtained after transforming the Hamiltonian in Eq.~\eqref{Pairing_Ham} to the quasiparticle basis according to the BCS transformation of Eq. \eqref{BCS_transformation},  
\begin{align}
P^{\dagger}_p &=  u_p \, v_p \, \left(1 - \mathcal{N}_p\right) + u_p^{2} \, \mathcal{P}_p^{\dagger} - v^{2}_p \, \mathcal{P}_p,
\\
N_p &= 2 \, v_p^{*} \, v_p + \big(u_p^{*} \, u_p - v_p^{*} \, v_p\big) \, \mathcal{N}_p + 2 \, u_p \, v_p^{*} \, \mathcal{P}^{\dagger}_p + 2 \, u_p^{*} \, v_p \, \mathcal{P}_p,
\end{align}

we obtain
\begin{widetext}
\begin{subequations}
    \begin{align}
        E_0 &= \sum_{pq} \Big(V_{pq} u_{q}^{*} v_{q}^{*} u_pv_{p} + 4W_{pq}v^{*}_pv_q^{*} v_{p} v_{q}\Big)   + \sum_{p} \Big(V_{pp} v_{p}^2v^{*2}_p + 4u_p^{*}v_p^{*}W_{pp} u_{p} v_{p} + 2 h_{p} v^{*}_p v_{p}\Big)\\
        H^{001}_{p}&=  \sum_{q} \Big(u_{p}^{*2} V_{qp} u_{q} v_{q} + 8 u_{p}^{*} v_{q}^{*} W_{qp} v_{p} v_{q} - u_{q}^{*} v_{q}^{*} V_{pq} v_{p}^{2}\Big)   + \Big(4 u_{p}^{*2} W_{pp} u_{p} v_{p} + 2 u_{p}^{*} v_{p}^{*} V_{pp} v_{p}^{2} - 4 u_{p}^{*} v_{p}^{*} W_{pp} v_{p}^{2} + 2 u_{p}^{*} h_{p} v_{p}\Big)  \\
        H^{100}_p &=  \sum_{q} \Big(u_{q} v_{q} V_{pq} u_{p}^{2} - v_{p}^{*2} V_{qp} u_{q} v_{q} + 8 v_{p}^{*} v_{q}^{*} W_{qp} u_{p} v_{q}\Big)   + \Big(4 u_{p}^{*} v_{p}^{*} W_{pp} u_{p}^{2} + 2 v_{p}^{*2} V_{pp} u_{p} v_{p} - 4 v_{p}^{*2} W_{pp} u_{p} v_{p} + 2 v_{p}^{*} h_{p} u_{p}\Big) \\
        H^{200}_{pq} &=\Big(-V_{pq} u_{p}^{2} v_{q}^{*2} + 4W_{pq} u_{p} u_{q} v_{p} ^{*}v_{q}^{*}\Big)\\
        H^{010}_p &=- \sum_{q} \Big(u_{p}^{*}v_{p}^{*}V_{qp} u_{q} v_{q} - 4 u_{p}^{*} v_{q}^{*} W_{qp} u_{p} v_{q} + u_{q}^{*} v_{q}^{*}V_{pq} u_{p} v_{p} + 4 v_{p}^{*} v_{q}^{*} W_{qp} v_{p} v_{q}\Big)\\ &- \Big(4 u_{p}^{*} v_{p}^{*} W_{p,p} u_{p} v_{p} - u_{p}^{*} h_{p} u_{p} + v_{p}^{*2} V_{p,p} v_{p}^{2} + v_{p}^{*} h_{p} v_{p}\Big)  \\
        H^{002}_{pq}&= \Big(-V_{pq} u_{q}^{*2} v_{p}^{2} + 4W_{pq} u_{p}^{*} u_{q}^{*} v_{p} v_{q}\Big)\\
        H^{020}_{pq}&= \Big(u_{p}^{*} u_{q}^{*} W_{pq} u_{p} u_{q} - 2 u_{p}^{*} v_{q}^{*} W_{pq} u_{p} v_{q} + u_{q}^{*} v_{q}^{*} V_{pq} u_{p} v_{p} + v_{p}^{*} v_{q}^{*} W_{pq} v_{p} v_{q}\Big) \\
        H^{101}_{pq}&= \Big(u_{q}^{*2} V_{pq} u_{p}^{2} + 8 u_{q}^{*} v_{p}^{*} W_{pq} u_{p} v_{q} + v_{p}^{*2} V_{qp} v_{q}^{2}\Big) \\
        H^{110}_{pq} &=\Big(4 u_{q}^{*} v_{p}^{*} W_{pq} u_{p} u_{q} - u_{q}^{*} v_{q}^{*} V_{pq} u_{p}^{2} + v_{p}^{*2} V_{qp} u_{q} v_{q} - 4 v_{p}^{*} v_{q}^{*} W_{pq} u_{p} v_{q}\Big)  \\
        H^{011}_{pq} &= \Big(4 u_{p}^{*} u_{q}^{*} W_{pq} u_{p} v_{q} + u_{p}^{*} v_{p}^{*} V_{qp} v_{q}^{2} - u_{q}^{*2} V_{pq} u_{p} v_{p} - 4 u_{q}^{*} v_{p}^{*} W_{pq} v_{p} v_{q}\Big) 
    \end{align}
\end{subequations}
\end{widetext}
We assume that the matrix elements $W_{pq}$ are symmetric.

\section{Parity symmetry in extreme bimodal BCS}\label{parity_symmetry}
Here we show that odd-rank coupled-cluster (CC) excitations do not contribute when the reference is an extreme-bimodal BCS state. The proof is symmetry-based and relies on the global $x$-parity operator
\begin{equation}
\label{eq:ref_parity}
P_x = \prod_{i=1}^{L} 2S_i^x,
\qquad
P_x^2 = \mathbb{I}\,,
\end{equation}
where the eigenvalues are $\pm 1$. We work in the spin representation, since this extreme bimodal structure is more common for spin systems than for pairing Hamiltonians.

For the proof, we consider the spin-$\tfrac12$ XXZ Hamiltonian
\begin{equation}
H
=
\sum_{\langle ij\rangle}
\left(
S_i^x S_j^x + S_i^y S_j^y + \Delta S_i^z S_j^z
\right),
\label{eq:xxz_app}
\end{equation}
where the sum may correspond to either open or periodic boundary conditions. The action of $P_x$ on the spins is
 \begin{equation}
P_x S_i^x P_x = S_i^x,
\qquad
P_x S_i^y P_x = -S_i^y,
\qquad
P_x S_i^z P_x = -S_i^z.
\end{equation}
Consequently, each bond term in Eq.~\eqref{eq:xxz_app} is invariant under conjugation by $P_x$, so that
\begin{equation}
P_x H P_x = H \Longrightarrow [P_x,H]=0.
\end{equation}
The same argument applies to any Hamiltonian built from bilinears $S_i^\alpha S_j^\alpha$, such as the $J_1$--${J_2}$ Heisenberg model.

\vspace{0.5em}
Introduce the local eigenstates of $S_i^x$,
\begin{equation}
|\rightarrow\rangle =
\frac{1}{\sqrt{2}}
\left(|\uparrow\rangle+|\downarrow\rangle\right),
\qquad
|\leftarrow\rangle =
\frac{1}{\sqrt{2}}
\left(|\uparrow\rangle-|\downarrow\rangle\right),
\end{equation}
which satisfy
\begin{equation}
2S_i^x|\rightarrow\rangle = |\rightarrow\rangle,
\qquad
2S_i^x|\leftarrow\rangle = -|\leftarrow\rangle.
\end{equation}
An extreme-bimodal BCS reference is a direct product of these local
states,
\begin{equation}
|\Phi_0\rangle =
\bigotimes_{i=1}^{L} |\phi_i\rangle,
\qquad
|\phi_i\rangle\in
\{|\rightarrow\rangle,|\leftarrow\rangle\}.
\end{equation}
Since every local state is an eigenstate of $2S_i^x$, the reference is an
eigenstate of the global parity operator,
\begin{equation}
    P_x|\Phi_0\rangle = p|\Phi_0\rangle,
    \qquad p=\pm1.
\end{equation}

Define the local excitation operator
\begin{equation}
    f_i^\dagger =
    |\bar{\phi}_i\rangle\langle\phi_i|,
\end{equation}
where $|\bar{\phi}_i\rangle$ is the local state with the opposite
$x$-parity. It follows that
\begin{equation}
    P_x f_i^\dagger P_x = -f_i^\dagger .
\end{equation}
Thus, a single local excitation is odd under $x$ parity.

For a cluster excitation on a set of sites $I$, define
\begin{equation}
C_I^\dagger = \prod_{i\in I} f_i^\dagger.
\end{equation}
Thus
\begin{equation}
P_x C_I^\dagger P_x
=
\prod_{i\in I} \left(P_x f_i^\dagger P_x\right)
=
(-1)^{|I|} C_I^\dagger,
\label{eq:cluster_parity}
\end{equation}
where $|I|$ is the number of local flips in the cluster. Thus, even-rank clusters are parity-even and odd-rank clusters are parity-odd.

Consider the parity-adapted CC solution in which the cluster operator is
restricted to the parity-even sector,
\begin{equation}
T = T_2 + T_4 + \cdots .
\end{equation}
We now show that this restriction is self-consistent: all odd
CC residuals vanish identically,
because each term in $T$ has even rank, Eq.\eqref{eq:cluster_parity} gives
\begin{equation}
[P_x,T]=0.
\end{equation}
It follows immediately that
\begin{equation}
[P_x,e^{\pm T}] = 0.
\end{equation}
Therefore, the similarity-transformed Hamiltonian
\begin{equation}
\bar{H} = e^{-T} H e^T
\end{equation}
also commutes with $P_x$:
\begin{equation}
[P_x,\bar{H}] = 0.
\label{eq:hbar_commutes}
\end{equation}

Consider now the projected CC residual associated with an odd cluster $I$,
\begin{equation}
R_I = \langle \Phi_0 | C_I \bar{H} | \Phi_0 \rangle,
\qquad |I| \text{ odd}.
\end{equation}
Insert $P_x^2=\mathbb{I}$ and use Eqs.~\eqref{eq:ref_parity}, \eqref{eq:cluster_parity}, and \eqref{eq:hbar_commutes}:
\begin{align}
R_I
&=
\langle \Phi_0 |
P_x
\left(P_x C_I P_x\right)
\left(P_x \bar{H} P_x\right)
P_x|\Phi_0\rangle
\nonumber\\
&= p^2 (-1)^{|I|}
\langle \Phi_0 | C_I \bar{H} | \Phi_0 \rangle
\nonumber\\
&=
- R_I,
\end{align}

because $P_x C_I P_x = (-1)^{|I|} C_I$ , $p^2=1$ and $|I|$ is odd. Hence,
\begin{equation}
R_I = 0.
\end{equation}

We conclude that, for an extreme-bimodal BCS reference, all odd-rank projected CC residual equations vanish identically in the parity-adapted sector. Consequently, the odd excitations do not contribute, and the cluster operator may be restricted to even ranks only without loss of generality.

\section{Orbital-relaxed reduced-density matrices}
\label{relaxed_RDMs}
In this section, we present the derivation and expressions for orbital-relaxed RDMs. The derivation is written for a generic $\mathfrak{su}(2)$ Hamiltonian expressed in the quasiparticle basis (Eq.\ref{quasiparticle_Hamiltonian}), and therefore applies to both the reduced BCS pairing Hamiltonian and the spin Hamiltonians considered in this work. 

For the mean-field reference used in this work, the parameters $u_p$ and $v_p$ are parametrized as 
\begin{equation}
u_p = \cos\theta_p,
\qquad
v_p = e^{i\phi_p}\sin\theta_p .
\end{equation}
We collect the independent real reference parameters into
\begin{equation}
\mathbf{q}
=
(\theta_1,\ldots,\theta_M,
 \phi_1,\ldots,\phi_M)^T .
\end{equation}
This parametrization enforces the normalization condition $|u_p|^2+|v_p|^2=1$.  The derivatives with respect to the BCS parameters $\theta_p$ and $\phi_p$ are therefore
\begin{subequations}
\begin{align}
\frac{\partial u_p}{\partial\theta_p}
&=-\sin\theta_p, &
\frac{\partial v_p}{\partial\theta_p}
&=e^{i\phi_p}\cos\theta_p, \\
\frac{\partial v_p^*}{\partial\theta_p}
&=e^{-i\phi_p}\cos\theta_p, &
\frac{\partial v_p}{\partial\phi_p}
&=i v_p, \\
\frac{\partial v_p^*}{\partial\phi_p}
&=-i v_p^* .
\end{align}
\label{eq:theta_chain_rule}
\end{subequations}

The Hamiltonian in the quasiparticle basis depends on the reference parameters \textbf{q} through the quasiparticle Hamiltonian tensors
$H(\textbf{q})
    =\left\{H^{abc}\right\}$
where the superscript $(abc)$ denotes the operator structure associated with the corresponding Hamiltonian tensor: $a$ counts the number of quasipair-creation operators $\mathcal{P}^\dagger$, $b$ counts the number of quasinumber operators $\mathcal{N}$, and $c$ counts the number of quasipair-annihilation operators $\mathcal{P}$. 
For example, $H^{101}_{pq}$ multiplies $\mathcal{P}_p^\dagger \mathcal{P}_q$, while $H^{020}_{pq}$ multiplies $\mathcal{N}_p\mathcal{N}_q$.

Thus, derivatives of the quasiparticle Hamiltonian tensors with respect to $q_a$ can be obtained by applying Eq.~\eqref{eq:theta_chain_rule} to the expressions for the tensors in Eq.~\eqref{quasiparticle_Hamiltonian}.

We define the orbital-gradient vector as the derivative of the stationary CC functional $\mathcal{E}$ with respect to the reference parameters $\mathbf{q}$ at the converged $T$ and $Z$ amplitudes:
\begin{equation}
    g_a
    =
    \left.
    \frac{\partial \mathcal{E}}{\partial q_a}
    \right|_{T,Z}
    =
    \left\langle 0 \left|
    (1+Z)e^{-T}
    \frac{\partial H}{\partial q_a}
    e^T
    \right|0\right\rangle ,
    \label{eq:g_q}
\end{equation}
where $q_a$ denotes either an amplitude parameter $\theta_p$
or a phase parameter $\phi_p$.

Because the CC functional $\mathcal{E}$ is linear in the Hamiltonian tensors, the gradient $g_a$ can be written as a contraction of tensor derivatives with the corresponding response RDMs:
\begin{align}
    g_a
    &=
    \frac{\partial E_0}{\partial q_a}
    +
    \sum_p
    \tilde{\gamma}^{010}_p
    \frac{\partial H^{010}_p}{\partial q_a}
    +
    \sum_p
    \tilde{\gamma}^{100}_p
    \frac{\partial H^{100}_p}{\partial q_a}
    +
    \sum_p
    \tilde{\gamma}^{001}_p
    \frac{\partial H^{001}_p}{\partial q_a}
    \nonumber \\
    &\quad
    + \sum_{pq}
    \tilde{\gamma}^{020}_{pq}
    \frac{\partial H^{020}_{pq}}{\partial q_a}
    + \sum_{pq}
    \tilde{\gamma}^{101}_{pq}
    \frac{\partial H^{101}_{pq}}{\partial q_a}
    + \sum_{pq}
    \tilde{\gamma}^{200}_{pq}
    \frac{\partial H^{200}_{pq}}{\partial q_a}
    \\&+ \sum_{pq}
    \tilde{\gamma}^{002}_{pq}
    \frac{\partial H^{002}_{pq}}{\partial q_a}
    \nonumber 
    +\sum_{pq}
    \tilde{\gamma}^{110}_{pq}
    \frac{\partial H^{110}_{pq}}{\partial q_a}
    +
    \sum_{pq}
    \tilde{\gamma}^{011}_{pq}
    \frac{\partial H^{011}_{pq}}{\partial q_a}.
    \label{eq:g_theta_tensor}
\end{align}
where $\tilde{\gamma}$ are the quasiparticle response RDMs:
\begin{align}
    \tilde{\gamma}^{010}_p
    &=
    \langle 0 |(1+Z)e^{-T}\mathcal{N}_p e^T|0\rangle ,
    \\
    \tilde{\gamma}^{100}_p
    &=
    \langle 0 |(1+Z)e^{-T}\mathcal{P}_p^\dagger e^T|0\rangle ,
    \\\tilde{\gamma}^{001}_p
    &=
    \langle 0 |(1+Z)e^{-T}\mathcal{P}_p e^T|0\rangle ,
    \\
    \tilde{\gamma}^{020}_{pq}
    &=\langle 0 |(1+Z)e^{-T}\mathcal{N}_p\mathcal{N}_q e^T|0\rangle ,
    \\\tilde{\gamma}^{101}_{pq}
    &=\langle 0 |(1+Z)e^{-T}\mathcal{P}_p^\dagger\mathcal{P}_q e^T|0\rangle ,
    \\
    \tilde{\gamma}^{200}_{pq}
    &=\langle 0 |(1+Z)e^{-T}\mathcal{P}_p^\dagger\mathcal{P}_q^\dagger e^T|0\rangle ,
    \\
    \tilde{\gamma}^{002}_{pq}
    &= \langle 0 |(1+Z)e^{-T}\mathcal{P}_p\mathcal{P}_q e^T|0\rangle ,
    \\
    \tilde{\gamma}^{110}_{pq}
    &=\langle 0 |(1+Z)e^{-T}\mathcal{P}_p^\dagger\mathcal{N}_q e^T|0\rangle ,
    \\\tilde{\gamma}^{011}_{pq}
    &=\langle 0 |(1+Z)e^{-T}\mathcal{N}_p\mathcal{P}_q e^T|0\rangle .
\end{align}

We now derive the reference-response term
$\frac{\partial C_a}{\partial x}$ in
Eq.~\eqref{response_equation}. In the present parametrization,
the reference parameters $\mathbf{C}$ are collected in the vector
\begin{equation}
    \mathbf{q}
    =
    (\boldsymbol{\theta},\boldsymbol{\phi}),
\end{equation}
where $\boldsymbol{\theta}$ and $\boldsymbol{\phi}$ denote the
amplitude and phase parameters of the BCS reference, respectively.
We define the first-order response of the reference parameters as
\begin{equation}
    q_a^{O}
    =
    \left.
    \frac{d q_a}{d x}
    \right|_{x=0}.
\end{equation}
The superscript $O$ indicates that the response depends on the
perturbing operator $O$.

The orbital-relaxation contribution to the corresponding property is
then
\begin{equation}
    O_{\mathrm{orb}}
    =
    \sum_a g_a q_a^{O},
\end{equation}
or, in matrix form,
\begin{equation}
    O_{\mathrm{orb}}
    =
    \mathbf{g}^{T}\mathbf{q}^{O}.
\end{equation}
Equivalently,
\begin{equation}
    O_{\mathrm{orb}}
    =
    \sum_r
    \left(
    g_r^{\theta}\theta_r^{O}
    +
    g_r^{\phi}\phi_r^{O}
    \right).
\end{equation}

The total derivative of the stationary CC functional is therefore
\begin{equation}
    \left.
    \frac{d\mathcal{E}}{dx}
    \right|_{x=0}
    =
    \left.
    \frac{\partial\mathcal{E}}{\partial x}
    \right|_{T,Z,\mathbf{q}}
    +
    \mathbf{g}^{T}\mathbf{q}^{O},
\end{equation}
which is the sum of the unrelaxed response and orbital relaxation
contributions. Hence, the relaxed RDM elements are given by
\begin{equation}
    \left(\gamma_p\right)_{\mathrm{rel}}
    =
    \left(\gamma_p\right)_{\mathrm{res}}
    +
    \mathbf{g}^{T}\mathbf{q}^{O},
\end{equation}
where the subscript ``res'' denotes the unrelaxed response RDM
defined in Eq.~\eqref{Response_RDMS}.

The remaining step is to determine $\mathbf{q}^{O}$, the response of
the BCS parameters to the perturbation. The BCS reference is optimized
subject to the constraint that the average particle number remains fixed,
\begin{equation}
N_{\mathrm{BCS}}(\mathbf q)
=
\langle 0(\mathbf q)|\hat N|0(\mathbf q)\rangle
=
N_0 .
\end{equation}
For the parametrization introduced above,
\begin{equation}
N_{\mathrm{BCS}}(\mathbf q)
=
2\sum_p |v_p|^2
=
2\sum_p \sin^2\theta_p ,
\end{equation}
and is therefore independent of the phase parameters $\phi_p$.

For a perturbed Hamiltonian
\begin{equation}
H(x)=H_0+xO,
\end{equation}
we define the constrained BCS Lagrangian
\begin{equation}
\mathcal{L}_{\mathrm{BCS}}
(\mathbf q,\lambda,x)
=
E_{\mathrm{BCS}}(\mathbf q,x)
-
\lambda
\left[
N_{\mathrm{BCS}}(\mathbf q)-N_0
\right],
\end{equation}
where $\lambda$ is the Lagrange multiplier associated with the
particle-number constraint.

The optimized reference satisfies the coupled conditions
\begin{align}
F_a(\mathbf q,\lambda,x)
&=
\frac{\partial\mathcal{L}_{\mathrm{BCS}}}
{\partial q_a}
=0,
\\
N_{\mathrm{BCS}}(\mathbf q)-N_0&=0.
\end{align}

We define the first-order responses
\begin{equation}
q_a^O
=
\left.
\frac{dq_a}{dx}
\right|_{x=0},
\qquad
\lambda^O
=
\left.
\frac{d\lambda}{dx}
\right|_{x=0}.
\end{equation}
Differentiating the stationarity conditions with respect to $x$ gives
\begin{equation}
\sum_b A^{\mathcal L}_{ab}q_b^O
-
n_a\,\lambda^O
=
-B_a^O,
\end{equation}
together with the differentiated number constraint
\begin{equation}
\sum_a n_a q_a^O=0.
\end{equation}
Here,
\begin{align}
A^{\mathcal L}_{ab}
&=
\left.
\frac{\partial^2\mathcal{L}_{\mathrm{BCS}}}
{\partial q_a\partial q_b}
\right|_{x=0},
\\
n_a
&=
\frac{\partial N_{\mathrm{BCS}}}{\partial q_a},
\\
B_a^O
&=
\left.
\frac{\partial O_{\mathrm{BCS}}}
{\partial q_a}
\right|_{x=0}.
\end{align}

Thus, the BCS response is obtained from the bordered linear system
\begin{equation}
\begin{pmatrix}
\mathbf A^{\mathcal L} & -\mathbf n \\
\mathbf n^{T} & 0
\end{pmatrix}
\begin{pmatrix}
\mathbf q^{O} \\
\lambda^{O}
\end{pmatrix}
=
-
\begin{pmatrix}
\mathbf B^{O} \\
0
\end{pmatrix}.
\end{equation}

Since $N_{\mathrm{BCS}}$ is independent of the phase parameters,
the components of the constraint gradient are
\begin{align}
\frac{\partial N_{\mathrm{BCS}}}{\partial\theta_r}
&=
4\sin\theta_r\cos\theta_r,
\\
\frac{\partial N_{\mathrm{BCS}}}{\partial\phi_r}
&=0.
\end{align}
Thus,
\begin{equation}
\mathbf n
=
\left(
4\sin\theta_1\cos\theta_1,\ldots,
4\sin\theta_M\cos\theta_M,
0,\ldots,0
\right)^T.
\end{equation}

The nonzero second derivatives of the number constraint are
\begin{equation}
\frac{\partial^2N_{\mathrm{BCS}}}
{\partial\theta_r\partial\theta_s}
=
4\left(
\cos^2\theta_r-\sin^2\theta_r
\right)\delta_{rs},
\end{equation}

Consequently, the constrained Hessian is
\begin{equation}
\mathbf A^{\mathcal L}
=
\mathbf A^{\mathrm{raw}}
-
\lambda\,
\nabla_{\mathbf q}^{2}N_{\mathrm{BCS}},
\end{equation}
where
\begin{equation}
A^{\mathrm{raw}}_{ab}
=
\frac{\partial^2E_{\mathrm{BCS}}}
{\partial q_a\partial q_b}.
\end{equation}

Solving the bordered system gives the orbital response
$\mathbf q^{O}$ while maintaining
$N_{\mathrm{BCS}}=N_0$ to first order in the perturbation. The
orbital-relaxation contribution to the coupled-cluster property is then
evaluated from $\mathbf g^T\mathbf q^O$.

For symmetry-broken references, continuous symmetries may give rise to zero modes of $\mathbf{A}^{\mathcal L}$. These modes are identified
by a singular-value decomposition and the corresponding null-space directions are removed from the response equations. Provided that the perturbation has no component along the null space, the response is obtained in the complementary nonsingular subspace. For complex references, a uniform shift of all phase parameters corresponds to a global $\mathfrak{u}(1)$ zero mode.

The reported relaxed pairing properties were obtained using the constrained relaxed-response formulation of Eq.~(C34). The constrained coupled-perturbed BCS equations were validated against direct finite differences in which the BCS reference was reoptimized at fixed average particle number for each perturbed Hamiltonian, with agreement to approximately \(10^{-9}\) in representative calculations.

\bibliographystyle{aipnum4-1}
\bibliography{sbcscc_ref}

\end{document}